\documentclass[a4paper,11pt]{article}
\usepackage[english]{babel}
\usepackage{jheppub}
\pdfoutput=1
\usepackage[T1]{fontenc} 
\usepackage{bbold}
\usepackage{amsmath}
\usepackage{empheq}
\usepackage{booktabs}
\usepackage{color}
\usepackage[utf8]{inputenc}
\usepackage{xspace}
\usepackage{scalerel}
\usepackage[most]{tcolorbox}
\usepackage{multirow}
\usepackage{scalefnt}
\usepackage{bold-extra}
\usepackage[tikz]{bclogo}
\usepackage{subcaption}
\usepackage{tikz-feynman}
\usepackage{cancel}
\usepackage{verbatim}
\usetikzlibrary{arrows,shapes}
\usepackage{afterpage}

\newcommand\F{${\rm F}$}
\newcommand\FJ{${\rm FJ}$}
\newcommand\FJJ{${\rm FJJ}$}
\newcommand\PhiBorn{\Phi_{\scriptscriptstyle \rm B}}
\newcommand\PhiReal{\Phi_{\scriptscriptstyle \rm R}}
\newcommand\PhiB{\Phi_{\scriptscriptstyle \rm F}}

\newcommand{\flav}{\ell}
\newcommand{\flavBorn}{\flav_{\scriptscriptstyle \rm B}}
\newcommand{\fullflavBorn}{\hat \flav_{\scriptscriptstyle \rm B}}

\newcommand{\fullflavprimeBorn}{\hat \flav'_{\scriptscriptstyle \rm B}}
\newcommand{\flavB}{\flav_{\scriptscriptstyle \rm F}}

\newcommand{\flavBJ}{\flav_{\scriptscriptstyle \rm FJ}}

\newcommand{\projflav}{\flavB\leftarrow\flavBJ}
\newcommand{\CF}{C_{\mathrm{F}}}
\newcommand{\CA}{C_{\mathrm{A}}}

\newcommand{\nf}{N_{\mathrm{f}}}

\newcommand{\flavZg}{\flav_{\scriptscriptstyle Z\gamma}}
\newcommand{\fullflavZg}{\hat \flav_{\scriptscriptstyle Z\gamma}}
\newcommand{\flavZgJ}{\flav_{\scriptscriptstyle Z\gamma {\rm J}}}
\newcommand{\fullflavZgJ}{\hat \flav_{\scriptscriptstyle Z\gamma {\rm J}}}

\newcommand{\projflavZg}{\flavZg\leftarrow\flavZgJ}

\newcommand\PhiBJ{\Phi_{\scriptscriptstyle \rm FJ}}

\newcommand\PhiZJ{\Phi_{\scriptscriptstyle \rm Z\gamma J}}

\newcommand\PhiZgam{\Phi_{\scriptscriptstyle \rm Z\gamma}}

\newcommand{\as}{\alpha_s}

\newcommand{\pt}{{p_{\text{\scalefont{0.77}T}}}}
\newcommand{\GZ}{{\Gamma_{\text{\scalefont{0.77}Z}}}}
\newcommand{\GW}{{\Gamma_{\text{\scalefont{0.77}W}}}}
\newcommand{\thW}{{\theta_{\text{\scalefont{0.77}W}}}}
\newcommand{\mtop}{{m_{\text{\scalefont{0.77}top}}}}
\newcommand{\qt}{{q_{\text{\scalefont{0.77}T}}}}

\newcommand{\ptg}{{p_{\text{\scalefont{0.77}T,$\gamma$}}}}
\newcommand{\ptgcut}{{p_{\text{\scalefont{0.77}T,$\gamma$}}^{\rm cut}}}
\newcommand{\ptjcut}{{p_{\text{\scalefont{0.77}T,$j$}}^{\rm cut}}}

\newcommand{\ptrad}{{p_{\text{\scalefont{0.77}T,rad}}}}

\newcommand{\ptllg}{{p_{\text{\scalefont{0.77}T,}\ell\ell\gamma}}}
\newcommand{\ptzg}{\ptllg}
\newcommand{\ptll}{{p_{\text{\scalefont{0.77}T,}\ell\ell}}}
\newcommand{\ptj}{{p_{\text{\scalefont{0.77}T,$j$}}}}
\newcommand{\ptjone}{{p_{\text{\scalefont{0.77}T,$j_1$}}}}

\newcommand{\ptl}{{p_{\text{\scalefont{0.77}T,$\ell$}}}}
\newcommand{\ptlone}{{p_{\text{\scalefont{0.77}T,$\ell_1$}}}}
\newcommand{\ptltwo}{{p_{\text{\scalefont{0.77}T,$\ell_2$}}}}

\newcommand{\mz}{{m_{\text{\scalefont{0.77}Z}}}}
\newcommand{\mw}{{m_{\text{\scalefont{0.77}W}}}}

\newcommand{\mgj}{{m_{\text{\scalefont{0.77}$\gamma j_1$}}}}
\newcommand{\mllg}{{m_{\text{\scalefont{0.77}$\ell\ell\gamma$}}}}
\newcommand{\mll}{{m_{\text{\scalefont{0.77}$\ell\ell$}}}}
\newcommand{\etal}{{\eta_{\text{\scalefont{0.77}$\ell$}}}}
\newcommand{\etallg}{{\eta_{\text{\scalefont{0.77}$\ell\ell\gamma$}}}}

\newcommand{\etaltwo}{{\eta_{\text{\scalefont{0.77}$\ell_2$}}}}
\newcommand{\etag}{{\eta_{\text{\scalefont{0.77}$\gamma$}}}}
\newcommand{\etaj}{{\eta_{\text{\scalefont{0.77}j}}}}
\newcommand{\detallgj}{{\Delta\eta_{\text{\scalefont{0.77}$\ell\ell\gamma, j_1$}}}}
\newcommand{\dphillg}{{\Delta\phi_{\text{\scalefont{0.77}$\ell\ell,\gamma$}}}}
\newcommand{\drlg}{{\Delta R_{\text{\scalefont{0.77}$\ell \gamma$}}}}
\newcommand{\drgjone}{{\Delta R_{\text{\scalefont{0.77}$\gamma j_1$}}}}
\newcommand{\drgjtwo}{{\Delta R_{\text{\scalefont{0.77}$\gamma j_2$}}}}

\newcommand{\drlj}{{\Delta R_{\text{\scalefont{0.77}$\ell,j$}}}}
\newcommand{\drgj}{{\Delta R_{\text{\scalefont{0.77}$\gamma j$}}}}

\newcommand{\fcl}{{E_{\text{\scalefont{0.77}T}}^{\text{\scalefont{0.77}cone$0.2$}}/p_{\text{\scalefont{0.77}T},\gamma}}}
\newcommand{\muF}{{\mu_{\text{\scalefont{0.77}F}}}}
\newcommand{\muR}{{\mu_{\text{\scalefont{0.77}R}}}}
\newcommand{\muFc}{{\mu_{\text{\scalefont{0.77}F},0}}}
\newcommand{\muRc}{{\mu_{\text{\scalefont{0.77}R},0}}}
\newcommand{\KF}{{K_{\text{\scalefont{0.77}F}}}}
\newcommand{\KR}{{K_{\text{\scalefont{0.77}R}}}}
\newcommand{\Q}{{Q_{\text{\scalefont{0.77}$0$}}}}
\newcommand{\Qc}{{Q_{\text{\scalefont{0.77}res},0}}}

\newcommand{\noun}[1]{{\scshape #1}}

\newcommand{\POWHEG}{\noun{POWHEG}}

\newcommand{\POWHEGBOX}{\noun{POWHEG-BOX}}
\newcommand{\POWHEGBOXRES}{\noun{POWHEG-BOX-RES}}

\newcommand{\minlosimple}{{\noun{MiNLO}}}
\newcommand{\minlo}{{\noun{MiNLO$^{\prime}$}}}
\newcommand{\minnlo}{{\noun{MiNNLO$_{\rm PS}$}}}
\newcommand{\Matrix}{{\noun{Matrix}}}
\newcommand{\PYTHIA}[1]{\noun{Pythia{#1}}}

\newcommand{\nnlops}{NNLO+PS}
\newcommand{\fnnlo}{NNLO}

\newcommand{\setupone}{{\tt ATLAS-setup-1}}
\newcommand{\setuptwo}{{\tt ATLAS-setup-2}}

\newcommand{\abar}{\frac{\as}{2\pi}}
\newcommand{\abarmu}[1]{\frac{\as(#1)}{2\pi}}

\newcommand{\citere}[1]{ref.\,\cite{#1}}
\newcommand{\citerec}[1]{Ref.\,\cite{#1}}
\newcommand{\citeres}[1]{refs.\,\cite{#1}}

\newcommand{\eqn}[1]{eq.\,(\ref{#1})}
\newcommand{\neqn}[1]{eqs.\,(\ref{#1})}
\newcommand{\fig}[1]{figure\,\ref{#1}}

\newcommand{\tab}[1]{table\,\ref{#1}}
\newcommand{\sct}[1]{section~\ref{#1}}

\newcommand{\app}[1]{appendix~\ref{#1}}
\newcommand{\LambdaPWG}{\Lambda_{\rm pwg}}

\usepackage{xcolor}
\newcommand{\mathd}{\mathrm{d}}
\newcommand{\tmop}[1]{\ensuremath{\operatorname{#1}}}

\newtcolorbox{empheqboxed}{colback=white!35, 
 colframe=black,
 width=\textwidth,
 sharpish corners,
 top=-2mm, 
 bottom=0pt
}

\title{{Advancing M{\scalefont{0.77}I}NNLO\boldmath{$_{\text{PS}}$} to diboson processes: \\\boldmath{$Z\gamma$} production at NNLO+PS}}
\author[a]{Daniele Lombardi,}
\author[a]{Marius Wiesemann,}
\author[a]{and Giulia Zanderighi}

\emailAdd{lombardi@mpp.mpg.de}
\emailAdd{marius.wiesemann@cern.ch}
\emailAdd{zanderi@mpp.mpg.de}

\affiliation[a]{Max-Planck-Institut f\"ur Physik, F\"ohringer Ring 6,
  80805 M\"unchen, Germany}

\abstract{We consider $Z \gamma$ production in hadronic collisions and
  present the first computation of next-to-next-to-leading order
  accurate predictions consistently matched to parton showers
  (NNLO+PS).  Spin correlations, interferences and off-shell effects
  are included by calculating the full process
  $pp\to\ell^+\ell^-\gamma$.  We extend the recently developed
  \minnlo{} method to genuine $2\to 2$ hard
  scattering processes at the LHC, which paves the way for NNLO+PS
  simulations of all diboson processes. This is the first $2\to2$
  NNLO+PS calculation that does not require an a-posteriori
  multi-differential reweighting.  We find that both NNLO corrections
  and matching to parton showers are crucial for an accurate
  simulation of the $Z \gamma$ process. Our predictions are in very
  good agreement with recent ATLAS data.  }


\preprint{MPP-2020-188}

\begin{document}

\maketitle

\section{Introduction}
\label{sec:intro}

Lacking clear hints for new-physics phenomena, particle phenomenology
at the Large Hadron Collider (LHC) has entered the precision era.  The
accurate measurement of Standard Model (SM) processes provides a
valuable alternative in the discovery of new-physics phenomena through
small deviations from SM predictions.  Many LHC reactions, in
particular colour-singlet processes, are not only measured, but also
predicted at a remarkable accuracy. For instance the recent $Z\gamma$
\cite{Aad:2019gpq} and $ZZ$ \cite{Sirunyan:2020pub} measurements
include the full Run-2 data and are hitting percent-level
uncertainties even for differential observables. The vast amount of
data collected at the LHC will continuously decrease experimental
uncertainties, thereby demanding accurate theory predictions in many
relevant physics processes.

The theoretical description of fiducial cross sections and kinematic
distributions has been greatly improved by the calculation of NNLO
corrections in QCD perturbation theory. Those have become the standard
for $2\to 1$ and $2\to 2$ colour-singlet processes
\cite{Ferrera:2011bk,Ferrera:2014lca,Ferrera:2017zex,Campbell:2016jau,Harlander:2003ai,Harlander:2010cz,Harlander:2011fx,Buehler:2012cu,Marzani:2008az,Harlander:2009mq,Harlander:2009my,Pak:2009dg,Neumann:2014nha,deFlorian:2013jea,deFlorian:2016uhr,Grazzini:2018bsd,Catani:2011qz,Campbell:2016yrh,Grazzini:2013bna,Grazzini:2015nwa,Campbell:2017aul,Gehrmann:2020oec,Cascioli:2014yka,Grazzini:2015hta,Heinrich:2017bvg,Kallweit:2018nyv,Gehrmann:2014fva,Grazzini:2016ctr,Grazzini:2016swo,Grazzini:2017ckn,Baglio:2012np,Li:2016nrr,deFlorian:2019app}
by now.  The recent NNLO calculation
of~$\gamma\gamma\gamma$~\cite{Chawdhry:2019bji,Kallweit:2020gcp}
production marks another milestone for precision calculations, since
it is the first $2\to 3$ genuine LHC process to be computed at this
level of precision. A comparison of theory predictions to LHC data
also highlights that a knowledge of NNLO corrections is crucial for
theory results to describe data within their experimental
uncertainties.

Vector-boson pair production processes in particular have become an
integral part of the rich precision programme at the LHC.  Being
measured by reconstructing the vector bosons from their leptonic decay
products, those processes offer clean experimental signatures with
rather small experimental uncertainties. Apart from the measurement of
their production rates and distributions, they provide a proxy for
both direct and indirect searches for beyond-the-SM (BSM)
physics. While very precise SM predictions of diboson processes are
not needed to find a light resonance structure in invariant mass
distributions, new-physics effects can give rise also to modifications
and distortions of kinematic distributions. These effects, which can
be parametrized through anomalous triple-gauge couplings or effective
operators, enter vector-boson pair processes already at the leading
order (LO). Constraining new-physics effects in these type of indirect
searches crucially relies on accurate theory predictions for event
rates and shapes of distributions.

$Z\gamma$ production in the $Z\to \ell^+\ell^-$ decay channel provides
a particularly pure experimental signature as the final state can be
fully reconstructed. In combination with its relatively large cross
sections this process is well suited for precision phenomenology.
Indeed, $Z\gamma$ production was measured extensively at the LHC at
7\,TeV~\cite{Chatrchyan:2011rr,Aad:2011tc,Aad:2012mr,Chatrchyan:2013fya,Chatrchyan:2013nda,Aad:2013izg},
8\,TeV~\cite{Aad:2014fha,Khachatryan:2015kea,Khachatryan:2016yro,Aad:2016sau},
and 13\,TeV~\cite{Aaboud:2018jst,Aad:2019gpq}, and
\citere{Aad:2019gpq} was the first diboson analysis to include the
full Run~II data set.  Even small deviations from the production rate
or distributions in this process would be a direct hint of BSM
physics.  So far, full agreement with the SM was found, which provides
a strong test of the gauge structure of electroweak (EW) interactions
and the mechanism of EW symmetry breaking.  On the other hand, the
measurement of a non-zero $ZZ\gamma$ coupling, which is absent in the
SM, would be direct evidence of physics beyond the SM (BSM). Moreover,
$Z\gamma$ final states are relevant in direct searches for BSM
resonances and in Higgs boson measurements, see
e.g.\ \citere{Sirunyan:2018tbk,Aad:2020plj}, with the SM production
being an irreducible background. Although the Higgs decay into a
$Z\gamma$ pair is rare, since it is loop-induced in the SM, effects
from new-physics may significantly enhance this decay channel.

A substantial effort has been made in fixed-order calculations for
$Z\gamma$ production in the past years.  Next-to-leading order (NLO)
QCD corrections were computed some time ago both for on-shell $Z$
bosons \cite{Ohnemus:1992jn} and including their leptonic decays
\cite{Baur:1997kz}.  The first contribution known at
next-to-next-to-leading order (NNLO) QCD was the loop induced
gluon-fusion contribution
\cite{Ametller:1985di,vanderBij:1988fb,Adamson:2002rm}.
\citerec{Campbell:2011bn} combined the NLO cross section, including
photon radiation off the leptons, with the loop-induced gluon fusion
contribution.  The full NNLO QCD cross section for
$\ell^+\ell^-\gamma$ production was calculated in
\citeres{Grazzini:2013bna,Grazzini:2015nwa} at the fully differential
level and it was later confirmed in \citere{Campbell:2017aul} by an
independent calculation. Also the NLO electroweak (EW) corrections are
known \cite{Hollik:2004tm,Accomando:2005ra}.

However, the validity of fixed-order calculations is limited to
observables dominated by hard QCD radiation.  In kinematical regimes
where soft and collinear QCD radiation becomes important, the
perturbative expansion in the strong coupling constant is challenged
by the appearance of large logarithmic contributions.  The analytic
resummation of those logarithms is usually restricted to a single
observable, see e.g.\ the recent next-to-next-to-next-to-logarithmic
(N$^3$LL) results for the $Z\gamma$ transverse momentum ($\ptzg$) in
\citeres{Wiesemann:2020gbm,Becher:2020ugp} for instance, or at most
two observables, such as the joint resummation of logarithms in
$\ptzg$ and in the leading jet transverse momentum ($\ptjone$) at
next-to-next-to-logarithmic (NNLL) \cite{Kallweit:2020gva}.
Parton showers, on the other hand, offer a numerical approach to
include all-order effects in all phase-space observables
simultaneously.
Although their all-order logarithmic accuracy is rather limited (see
\citere{Dasgupta:2020fwr} for a recent discussion on this topic),
parton-shower simulations are extremely important for experimental
analysis since they allow for an exclusive description of the final
state.
Moreover, as measurements operate at the level of hadronic events,
they require full-fledged parton-shower Monte Carlo simulations.
In fact, any BSM analysis searching for small deviations from SM
predictions at event level requires parton-shower predictions that
include the highest possible fixed-order accuracy.  While matching of
NLO QCD predictions and parton showers (NLO+PS) has been worked out a
while ago in seminal papers
\cite{Frixione:2002ik,Nason:2004rx,Frixione:2007vw},\footnote{For the
  processes considered in this work, in \citere{Krause:2017nxq} NLO
  QCD predictions for $Z+\gamma$ + jets with different
  jet-multiplicities have been merged using the MEPS@NLO approach,
  which separates samples using a merging scale, and then interfaced
  to a parton shower.} current experimental measurements demand the
inclusion of NNLO QCD corrections in event generators to fully exploit
LHC data.

So far four different NNLO+PS approaches have been presented in the
literature
\cite{Hamilton:2012rf,Alioli:2013hqa,Hoeche:2014aia,Monni:2019whf,Monni:2020nks},
and all of them are formulated for colour singlet processes only.
The methods of \citeres{Hamilton:2012rf,Monni:2019whf,Monni:2020nks}
originate from the \minlo{}
procedure~\cite{Hamilton:2012np,Hamilton:2012rf}, which uses a
transverse momentum resummation to upgrade an NLO calculation for
$X+1$ jet to become NLO accurate both for $X$ and $X+1$ jet.  A
numerical method which allows to extend the \minlo{} procedure also to
more complex final states was presented in \citere{Frederix:2015fyz},
where in particular the case of Higgs production in association with
up to two jets was worked out.
NNLO+PS approaches have been mostly applied to the simple $2\to1$ LHC
processes, such as Higgs-boson production
\cite{Hamilton:2013fea,Hoche:2014dla,Monni:2019whf,Monni:2020nks}, the
Drell-Yan process
\cite{Hoeche:2014aia,Karlberg:2014qua,Alioli:2015toa,Monni:2019whf,Monni:2020nks},
Higgs-strahlung \cite{Astill:2016hpa,Astill:2018ivh,Alioli:2019qzz},
which in terms of QCD corrections is still a $2\to1$ process, and to
the $1\to 2$ process $H\to b \bar b$ decay~\cite{Bizon:2019tfo}.

Up to now, only the approach of \citere{Hamilton:2012rf}, which relies on
a numerically highly demanding multi-dimensional reweighting in the
Born phase space, has been applied to a genuine $2\to 2$ process,
namely $W^+W^-$ production \cite{Re:2018vac}.\footnote{The $W^+W^-$
  simulation is based on the \minlo{} calculation of
  \citere{Hamilton:2016bfu}, and the NNLO calculation of
  \citere{Grazzini:2016ctr} performed within the \Matrix{}
  framework~\cite{Grazzini:2017mhc}.}  This calculation has taken the
reweighting procedure to its extreme.
In fact, the Born phase space for $W^+W^-\to e^+ \mu^- \nu_e\bar
\nu_\mu$ involves nine variables (after taking the azimuthal symmetry
into account). \citerec{Re:2018vac} had to recur to a number of
features of the $W$-decays, such as the fact that the full angular
dependence of boson decays can be parametrized through eight
Collins-Soper functions, and used the fact that QCD corrections are
largely independent of the off-shellness of the vector bosons to
reduce the number of kinematic variables in the parametrization of the
Born phase space. Still, the residual Born-phase space dependence had
to be discretized. The finite bin sizes used in the reweighting limits
the accuracy of the results in regions of phase space characterized by
coarse binning. This is typically the case in regions close to the
kinematic edges and limits of the phase space, such as tails of
kinematic distributions, which, on the other hand, are particularly
interesting for BSM searches.
Actually, the numerical limitations related to an a-posteriori
reweighting constitutes a problem already for a way simpler process
such as Drell-Yan, since in this case, given the very large amount of
data available, experiments require a high theoretical precision over
the whole phase space.

In this paper, we consider $Z\gamma$ production and present the first
NNLO+PS calculation for a genuine $2\to 2$ process that includes NNLO
corrections directly during event generation, without any
post-processing or reweighting of the events being required.  In fact,
this is also the first time a NNLO $Z\gamma$ calculation 
independent of a slicing cutoff is performed
(cf. \citeres{Grazzini:2013bna,Grazzini:2015nwa,Campbell:2017aul}).
To this end, we have extended the just recently developed \minnlo{}
method \cite{Monni:2019whf} to deal with genuine $2 \to 2$ reactions,
which paves the way for NNLO+PS simulations of all other diboson
processes. As anticipated already in \citere{Monni:2019whf}, \minnlo{}
is a very powerful approach as its underlying idea applies beyond $2
\to 1$ processes and beyond colour-singlet production, with the
following features:
\begin{itemize}
\item NNLO corrections are calculated directly during the generation
  of the events, with no need for further reweighting.
\item No merging scale is required to separate different
  multiplicities in the generated event samples.
\item When combined with transverse-momentum ordered parton showers,
  the matching preserves the leading logarithmic structure of the
  shower simulation.\footnote{We note that, while maintaining the
    logarithmic accuracy of the shower is sometimes taken for granted,
    this is instead a crucial and subtle point in any NNLO+PS
    approach. When \minnlo{} predictions are interfaced to a
    transverse-momentum ordered shower, this requirement is
    automatically met since the two hardest emissions are generated by
    \POWHEG{} and the remaining ones by the parton shower. However, if
    the shower ordering variable differs from the NLO+PS one, then
    maintaining the leading logarithmic accuracy of the shower can
    required the introduction of vetos to the shower radiation and of
    additional contributions, such as truncated
    showers~\cite{Nason:2004rx}.}
\end{itemize}
We consider all topologies leading to the final state
$\ell^+\ell^-\gamma$ in our calculation, including off-shell effects
and spin correlations.  Since, neither a $Z\gamma$ nor a $Z\gamma$+jet
generator was available within the \POWHEGBOX{} framework
\cite{Alioli:2010xd}, we also present new calculations of these two
processes using the \POWHEG{} method~\cite{Nason:2004rx}.  Our
implementation of the $Z\gamma$+jet generator builds the basis for the
inclusion of NNLO QCD corrections to $Z\gamma$ production through the
\minnlo{} method. The ensuing calculation allows us to retain NNLO QCD
accuracy in the event generation and to interface it to a parton
shower, which is a necessary step for a complete and realistic event
simulation.  In particular, multiple photon emissions through the QED
shower, as well non-perturbative QCD effects using hadronization and
underlying event models can be included.  While these corrections have
only very mild effects on fully inclusive quantities, they can have a
substantial impact on jet-binned cross-sections and other more
exclusive observables measured at the LHC.

This manuscript is organized as follows: in \sct{sec:description} we
discuss the implementation of $Z\gamma$ and $Z\gamma$+jet generators
within the \POWHEGBOXRES{} framework, where we introduce the two
processes (\sct{sec:process}), some basics about \POWHEGBOXRES{}
(\sct{sec:powhegboxres}), and give details about the treatment of
photon isolation and its practical implementation
(\sct{sec:photon}). The inclusion of NNLO corrections to $Z\gamma$
production into $Z\gamma$+jet generator is discussed in
\sct{sec:reachingNNLO}, including details on the extraction of the the
two-loop amplitude (\sct{sec:nnlo}), a general discussion on how we
extended the \minnlo{} approach to $2\to2$ processes
(\sct{sec:MiNNLO}), and further practical details relevant for the
specific case of the $Z\gamma$ \minnlo{} generator
(\sct{sec:details}).  In \sct{sec:phenomenology}, after describing
  the setup used in our calculation and the set of fiducial cuts used
  in the analysis (\sct{sec:setup}), we first present some validation
  plots (\sct{sec:cmpnnlo}), compare to the most accurate predictions
  for the transverse momentum distribution of the colourless system
  (\sct{sec:cmpNNLON3LL}), and finally compare our results for
  fiducial cross sections and distributions to ATLAS data
  (\sct{sec:data}). We conclude and summarize in
  \sct{sec:summary}. Some technical aspects are discussed in more
  details in the appendices.

\section{NLO+PS simulation of \boldmath{$Z\gamma$} and \boldmath{$Z\gamma$+jet} production}
\label{sec:description}
In this section, we discuss the implementation of NLO+PS generators
for $Z\gamma$ and $Z\gamma$+jet production in the \POWHEGBOX{}
framework~\cite{Alioli:2010xd}.  Both processes were not yet available
in this framework and we present their first calculation in the
\POWHEG{} \cite{Frixione:2007vw} approach at NLO+PS.  Moreover, the
$Z\gamma$+jet process serves as starting point to reach NNLO accuracy
for $Z\gamma$ production through the \minnlo{} method, as detailed in
\sct{sec:reachingNNLO}.  Since these processes involve an EW
resonance, we exploit the \POWHEGBOXRES{} code~\cite{Jezo:2015aia},
which is specifically designed to deal with intermediate resonances.
In the following, we first introduce the two processes under
consideration, then we recall some relevant features of the
\POWHEGBOXRES{} framework, and finally discuss details regarding the
treatment of the photon in the final state and the QED singularities
associated to it.

\subsection{Description of the processes}
\label{sec:process}

\begin{figure}[t]
  \begin{center}
    \begin{subfigure}[b]{.5\linewidth}
      \centering
\begin{tikzpicture}
  \begin{feynman}
    \vertex (a1) {\( q\)};
    \vertex[below=1.6cm of a1] (a2){\(\overline q\)};
    \vertex[right=2cm of a1] (a3);
    \vertex[right=2cm of a2] (a4);
    \vertex[right=1.65cm of a3] (a5){\(\gamma\)};
    \vertex[right=2cm of a3] (a9);
    \vertex[right=1cm of a4] (a6);
    \vertex[below=0.7cm of a9] (a7){\(\ell^+\)} ;
    \vertex[below=1.7cm of a9] (a8){\(\ell^-\)};
 
    \diagram* {
      {[edges=fermion]
        (a1)--(a3)--(a4)--(a2),
        (a7)--(a6)--(a8),
      },
      (a3) -- [ boson] (a5),
      (a4) -- [boson, edge label'=\(Z/\gamma^*\)] (a6),
       };

  \end{feynman}
\end{tikzpicture}
\caption{$q$-type diagram}
        \label{subfig:qtypeZgam}
\end{subfigure}%
\begin{subfigure}[b]{.5\linewidth}
  \centering
\begin{tikzpicture}
  \begin{feynman}
    \vertex (a1) {\(  q\)};
    \vertex[below=1.6cm of a1] (a2){\(\overline q\)};
    \vertex[below=0.8cm of a1] (a3);
    \vertex[right=1.5cm of a3] (a4);
    \vertex[right=1.5cm of a4] (a5);
    \vertex[right=0.4cm of a5] (a6);
    \vertex[right=1.1cm of a6] (a7);
    \vertex[below=0.8cm of a7](a8){\(\ell^+\)};
    \vertex[above=0.5cm of a6](a9);
    \vertex[above=0cm of a7] (a10){\(\gamma\)}; ;
    \vertex[above=0.6cm of a7] (a11){\(\ell^-\)};
 
    \diagram* {
      {[edges=fermion]
        (a1)--(a4)--(a2),
        (a8)--(a5)--(a9)--(a11),
      },
      (a4) -- [boson, edge label=\(Z/\gamma^*\)] (a5),
      (a9) -- [boson] (a10),
       };

  \end{feynman}

\end{tikzpicture}
\caption{$\ell$-type diagram}
        \label{subfig:ltypeZgam}
\end{subfigure}
\end{center}
\caption{\label{DiagramsZgamma} Sample LO diagrams for the
  $\ell^+\ell^-\gamma$ production with two different resonance
  stuctures.}
\end{figure}
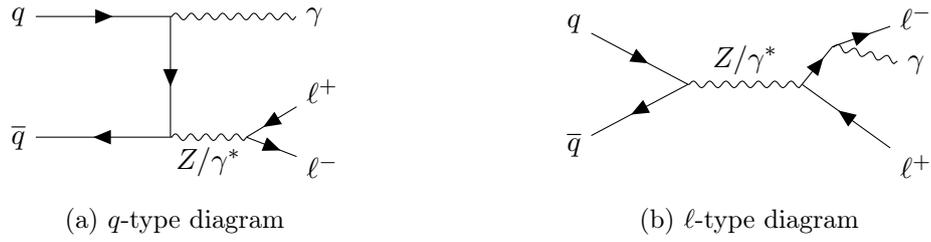

We consider the production processes
\begin{align}
    p p \to \ell^+ \ell^- \gamma \quad \textrm{and}\quad p p \to
    \ell^+ \ell^- \gamma + {\rm jet}\,,
\end{align}
where $\ell\in\{e,\mu\}$ is a massless charged lepton. For brevity, we
refer to these processes as $Z\gamma$ and $Z\gamma$+jet production in
the following.

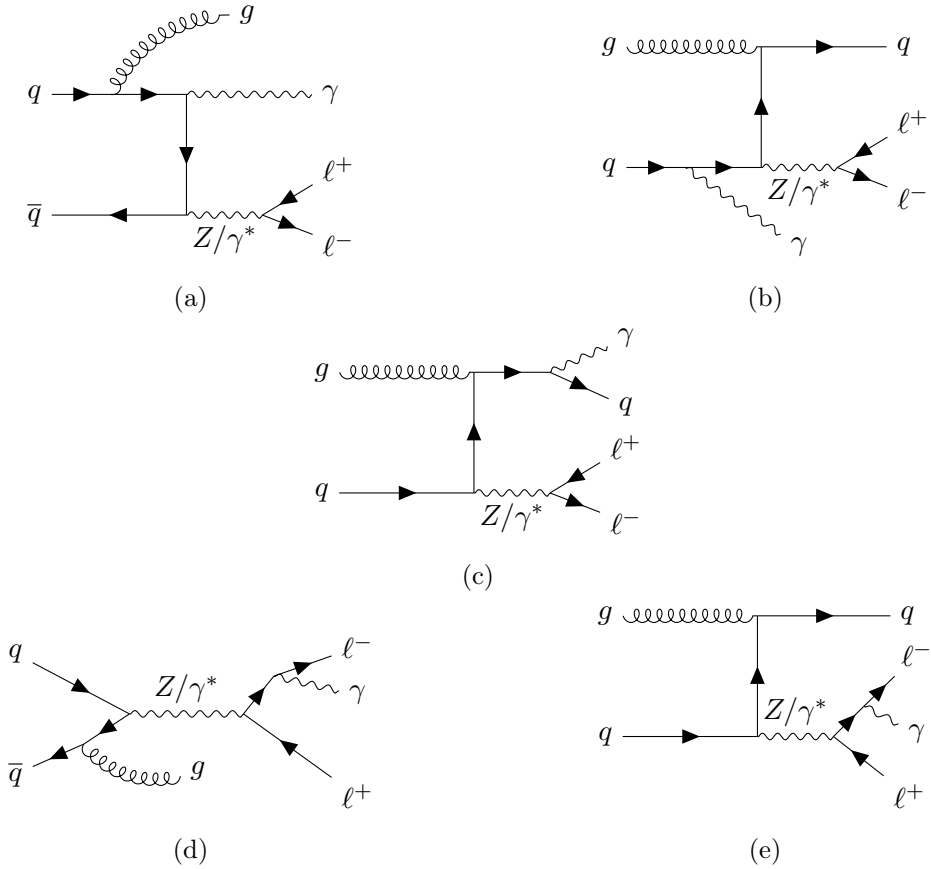
\begin{figure}[t]
  \begin{center}
    \begin{subfigure}[b]{0.5\linewidth}
      \centering
\begin{tikzpicture}
  \begin{feynman}
    \vertex (a1) {\( q\)};
    \vertex[below=1.6cm of a1] (a2){\(\overline q\)};
    \vertex[right=1cm of a1] (a3);
    \vertex[right=1cm of a3] (a9);
    \vertex[right=0.8cm of a9] (a11);
    \vertex[above=0.8cm of a11] (a10){\(g\)};
    \vertex[right=2cm of a2] (a4);
    \vertex[right=1.65cm of a9] (a5){\(\gamma\)};
    \vertex[right=2cm of a9] (a12);
    \vertex[right=1cm of a4] (a6);
    \vertex[below=0.7cm of a12] (a7){\(\ell^+\)} ;
    \vertex[below=1.7cm of a12] (a8){\(\ell^-\)};

    \diagram* {
      {[edges=fermion]
        (a1)--(a3)--(a9)--(a4)--(a2),
        (a7)--(a6)--(a8),
      },
      (a9) -- [ boson] (a5),
      (a4) -- [boson, edge label'=\(Z/\gamma^*\)] (a6),
      (a3) -- [gluon, bend left] (a10),
       };

  \end{feynman}
\end{tikzpicture}
\caption{}
        \label{subfig:qtype}
    \end{subfigure}%
\begin{subfigure}[b]{.5\linewidth}
      \centering
\begin{tikzpicture}
  \begin{feynman}
    \vertex (a1) {\( g\)};
    \vertex[below=1.6cm of a1] (a2){\(  q\)};
    \vertex[right=2cm of a1] (a3);
    \vertex[right=1cm of a2] (a9);
    \vertex[right=1cm of a9] (a4);
    \vertex[right=0.5cm of a4] (a10);
    \vertex[below=0.8cm of a10] (a11){\(\gamma\)};
    \vertex[right=1.65cm of a3] (a5) {\(q\)};
    \vertex[right=2cm of a3] (a12);
    \vertex[right=1cm of a4] (a6);
    \vertex[below=0.7cm of a12] (a7){\(\ell^+\)} ;
    \vertex[below=1.7cm of a12] (a8){\(\ell^-\)};

    \diagram* {
      {[edges=fermion]
        (a2)--(a9)--(a4)--(a3)--(a5),
        (a7)--(a6)--(a8),
      },
      (a1) -- [ gluon ] (a3),
      (a9) -- [ boson] (a11),
      (a4) -- [boson, edge label'=\(Z/\gamma^*\)] (a6),
       };

  \end{feynman}
\end{tikzpicture}
\caption{}
        \label{subfig:qtype2}
\end{subfigure}%

\begin{subfigure}[b]{1\linewidth}
      \centering
\begin{tikzpicture}
  \begin{feynman}
    \vertex (a1) {\( g\)};
    \vertex[below=1.6cm of a1] (a2){\(  q\)};
    \vertex[right=2cm of a1] (a3);
    \vertex[right=2cm of a2] (a4);
    \vertex[right=2cm of a3] (a5);
    \vertex[right=1cm of a3] (a9);
    \vertex[below=0.2cm of a5] (a10){\(  q\)};
    \vertex[above=0.2cm of a5] (a11){\(\gamma\)};
    \vertex[right=1cm of a4] (a6);
    \vertex[below=0.7cm of a5] (a7){\(\ell^+\)} ;
    \vertex[below=1.7cm of a5] (a8){\(\ell^-\)};

    \diagram* {
      {[edges=fermion]
        (a2)--(a4)--(a3)--(a9)--(a10),
        (a7)--(a6)--(a8),
      },
      (a1) -- [ gluon ] (a3),
      (a9) -- [ boson] (a11),
      (a4) -- [boson, edge label'=\(Z/\gamma^*\)] (a6),
       };

  \end{feynman}
\end{tikzpicture}
\caption{}
        \label{subfig:qtype3}
\end{subfigure}%

\begin{subfigure}[b]{0.5\linewidth}
  \centering
\begin{tikzpicture}
  \begin{feynman}
    \vertex (a1) {\( q\)};
    \vertex[below=1.6cm of a1] (a2){\(\overline q\)};
    \vertex[below=0.8cm of a1] (a3);
    \vertex[right=1.5cm of a3] (a4);
    \vertex[right=0.9cm of a3] (a13);
    \vertex[below=0.4cm of a13] (a14);
    \vertex[right=1.5cm of a4] (a5);
    \vertex[right=0.9cm of a4] (a16);
    \vertex[below=0.5cm of a16] (a15){\(g\)};
    \vertex[right=0.4cm of a5] (a6);
    \vertex[right=1.1cm of a6] (a7);
    \vertex[below=0.8cm of a7](a8){\(\ell^+\)};
    \vertex[above=0.5cm of a6](a9);
    \vertex[above=0cm of a7] (a10){\(\gamma\)}; 
    \vertex[above=0.6cm of a7] (a11){\(\ell^-\)};

    \diagram* {
      {[edges=fermion]
        (a1)--(a4)--(a14)--(a2),
        (a8)--(a5)--(a9)--(a11),
      },
      (a4) -- [boson, edge label=\(Z/\gamma^*\)] (a5),
      (a9) -- [boson] (a10),
      (a14)-- [gluon,bend right](a15)
       };

  \end{feynman}

\end{tikzpicture}
\caption{}
        \label{subfig:ltype}
\end{subfigure}%
\begin{subfigure}[b]{0.5\linewidth}
  \centering
\begin{tikzpicture}
  \begin{feynman}
   \vertex (a1) {\( g\)};
    \vertex[below=1.6cm of a1] (a2){\(  q\)};
    \vertex[right=2cm of a1] (a3);
    \vertex[right=2cm of a2] (a4);
    \vertex[right=1.75cm of a3] (a5){\(  q\)};
    \vertex[right=1cm of a4] (a6);
    \vertex[right=1cm of a6] (a7);
    \vertex[right=0.4cm of a6] (a9);
    \vertex[below=0.5cm of a7](a8){\(\ell^+\)};
    \vertex[above=0.4cm of a9](a10);
    \vertex[right=0.7cm of a10](a11);
    \vertex[below=0.1cm of a11] (a12){\(\gamma\)}; ;
    \vertex[above=0.4cm of a11] (a13){\(\ell^-\)};

    \diagram* {
      {[edges=fermion]
        (a2)--(a4)--(a3)--(a5),
        (a8)--(a6)--(a10)--(a13),
      },
      (a4) -- [boson, edge label=\(Z/\gamma^*\)] (a6),
      (a10) -- [boson] (a12),
      (a1) -- [gluon] (a3)
       };

  \end{feynman}

\end{tikzpicture}
\caption{}
        \label{subfig:ltype2}
\end{subfigure}
\end{center}
  \caption{\label{DiagramsZgammaj} Sample LO diagrams for
    $\ell^+\ell^-\gamma $+jet production including $q$-type diagrams
    (a-c) and $\ell$-type diagrams (d-e).}
\end{figure}

As illustrated in \fig{DiagramsZgamma}, $Z\gamma$ production is
initiated by quark-antiquark annihilation at LO. The photon can be
emitted either by the quark line ($q$-type diagrams) or by the lepton
line ($\ell{}$-type diagrams), each of which yields a different
resonance structure of the respective amplitudes.  Sample LO diagrams
for $Z\gamma$+jet production are shown in \fig{DiagramsZgammaj}, with
the same classification into $q$-type and $\ell{}$-type diagrams. The
distinction between those two resonance structures will be relevant
when treating them as two different resonance histories within the
\POWHEGBOXRES{} framework, discussed in \sct{sec:powhegboxres}.  In
addition to the tree-level Born amplitudes, the NLO calculation of the
$Z\gamma$ ($Z\gamma$+jet) process requires the respective one-loop
contributions as well as the tree-level real emission $Z\gamma$+jet
($Z\gamma$+2-jet) amplitudes.

The NLO corrections to $Z\gamma$ and $Z\gamma$+jet production have
been implemented within the \POWHEGBOXRES{} framework.  For the
$Z\gamma$ generator the relevant amplitudes have been extracted from
\noun{MCFM}~\cite{Campbell:2019dru}, while for the $Z\gamma$+jet
generator those have been implemented both using \noun{MCFM} and via
an interface to \noun{OpenLoops} $2$~\cite{Buccioni:2019sur}.  The
helicity amplitudes of \noun{MCFM} are implemented from the analytic
expressions computed in \citeres{Dixon:1998py,DeFlorian:2000sg} for
$Z\gamma$ production and in \citere{Campbell:2012ft} for $Z\gamma$+jet
production. The width of the $Z$ boson is included in the fixed-width
scheme.  For $Z\gamma$+jet, the contribution from third generation
quarks inside loops has been entirely removed for those diagrams where
the Z boson attaches to a fermion loop through an axial-vector
coupling, while the massless bottom loop has been retained for those
contributions where the corresponding top effects decouple as
$1/\mtop^4$~\cite{Campbell:2017aul}.  The impact of this approximation
is expected to be rather small as shown in \citere{Campbell:2016tcu},
where the leading heavy-quark loop contribution has been evaluated in
the $1/\mtop^2$ expansion in the context of $Zj$ and $Zjj$.  We
further note that, in view of the NNLO calculation for $Z\gamma$
production discussed in \sct{sec:reachingNNLO}, omitting the
contribution of third generation quarks is in line with the fact that
the heavy-quark loop contributions at two loops are currently not
known, and therefore not included throughout our \nnlops{} results.

When using the \noun{OpenLoops} interface for the $Z\gamma$+jet
computation of the NLO amplitudes, the complex mass scheme can be used
and the full top-mass effects can be accounted for, while in the MCFM
amplitudes the width is implemented only in a fixed-width scheme and
heavy-quark loop effects are included only approximately.
Since QED effects are included just at LO, the difference between the
complex-mass and the fixed-width scheme amounts to an overall
normalization, whose impact is below $0.1\%$.
When comparing results with full top-mass effects as available in
\noun{Openloops} to approximate results as implemented in \noun{MCFM}
we found per mille effects for quantities inclusive on QCD
radiation. This is expected, since heavy-quark effects at one loop are
non-vanishing only in the presence of final-state radiation.
For jet-related quantities, the discrepancy between the two approaches
can range from a few percent at low transverse momentum up to
approximately $10$-$20\%$ in highly boosted regions ($\pt > 2 \mtop$)
or high-invariant mass regions.
This is not surprising, since the process at hand involves $s$-channel
fermion-loop contributions, which become more important in these phase
space regions.
For observables involving a jet our results are NLO accurate only,
hence characterized by larger theoretical uncertainties. In summary,
we find that mass effects are always much smaller than our quoted
theoretical uncertainties and for the numerical studies performed in
this paper, which are not devoted to boosted regions, it is perfectly
fine to use approximate results for the heavy-quark mass effects.
Accordingly, because of the better numerical performance of
\noun{MCFM}, we choose to use these amplitudes to obtain the results of
this paper.
Specifically, we find that \noun{MCFM} virtual amplitudes are about
ten times faster than \noun{Openloop} ones. On the other hand, one can
make use of the folding option in \POWHEG{}, where the real
contribution is evaluated multiple times for each virtual one. This
improves the numerical performance whenever the virtual amplitudes
constitute the bottleneck in the numerical evaluation.
For greater flexibility, in the release of the numerical code, the
option to choose between the \noun{OpenLoops} and the \noun{MCFM}
implementations has been made available.

\subsection{The \POWHEGBOXRES{} framework}
\label{sec:powhegboxres}

We calculate NLO+PS predictions for $Z\gamma$ and $Z\gamma$+jet using
the \POWHEG{} method, which is based on the following master formula
\cite{Nason:2004rx,Frixione:2007vw,Alioli:2010xd}:
\begin{align}
\frac{\mathd\sigma}{\mathd\PhiBorn}={\bar B}(\PhiBorn) \times
\bigg\{\Delta_{\rm pwg} (\LambdaPWG) + \int\mathd \Phi_{\tmop{rad}} 
  \Delta_{\rm pwg} (\ptrad)  \frac{R (\PhiBorn{}, \Phi_{\tmop{rad}})}{B
  (\PhiBorn{})}\bigg\}\,,
\label{eq:POWHEGmaster}
\end{align}
where $\PhiBorn$ is the Born phase space of the process under
consideration.  The function ${\bar B}(\PhiBorn)$ describes the
inclusive NLO process, where extra QCD emissions are integrated out;
the content of the curly brackets is responsible for the exclusive
generation of one extra QCD radiation with respect to the Born process
according to the \POWHEG{}
method~\cite{Nason:2004rx,Frixione:2007vw,Alioli:2010xd}.  $B$ and $R$
denote the squared tree-level matrix elements for the process at Born
level and for its real radiation, respectively.  The evaluation of the
\POWHEG{} Sudakov form factor $\Delta_{\rm pwg}$~\cite{Nason:2004rx}
depends on the real phase space $\Phi_{\tmop{rad}}$ through the
transverse momentum $\ptrad$ of the extra radiation.  The \POWHEG{}
cutoff $\LambdaPWG$ is used to veto QCD emissions in the
non-perturbative regime and its default value is
$\LambdaPWG=0.89$\,GeV. The parton shower then adds additional
radiation to \eqn{eq:POWHEGmaster} that contributes beyond NLO with
respect to the underlying Born at all orders in perturbation
theory. We refer to the original publications for explicit
fomulae~\cite{Nason:2004rx,Frixione:2007vw,Alioli:2010xd}.

For the practical implementation of the $Z\gamma$ and $Z\gamma$+jet
generators we exploit the \POWHEGBOXRES{}
framework~\cite{Jezo:2015aia}, which takes into account the different
resonance structures of each process.  All possible resonance
histories are automatically identified and the code performs a
resonance-aware phase space sampling.  The efficiency of the infrared
subtraction is improved by means of its resonance-aware subtraction
algorithm, where the mapping from a real to the underlying Born
configuration preserves the virtuality of intermediate
resonances~\cite{Jezo:2015aia,Jezo:2016ypn}. \ Moreover, in a
parton-shower context the distortion of resonances through recoil
effects is avoided by supplying it with details on the resonance
cascade chain.  The resonance awareness of the phase space sampling
and the subtraction improve the numerical stability of the evaluation
of the $\bar B$ function in the \POWHEG{} formula of
\eqn{eq:POWHEGmaster}.

The key idea behind the algorithm used in the \POWHEGBOXRES{}
framework is to decompose the cross section into contributions
associated to a well-defined resonance structure, which are enhanced
on that particular cascade chain.  As discussed in \sct{sec:process},
both $Z\gamma$ and $Z\gamma$+jet production have two different
resonance histories, which can be associated to $q$-type diagrams,
where the photon is emitted from the quark/antiquark line, and
$\ell$-type ones, where the final state photon is radiated off one of
the two leptons.

The \POWHEGBOXRES{} framework takes as input {\it bare} flavour
structures ${\flavBorn}$ of the Born process
(e.g.\ $\flavBorn=\{u\bar{u}\to \ell^+\ell^-\gamma\}$), which only
contain the information on the initial- and final-state flavour
structure. The full Born cross section can be written as the sum over
all bare flavour structures of the corresponding Born contribution
$B_{\flavBorn}$
\begin{align}
  B=\sum_{\flavBorn} B_{\flavBorn}.
  \label{eq:born}
\end{align}
After introducing the {\it full} flavour and resonance structure
$\fullflavBorn$ of the Born process
(e.g.\ $\fullflavBorn=\{u\bar{u}\to Z\to \ell^+\ell^-\gamma\}$ or
$\fullflavBorn=\{u\bar{u}\to Z\gamma\to \ell^+\ell^-\gamma\}$), which
embodies details on the entire resonance history, we can further
decompose $B_{\flavBorn}$ as a weighted sum over $\fullflavBorn$ using
weight functions ${\cal P}_{\fullflavBorn}$
\begin{align}
  B_{\flavBorn}=\sum_{\fullflavBorn\in T(\flavBorn)} {\cal
    P}_{\fullflavBorn}B_{\flavBorn}, \quad \text{with}
  \sum_{\fullflavBorn\in T(\flavBorn)} {\cal P}_{\fullflavBorn}=1,
  \label{eq:resdecomposition}
\end{align}
where $T(\flavBorn)$ is named a {\it tree} and denotes all graphs with
the given initial- and final-state flavour configuration $\flavBorn$.
The weight functions ${\cal P}_{\fullflavBorn}$ are chosen such that
\eqn{eq:resdecomposition} expresses $B_{\flavBorn}$ as a sum over
resonance-peaked terms ${\cal P}_{\fullflavBorn} B_{\fullflavBorn}$,
which develop the expected resonance enhancement of $\fullflavBorn$.
There is a certain freedom in their explicit expression, and in the
\POWHEGBOXRES{} code the following choice is made:
\begin{align}
  {\cal
    P}_{\fullflavBorn}=\frac{P_{\fullflavBorn}}{\sum_{\fullflavprimeBorn\in
      T(\flavBorn(\fullflavBorn))} P_{\fullflavprimeBorn} }\,,
  \label{eq:pifactor}
\end{align}
where the sum in the denominator runs over all configurations in the
tree $T(\flavBorn(\fullflavBorn))$ of graphs having a bare flavour
structure $\flavBorn(\fullflavBorn)$ compatible with $\fullflavBorn$.
In the specific case of $Z\gamma$ and $Z\gamma$+jet production, ${\cal
  P}_{\fullflavBorn}$ assumes two different functional forms depending
on whether $\fullflavBorn$ refers to $q$-type or $\ell$-type diagrams,
and they read
\begin{align}
  P_{\fullflavBorn}=
  \begin{cases}
    \frac{\mz^2}{(s_{\text{\scalefont{0.77}$\ell\ell$}}-\mz^2)^2+\GZ^2\mz^2}   & \text{$\fullflavBorn$ is of $q$-type}\,,\\
    \frac{\mz^2}{(s_{\text{\scalefont{0.77}$\ell\ell\gamma$}}-\mz^2)^2+\GZ^2\mz^2}   & \text{$\fullflavBorn$ is of $\ell$-type}\,,
    \end{cases}
  \label{eq:pifactorZa}
\end{align}
where $s_{\text{\scalefont{0.77}$\ell\ell$}}$ is the invariant mass of
the lepton pair and $s_{\text{\scalefont{0.77}$\ell\ell\gamma$}}$ that
of the produced colour-singlet system.  The same discussion applies to
the virtual corrections, while for the real-emission contribution a
similar decomposition is performed taking into account the different
singular regions. As described in detail in \citere{Jezo:2015aia}, the
concept of resonance history directly affects the definition of QCD
singular regions: only soft/collinear singular regions compatible with
a given full real flavour structure $\hat\ell_R$ are considered in the
FKS decomposition of the real amplitude.
It is important to note that, for each of the full real flavour
structures $\hat\ell_R$ the mappings from the real to the Born
configurations preserve the virtualities of the intermediate
resonances, which is crucial to guarantee a cancellation of
singularities between real corrections and their counterterms.

\subsection{Treatment of the isolated photon and details of the implementation}
\label{sec:photon}

The emission of a soft or collinear photon from a quark or a charged
lepton induces QED singularities. Processes with final-state photons
therefore require not only suitable criteria to isolate photons in the
experimental analyses, but they also call for an IR-safe isolation
procedure on the theory side. Since in the \POWHEGBOX{} framework
fiducial cuts are usually applied at analyses level after parton
showering, which modifies the kinematics of the final states, we
discuss how to include photon-isolation requirements already at the
event generation level in this framework to obtain IR-safe
predictions.

To produce isolated photons in the final state there are two relevant
mechanisms: the {\it direct} production in the hard process, which can
be described perturbatively, and the production through {\it
  fragmentation} of a quark or a gluon, which is non-perturbative.
The separation between the two production mechanisms in theoretical
predictions is quite delicate, as sharply isolating the photon from
the partons would spoil infrared (IR) safety.  So-called {\it
  fragmentation functions} are required to absorb singularities
related to collinear photon emissions in the latter production
mechanism. Those functions are determined from data with relatively
large uncertainties.  On the other hand, Frixione's {\it smooth-cone
  isolation} of the photons~\cite{Frixione:1998jh} offers an IR-safe
alternative that completely removes the fragmentation component.  This
substantially simplifies theoretical calculations of processes with
isolated photons at higher orders in perturbation theory. Although the
direct usage of smooth-cone isolation in experimental analyses is not
possible due to the finite granularity of the calorimeter,
data--theory comparisons are facilitated by tuning the smooth-cone
parameters to mimic the fixed-cone isolation used by the experiments,
see e.g. \citere{Catani:2018krb}.

So far, only few processes involving final-state photons have been
implemented in the \POWHEGBOX{} framework:
$W\gamma$~\cite{Barze:2014zba} and the direct
photon~\cite{Jezo:2016ypn, Klasen:2017dsy}. These two generators make
use of the photon fragmentation component. In particular in
\citere{Barze:2014zba} the hadron fragmentation into photons is
modelled within POWHEG in combination with a QCD+QED shower.
In this case the theoretical prediction can apply directly the photon
isolation criteria imposed by the experiments to distinguish prompt
photons taking part in the hard scattering process from possible
background sources (such as photons from decay of $\pi^0$ mesons or
from quark fragmentation). This facilitates a direct comparison
between experimental and theoretical results. Those isolation criteria
limit the hadronic activity in the vicinity of the photon by imposing
\begin{align}
  \sum_{\text{had}\in R_0}E_{\text{\scalefont{0.77}T}}^{\text{had}}<E_{\text{\scalefont{0.77}T}}^{\text{max}}\quad\text{with}\quad R_0=\sqrt{\Delta\eta^2+\Delta\phi^2}\,,
  \label{eq:isolation}
\end{align}
where the sum of the transverse energy $E_{\text{\scalefont{0.77}T}}^{\text{had}}$ of the hadrons
inside a fixed cone of radius $R_0$ around the photon is constrained
to be less than $E_{\text{\scalefont{0.77}T}}^{\text{max}}$.

In view of the NNLO extension considered in this paper, presented in
\sct{sec:reachingNNLO}, we instead rely on smooth-cone
isolation~\cite{Frixione:1998jh} to turn off the fragmentation
component and to deal with QED collinear singularities perturbatively
in an IR-safe manner.  In this case, phase-space configurations where
the photon becomes collinear to a quark are removed while preserving
IR safety by allowing arbitrarily soft QCD radiation within smoothly
decreasing cones around the photon direction.  In practice, this means
that the smooth-cone isolation is implemented by restricting the
hadronic (partonic) activity within every cone of radius
\mbox{$\delta=\sqrt{(\Delta \eta)^2+(\Delta
    \phi)^2}<\delta_{\text{\scalefont{0.77}$0$}}$} around a
final-state photon, where $\delta_{\text{\scalefont{0.77}$0$}}$ sets
the maximal cone size, by imposing the following condition
\begin{align}
  \sum_{\text{had/part}\in\delta}E_{\text{\scalefont{0.77}T}}^{\text{had/part}}
  \;\leq\;E_{\text{\scalefont{0.77}T}}^{\rm max}(\delta)\;=\;
  E_{\text{\scalefont{0.77}T}}^{\mathrm{ref}} \left(\frac{1-\cos
    \delta}{1- \cos \delta_{\text{\scalefont{0.77}$0$}}}\right)^n\,,
  \qquad \forall\; \delta\;\leq\;
  \delta_{\text{\scalefont{0.77}$0$}}\, ,
\label{eq:frixione}
\end{align}
such that the total hadronic (partonic) transverse energy inside the
cone is smaller than $E_{\text{\scalefont{0.77}T}}^{\rm max}(\delta)$.
The parameter $n$ controls the smoothness of the isolation function
and $E_{\text{\scalefont{0.77}T}}^{\mathrm{ref}}$ is a reference
transverse-momentum scale that can be chosen to be either a fraction
$\epsilon_{\text{\scalefont{0.77}$\gamma$}}$ of the transverse
momentum of the respective photon ($\ptg$) or a fixed value
($p_{\text{\scalefont{0.77}T}}^0$),
\begin{align}
E_{\text{\scalefont{0.77}T}}^{\mathrm{ref}}\;=\;\epsilon_{\text{\scalefont{0.77}$\gamma$}} \,\ptg{} \qquad\textnormal{or}\qquad E_{\text{\scalefont{0.77}T}}^{\mathrm{ref}}\;=\;p_{\text{\scalefont{0.77}T}}^0 \, .
\label{eq:frixionepTscale}
\end{align}

In our calculations we impose smooth-cone isolation on the phase space
of all $Z\gamma$+jet and $Z\gamma$+2-jet configurations. Furthermore,
various technical phase-space cuts at event generation level are
necessary in order to obtain IR safe results. Those generation cuts
and parameters of the smooth-cone isolation are given in
\app{app:cuts}. They are chosen to be much looser than the ones
eventually applied at analyses level after parton showering.
We stress that, since we also employ suppression factors for the NLO
squared amplitudes (as discussed in detail below), the resulting
differential cross section times suppression factors vanishes in the
singular regions, which will not pass fiducial cuts.

As commonly used in many \POWHEGBOX{} generators, we exploit the
possibility to split the real squared matrix element $R$ into a
singular and a finite (remnant) contribution. Such splitting improves
the numerical performance of the code, especially as far as the
efficiency of the event generation is concerned, in cases where the
ratio $R/B$ departs from its corresponding soft/collinear
approximation, for instance in presence of Born zeros
\cite{Alioli:2008gx}.  Following section 5 of \citere{Alioli:2010xd},
we write the splitting into a singular and a remnant contribution for
each singular region $\alpha$ of the real amplitude as:
\begin{align}
  \begin{split}
 &R = \sum_\alpha R^\alpha(\PhiReal{})=\sum_\alpha \left[R^\alpha_{\rm
        sing.}(\PhiReal{})+R^\alpha_{\rm remn.}(\PhiReal{})\right]\,,
    \\ &R^\alpha_{\rm sing.}(\PhiReal{}) = {\cal S}
    R^\alpha(\PhiReal{})\,,\quad R^\alpha_{\rm remn.}(\PhiReal{}) =
    (1-{\cal S}) R^\alpha(\PhiReal{})\,,\
  \label{eq:singremn}
\end{split}
  \end{align}
where $\PhiReal{}$ is the real phase-space and ${\cal S}$ is called
damping factor.
Only the singular contribution $\sum_\alpha R^\alpha_{\rm
  sing.}(\PhiReal{})$ is exponentiated in the \POWHEG{} Sudakov
$\Delta_{\rm pwg}$ and used to generate the first emission according
to the \POWHEG{} method in \eqn{eq:POWHEGmaster}, while the finite
remnant contribution $\sum_\alpha R^\alpha_{\rm remn.}(\PhiReal{})$
can be treated separately, by generating it with standard techniques
and feeding it directly into the parton shower.

A standard damping factor \cite{Alioli:2010xd} is used in both
$Z\gamma$ and $Z\gamma$+jet generators, where ${\cal S}$=0 when the
real squared amplitude in a singular region is greater than five times
its \mbox{soft/collinear} approximation, and ${\cal S}$=1
otherwise. Additionally, to improve the numerical convergence, the
$Z\gamma$+jet generator requires a special setting of the damping
factor, which ensures that QED singularities appearing in the real
squared matrix element are moved into the remnant contribution.
Indeed, not all of the QED singular regions appearing in the real
matrix elements have a singularity in their underlying
Born. Accordingly, the associated real singularity is not mitigated by
an overall Born suppression factor (as described in more detail
below).  To deal with this issue, we define, in each singular region,
the invariant mass $m_{\rm rad}$ of the emitter-emitted pair of that
singular region, which is the quantity that becomes small close to QCD
singularities, and we use as a damping factor\footnote{We thank Carlo Oleari for suggesting to absorb the QED singularities into the remnant contribution.}
\begin{equation}
{\cal S'} = \frac{(m^2_{\rm rad})^{-1}}{(m^2_{\rm rad})^{-1}+c \sum_{i
    \in (q, \bar q) } d_{i\gamma}^{-1}} {\cal S}\,,
\label{eq:calS}
\end{equation}
where $c\in[0,1]$ is a free parameter that we choose below, and the
sum runs over all (initial- and final-state) quarks with
\begin{equation}
\begin{split}
  d_{i\gamma} &= p_{\text{\scalefont{0.77}T},\gamma}^2\quad\;\;\;\textrm{when $i$ is a quark in the initial state}\,,\\
  d_{i\gamma} &= p_i \cdot p_{\gamma}\quad\textrm{when $i$ is a quark in the final state}\,. 
\end{split}
\end{equation}
The splitting induced by the suppression factor in \eqn{eq:calS} is
such that, when a QED singularity dominates, the event is included in
$R^\alpha_{\rm remn.}(\PhiReal{})$ (${\cal S'} \to 0$), and, when the
QCD singularity is dominant, the event is moved into $R^\alpha_{\rm
  sing.}(\PhiReal{})$ (${\cal S'} \to 1$).  The numerical constant $c$
controls the transition region between QED singularities in
$d_{i\gamma}$ and QCD singularities in $m^2_{\rm rad}$. Since
$\alpha_{\rm e}/\alpha_{\rm s}\sim 0.1$, we use the value $c=0.1$,
which we have checked to be suitable for an efficient generation of
events.

Finally, we exploit another tool of the \POWHEGBOX{} framework that
allows us to improve the numerical convergence in the relevant
phase-space regions. By introducing suppression factors, which
multiply the cross section during integration and are a posteriori
divided out again, it is possible to redirect the numerical sampling
of events into certain regions in phase space. This is mandatory for
processes that have singularities at Born level, such as the QED
singularities in $Z\gamma$ and $Z\gamma$+jet production, to avoid
sampling large statistics in phase-space regions which are eventually
removed by the fiducial cuts at analysis level.  In order to obtain
suitable integration grids that give more weight to the phase-space
regions selected by the fiducial cuts, we have introduced a Born
suppression factor that vanishes in singular regions related both to
QCD and QED emissions.  Its precise form is given in
\app{app:cuts}. Since the real phase space is generated directly from the
Born one in the \POWHEGBOX{}, the same factor is also applied to
$R^\alpha_{\rm sing.}(\PhiReal{})$.  For the remnant contribution
$R^\alpha_{\rm remn.}(\PhiReal{})$, on the other hand, which is QCD
regular, but is in our case QED singular, an analogous suppression
factor has been chosen, that is given in \app{app:cuts} as well.

We stress that the implementational details discussed throughout this
section are by no means standard, despite the fact that the tools we
are using existed already in the \POWHEGBOX{} framework.  The proper
adjustment of those tools and their related parameters required a
great effort, which was necessary to obtain a generator for
$Z\gamma$+jet production with sufficient numerical efficiency to be
extended to NNLO corrections to $Z\gamma$ production discussed in the
next section.

\section{Reaching NNLO accuracy for \boldmath{$Z\gamma$} production using \minnlo{}}
\label{sec:reachingNNLO}

We use our implementation of the $Z\gamma$+jet generator in the \POWHEGBOXRES{} framework discussed
in \sct{sec:description} as a starting point. 
To include NNLO corrections for $Z\gamma$ production in this calculation 
we employ the recently developed \minnlo{} method~\cite{Monni:2019whf}.
To this end, we extend the \minnlo{} method to processes with non-trivial two-loop corrections, i.e.\ genuine $2\to 2$ hard-scattering 
processes such as vector-boson pair production.
After a general discussing of the ingredients required for a NNLO calculation, we recall the details of the \minnlo{} method and 
present its extension to $2\to 2$ processes. Finally, we provide practical details on how that calculation is embedded in the \POWHEGBOXRES{} framework.
 
\subsection{Ingredients of a NNLO calculation}
\label{sec:nnlo}

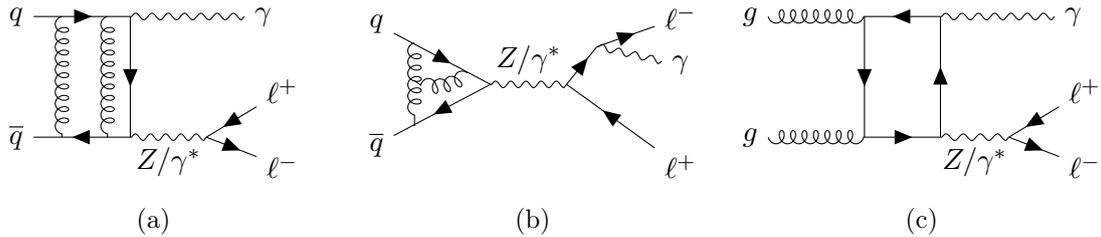
\begin{figure}[t]
  \begin{center}
\hspace{-0.4cm}
    \begin{subfigure}[b]{.33\linewidth}
      \centering
\begin{tikzpicture}
  \begin{feynman}
    \vertex (a1) {\( q\)};
    \vertex[below=1.6cm of a1] (a2){\(\overline q\)};
    \vertex[right=1.5cm of a1] (a3);
    \vertex[right=1.5cm of a2] (a4);
    \vertex[right=1.5cm of a3] (a5){\(\gamma\)};
    \vertex[right=2cm of a3] (a9);
    \vertex[right=1cm of a4] (a6);
    \vertex[below=0.7cm of a9] (a7){\(\ell^+\)} ;
    \vertex[below=1.7cm of a9] (a8){\(\ell^-\)};
    \vertex[right=0.6cm of a1] (a21);
    \vertex[right=0.6cm of a2] (a22);
    \vertex[right=0.6cm of a21] (a31);
    \vertex[right=0.6cm of a22] (a32);
    
    \diagram* {
      {[edges=fermion]
        (a1)--(a3)--(a4)--(a2),
        (a7)--(a6)--(a8),
      },
      (a3) -- [ boson] (a5),
      (a4) -- [boson, edge label'=\(Z/\gamma^*\)] (a6),
      (a21) -- [gluon] (a22),
      (a31) -- [gluon] (a32),
       };

  \end{feynman}
\end{tikzpicture}
\caption{}
        \label{subfig:t}
\end{subfigure}%
\begin{subfigure}[b]{.33\linewidth}
  \centering
\begin{tikzpicture}
  \begin{feynman}
    \vertex (a1) {\(  q\)};
    \vertex[below=1.6cm of a1] (a2){\(\overline q\)};
    \vertex[below=0.8cm of a1] (a3);
    \vertex[right=1.5cm of a3] (a4);
    \vertex[right=1cm of a4] (a5);
    \vertex[right=0.4cm of a5] (a6);
    \vertex[right=1.1cm of a6] (a7);
    \vertex[below=0.8cm of a7](a8){\(\ell^{+}\)};
    \vertex[above=0.5cm of a6](a9);
    \vertex[above=0cm of a7] (a10){\(\gamma\)};
    \vertex[above=0.6cm of a7] (a11){\(\ell^{-}\)};
    \vertex[below=0.25cm of a1] (a31);
    \vertex[above=0.25cm of a2] (a32);
    \vertex[right=0.5cm of a31] (a21);
    \vertex[right=0.5cm of a32] (a22);
    \vertex[below=0.65cm of a1] (a132);
    \vertex[right=1.2cm of a132] (a122);
    \vertex[below=0.5cm of a21] (a121);
 
    \diagram* {
      {[edges=fermion]
        (a1)--(a4)--(a2),
        (a8)--(a5)--(a9)--(a11),
      },
      (a4) -- [boson, edge label=\(Z/\gamma^*\)] (a5),
      (a9) -- [boson] (a10),
      (a21) -- [gluon] (a22),
      (a122) -- [gluon, bend left] (a121),
       };

  \end{feynman}

\end{tikzpicture}
\caption{}
        \label{subfig:s}
\end{subfigure}
\begin{subfigure}[b]{.33\linewidth}
  \centering
\begin{tikzpicture}
  \begin{feynman}
    \vertex (a1) {\( g\)};
    \vertex[below=1.6cm of a1] (a2){\(g\)};
    \vertex[right=1.5cm of a1] (a3);
    \vertex[right=1.5cm of a2] (a4);
    \vertex[right=1cm of a3] (a5);
    \vertex[right=1cm of a4] (a6);
    \vertex[right=1.5cm of a5] (a7){\(\gamma\)};
    \vertex[right=1.9cm of a5] (a11);
    \vertex[right=0.9cm of a6] (a8);
    \vertex[below=0.7cm of a11] (a9){\(\ell^+\)} ;
    \vertex[below=1.7cm of a11] (a10){\(\ell^-\)};
 
    \diagram* {
      {[edges=fermion]
        (a3)--(a4)--(a6)--(a5)--(a3),
        (a9)--(a8)--(a10),
      },
      (a1) -- [ gluon] (a3),
      (a2) -- [ gluon] (a4),
      (a5) -- [ boson] (a7),
      (a6) -- [boson, edge label'=\(Z/\gamma^*\)] (a8),
       };

  \end{feynman}
\end{tikzpicture}
\caption{}
        \label{subfig:gg}
\end{subfigure}
\end{center}
\caption{\label{fig:diagNNLO} Sample Feynman diagrams entering the $\ell^+\ell^-\gamma$ process at NNLO: (a) $q$-type and (b) $\ell$-type two-loop diagrams; (c) loop-induced gluon-fusion contribution.}
\end{figure}

In \sct{sec:process} we have discussed the contributions relevant to evaluate NLO corrections to
$Z\gamma$ and $Z\gamma$+jet production. Those involve 
tree-level and one-loop amplitudes for the processes $pp\to Z\gamma$ and  $pp\to Z\gamma$+jet 
as well as the tree-level amplitude for $pp\to Z\gamma$+2-jet.
The same amplitudes enter the NNLO calculation for $Z\gamma$ production, i.e.
at the Born level and as real, virtual one-loop, real-virtual and double-real corrections.
The only missing ingredients for the NNLO calculation are the two-loop corrections in the $q\bar{q}$ channel (sample diagrams are shown in \fig{fig:diagNNLO}\,(a-b)), 
and the loop-induced contribution in the gluon-fusion channel (with a sample diagram shown in \fig{fig:diagNNLO}\,(c)).
The latter can be separated from the others since it is effectively only LO accurate. 
Its size is rather small, being less than $10\%$ of the NNLO corrections and below $1\%$ of 
the full $Z\gamma$ cross section at NNLO \cite{Grazzini:2017mhc}.
We thus refrain from including the loop-induced gluon-fusion contribution in our calculation. 
We note, however, that while its calculation at LO+PS is quite straightforward and easily done 
with current standard tools, see e.g. \citeres{Alioli:2010xd,Hirschi:2015iia}, a more sophisticated treatment would require to match the NLO
corrections to the loop-induced gluon-fusion contribution with parton showers. Despite being feasible with 
current technology, this is beyond the scope of this paper and left for future studies.

For the two-loop corrections we use the $q\bar{q}\to \ell\ell\gamma$ helicity amplitudes calculated in \citere{Gehrmann:2011ab}. Those have been
fully implemented into the \Matrix{} framework~\cite{Grazzini:2017mhc,Matrixurl} using the results of \citere{Gehrmann:2011ab}. 
In order to exploit this implementation and evaluate the two-loop helicity amplitudes within our \minnlo{} calculation presented in this paper we 
have compiled \Matrix{} as a {\tt C++} library and linked it to our \POWHEGBOXRES{} $Z\gamma$+jet generator.

We further exploit \Matrix{} for all fixed-order NNLO
results used for comparisons throughout this paper.
The \Matrix{} framework features NNLO QCD corrections to a large number of colour-singlet processes at hadron colliders. Several state-of-the-art NNLO QCD predictions \cite{Grazzini:2013bna,Grazzini:2015nwa,Cascioli:2014yka,Grazzini:2015hta,Gehrmann:2014fva,Grazzini:2016ctr,Grazzini:2016swo,Grazzini:2017ckn,Kallweit:2018nyv,Kallweit:2020gcp}\footnote{It was also used in the NNLO+NNLL computation of \citere{Grazzini:2015wpa}, and in the NNLOPS computations of \citeres{Re:2018vac,Monni:2019whf,Alioli:2019qzz,Monni:2020nks}.} have been obtained with this framework, and for massive diboson processes
it has been extended to combine NNLO QCD both with NLO EW corrections \cite{Kallweit:2019zez} and with NLO QCD corrections to the loop-induced gluon fusion contribution \cite{Grazzini:2018owa,Grazzini:2020stb}. 
Through the recently implemented \textsc{Matrix+RadISH} interface \cite{Kallweit:2020gva,Wiesemann:2020gbm} the code now also
combines NNLO calculations with high-accuracy resummation through the \textsc{RadISH} formalism \cite{Monni:2016ktx,Bizon:2017rah,Monni:2019yyr}
for different transverse observables, such as the transverse momentum of the colour singlet.

\subsection{Generalization of \minnlo{} to $2\to 2$ colour-singlet processes}
\label{sec:MiNNLO}

In the following we recall the central aspects of the \minnlo{} method of \citere{Monni:2019whf} as well as 
some of its improvements presented in \citere{Monni:2020nks}, and we discuss in detail
all the changes required to apply the method to a genuine $2\to 2$ colour-singlet process, such as vector-boson pair production.
The main difference compared to $2\to 1$ hadronic processes, such as Higgs and Drell-Yan production, is that the 
one- and two-loop virtual corrections for a general process can not be written as a simple form factor times the Born 
amplitude, and thereby receive a dependence on the respective flavour configuration.
To this end, we will recast, where appropriate, the relevant formulae of \citeres{Monni:2019whf,Monni:2020nks} in a flavour-dependent notation.

\minnlo{} is a method to perform a NNLO calculation for a produced colour singlet \F{} with invariant mass $Q$
fully differential in the respective Born phase space $\PhiB$, in such a way that it can be subsequently matched to a 
parton shower. In the context of this paper, \F{} 
would be the $Z\gamma$ (or more precisely $\ell^+\ell^-\gamma$) final state, but we prefer 
to keep the discussion general throughout this section.
We employ the same notation as in \citeres{Monni:2019whf,Monni:2020nks} in the following.
The starting point of \minnlo{} is a \POWHEG{} implementation of colour singlet production in association with one jet (\FJ{}),
whose phase space we denote by $\PhiBJ$. We thus write \eqn{eq:POWHEGmaster} explicitly with the Born process being the \FJ{} production:\begin{align}
\frac{\mathd\sigma}{\mathd\PhiBJ}={\bar B}(\PhiBJ) \times
\bigg\{\Delta_{\rm pwg} (\LambdaPWG) + \int\mathd \Phi_{\tmop{rad}} 
  \Delta_{\rm pwg} (\ptrad)  \frac{R (\PhiBJ{}, \Phi_{\tmop{rad}})}{B
  (\PhiBJ{})}\bigg\}\,.
\label{eq:master}
\end{align}
Here, ${\bar B}(\PhiBJ)$ describes the \FJ{} process, i.e. including the first radiation, using the full
NLO cross section while the
content of the curly brackets accounts for the second QCD emission through the \POWHEG{} mechanism, with
$B$ and $R$ being the squared tree-level matrix elements for \FJ{} and \FJJ{} production, respectively,
and $\Phi_{\tmop{rad}} $ ($\ptrad$) referring to the phase space (transverse momentum) of the second emission.
Radiation beyond the second one is generated by the parton shower, which adds corrections of $\mathcal{O}(\as^3(Q))$ and higher at all orders in perturbation theory.
In order to reach NNLO accuracy in the phase space of the color singlet \F{} in \eqn{eq:master}, we modify the content of the ${\bar B}(\PhiBJ)$ function, 
which is the central ingredient of the \minnlo{} method.

The derivation of the ${\bar B}(\PhiBJ)$ function in \minnlo{}~\cite{Monni:2019whf} 
is based on the following formula that describes the transverse momentum of the color singlet ($\pt$)
up to NNLO and is fully differential in the Born phase space $\PhiB$:
\begin{align}
\label{eq:start}
  \frac{\mathd\sigma}{\mathd\PhiB\mathd \pt} &= \frac{\mathd}{\mathd \pt}
     \Bigg\{\sum_{\flavB}\exp[-\tilde{S}_{\flavB}(\pt)] {\cal L}_{\flavB}(\pt)\Bigg\} +
                                               R_f(\pt) \\
                                               &= \sum_{\flavB}\exp[-\tilde{S}_{\flavB}(\pt)] \, D_{\flavB}(\pt) +
                                               R_f(\pt) \nonumber\,,
\end{align}
where $R_f$ includes only non-singular contributions at small $\pt$, and 
\begin{equation}
\label{eq:Dterms}
  D_{\flavB}(\pt)  \equiv - \sum_{\flavB} \frac{\mathd \tilde{S}_{\flavB}(\pt)}{\mathd \pt} {\cal L}_{\flavB}(\pt)+\frac{\mathd {\cal L}(\pt)}{\mathd \pt}\,.
\end{equation}
At variance with the formulas in \citeres{Monni:2019whf,Monni:2020nks}, we have introduced an 
explicit sum over the flavour structures $\flavB$ of the Born process $pp\to\text{\F}$. The quantities 
without index $\flavB$ should be understood as having been summed implicitly over $\flavB$ as in the 
original formulation of \citeres{Monni:2019whf,Monni:2020nks}, in particular
\begin{align}
{\cal L}(\pt)\equiv \sum_{\flavB}  {\cal L}_{\flavB}(\pt)\,.
\end{align}
Introducing the flavour sum in \eqn{eq:Dterms} becomes necessary for a general 
colour-singlet process, because not only the luminosity factor ${\cal L}_{\flavB}$, but 
also the Sudakov form factor $\tilde{S}_{\flavB}(\pt)$ becomes flavour dependent.
Their expressions read, cf. eqs.~(4.31) and (2.9) of \citere{Monni:2019whf},\footnote{The convolution between two
  functions $f(z)$ and $g(z)$ is defined as
  $(f\otimes g)(z) \equiv \int_z^1 \frac{\mathd x}{x} f(x)
  g\left(\frac{z}{x}\right)$.}
\begin{align}
{\cal L}_{\flavB=cc'}(\pt)&= \sum_{i, j}
\bigg\{\left(\tilde{C}^{[a]}_{c i}\otimes f_i^{[a]}\right) \frac{\mathd\big[|M^{\scriptscriptstyle\rm F}|_{cc'}^2\, \tilde{H}_{\flavB=cc'}(\pt)\big]}{\mathd\PhiB}
\left(\tilde{C}^{[b]}_{c' j}\otimes f_j^{[b]}\right)  \notag\\
 &\hspace{1.02cm}+ \left(G^{[a]}_{c i}\otimes f_i^{[a]}\right) \frac{\mathd\big[|M^{\scriptscriptstyle\rm F}|_{cc'}^2\, \tilde{H}_{\flavB=cc'}(\pt)\big]}{\mathd\PhiB} \left(G^{[b]}_{c'
j}\otimes f_j^{[b]}\right)\bigg\}\,,
\label{eq:luminosity}\\
\tilde{S}_{\flavB}(\pt) &= 2\int_{\pt}^{Q}\frac{\mathd q}{q}
                    \left(A(\as(q))\ln\frac{Q^2}{q^2} +
                    \tilde{B}_{\flavB}(\as(q))\right)\,,\label{eq:minnlops-Sudakov}
\end{align}
with
\begin{align}
A(\as)=& \left(\abar\right) A^{(1)} + \left(\abar\right)^2 A^{(2)}+ \left(\abar\right)^3 A^{(3)}\,,\notag\\
\tilde{B}_{\flavB}(\as)=& \left(\abar\right) B^{(1)} + \left(\abar\right)^2 \tilde{B}_{\flavB}^{(2)}\,.
\end{align}

The flavour dependence of these quantities originates entirely from the hard-virtual coefficient 
function $H_{\flavB}$,\footnote{Note that $\tilde{H}_{\flavB}$ in \eqn{eq:luminosity} just includes an additional shift with respect to $H_{\flavB}$, see eq. (4.26) of \citere{Monni:2019whf}.} which contains the virtual corrections that, for 
a general $2\to 2$ hadronic process, become dependent on the flavour and on the Born 
phase-space $\PhiB$. 
Up to two loops it is given by
\begin{align}
\label{eq:Hdef}
  H_{\flavB}(\pt) =&  \,1  +  \left( \frac{\as(\pt)}{2\pi} \right)\, H_{\flavB}^{(1)} + \left( \frac{\as(\pt)}{2\pi} \right)^2\, H_{\flavB}^{(2)}\,.
\end{align}
This dependence propagates through the $\tilde{B}_{\flavB}^{(2)}$ coefficient to the Sudakov form 
factor, since the derivation of the \minnlo{} formalism is based on setting the renormalization scale 
$\muR\sim \pt$, which exponentiates the $H_{\flavB}^{(1)}$ coefficient and requires a redefinition
  of $\tilde{B}_{\flavB}^{(2)}$, 
as discussed in the derivation of the replacement in eq.~(4.26) of \citere{Monni:2019whf}:
\begin{align}
\tilde{B}_{\flavB}^{(2)} =& B^{(2)} + 2\zeta_3 (A^{(1)})^2 + 2 \pi \beta_0
                   H_{\flavB}^{(1)}\,,
\end{align}
where $\beta_0 = \frac{11 \CA{} - 2 \nf}{12\pi}$.
A few comments are in order: all quantities with index $\flavB$ also depend 
on the Born kinematics. For ease of notation we do not indicate explicitly their $\PhiB$ 
dependence. 
The $G$ functions~\cite{Catani:2010pd} in \eqn{eq:luminosity} are present only in the case of 
gluon-induced reactions, i.e.\ they are zero for $Z\gamma$ production and kept here 
only for completeness. 
For a colour-singlet process the Born flavours $\flavB$ correspond to the two initial-state partons, which 
have been denoted with $cc'$ in \eqn{eq:luminosity}. In that equation
$|M^{\scriptscriptstyle\rm F}|_{cc'}^2$ denotes the Born matrix element squared, $\tilde{C}$ are the collinear coefficient functions, and $f_{i/j}^{[a/b]}$ are the parton densities. 
A crucial feature of \minnlo{} is that both the renormalization
and the factorization scales are set to $\muR\sim\muF\sim \pt$.
The precise definition of the first- and second-order hard functions $H^{(1)}_{\flavB}$ and  $H^{(2)}_{\flavB}$ is given in \sct{sec:details}.

Writing also the differential NLO cross section for \FJ{} production as a sum over its flavour structures $\flavBJ$
\begin{equation}
\label{eq:NLO}
\frac{\mathd\sigma^{\rm (NLO)}_{\scriptscriptstyle\rm FJ}}{\mathd\PhiB\mathd
      \pt} =  \sum_{\flavBJ} \left\{\abarmu{\pt}\left[\frac{\mathd\sigma_{{\scriptscriptstyle\rm FJ}}}{\mathd\PhiB\mathd
      \pt}\right]^{(1)}_{\flavBJ}  + \left(\abarmu{\pt}\right)^2\left[\frac{\mathd\sigma_{\scriptscriptstyle\rm FJ}}{\mathd\PhiB\mathd
      \pt}\right]^{(2)}_{\flavBJ}\,\right\},
\end{equation}
where $[X]^{(i)}$ denotes the coefficient of the
$i^{\rm th}$ term in the perturbative expansion of the quantity $X$,
allows us to recast our starting formula in \eqn{eq:start} as
\begin{align}
\label{eq:minnlo}
&  \frac{\mathd\sigma}{\mathd\PhiB\mathd \pt}  = \sum_{\flavBJ} \Bigg\{\exp[-\tilde{S}_{\projflav}(\pt)] \bigg\{\abarmu{\pt}\left[\frac{\mathd\sigma_{\scriptscriptstyle\rm FJ}}{\mathd\PhiB\mathd \pt}\right]^{(1)}_{\flavBJ} \left(1+\abarmu{\pt} [\tilde{S}_{\projflav}(\pt)]^{(1)}\right)\notag
\\
&\quad+ \left(\abarmu{\pt}\right)^2\left[\frac{\mathd\sigma_{\scriptscriptstyle\rm FJ}}{\mathd\PhiB\mathd \pt}\right]^{(2)}_{\flavBJ}\bigg\} + \sum_{\flavB} \exp[-\tilde{S}_{\flavB}(\pt)] \,\mathcal{D}_{\flavB}(\pt)+ {\rm
  regular~terms~of~{\cal O}(\as^3)}\Bigg\}\,,
\end{align}
where $\projflav$ denotes the projection from the flavour structure of \FJ{} production to the one of the Born process \F{}.
This projection is trivial in the case of $Z\gamma$ production, as the Born is always $q\bar{q}$ initiated and only the 
respective quark-flavour is of relevance in the $\flavB$ dependence of $\tilde{S}_{\flavB}(\pt)$.
Furthermore, we have introduced a new symbol $\mathcal{D}_{\flavB}(\pt)$ that accounts 
for the relevant NNLO corrections with the following two choices of treating 
terms beyond accuracy. In the original 
\minnlo{} formulation of \citere{Monni:2019whf} we have truncated \eqn{eq:minnlo} to third order in $\as(\pt)$, i.e.\
\begin{align}
\label{eq:Dold}
\mathcal{D}_{\flavB}(\pt) \equiv \left(\abarmu{\pt}\right)^3 [D_{\flavB}(\pt)]^{(3)}+\mathcal{O}(\as^4)\,.
\end{align}
With this truncation at the differential level \eqn{eq:minnlo} does not 
reproduce anymore the exact total derivative that we started with in \eqn{eq:start}. In order to preserve the total 
derivative and keep into account additional terms beyond accuracy, a new prescription has been 
suggested in \citere{Monni:2020nks}
\begin{align}
\label{eq:Dnew}
\mathcal{D}_{\flavB}(\pt) \equiv D_{\flavB}(\pt) -\abarmu{\pt} [D_{\flavB}(\pt)]^{(1)} -\left(\abarmu{\pt}\right)^2 [D_{\flavB}(\pt)]^{(2)}\,,
\end{align}
which will be our default choice throughout this paper. 
The relevant expressions for its evaluation, including the ones of 
the $[D_{\flavB}(\pt)]^{(i)}$ coefficients, 
are reported in appendix C and D of \citere{Monni:2019whf} and in appendix A of \citere{Monni:2020nks}, where the flavour dependence can be simply included 
through the replacements $H^{(1)}\rightarrow H^{(1)}_{\flavB}$, $H^{(2)}\rightarrow H^{(2)}_{\flavB}$, and $\tilde B^{(2)}\rightarrow \tilde B^{(2)}_{\flavB}$.
Eq.~\eqref{eq:minnlo}, upon integration over $\pt$, is NNLO accurate with both 
choices of $\mathcal{D}_{\flavB}(\pt)$, as they differ only by
terms of ${\cal O}(\as^4)$ and higher. As discussed in detail in
\citeres{Monni:2019whf,Monni:2020nks}, this is achieved by the consistent inclusion of all singular 
terms to third order in $\as(\pt)$.

We can now make the connection to \POWHEG{} and the parton-shower matching formula in \eqn{eq:master}. 
The formulation of \eqn{eq:minnlo} applies also to the fully differential cross section in the $\PhiBJ$ 
phase space, and it can be used to achieve NNLO accuracy for the
${\bar B}(\PhiBJ)$ function~\cite{Monni:2019whf,Monni:2020nks}
\begin{align}
\label{eq:Bbar}
{\bar B}(\PhiBJ)&\equiv \sum_{\flavBJ} \Bigg\{\exp[-\tilde{S}_{\projflav}(\pt)] \bigg\{\abarmu{\pt}\left[\frac{\mathd\sigma_{\scriptscriptstyle\rm FJ}}{\mathd\PhiBJ}\right]^{(1)}_{\flavBJ} \left(1+\abarmu{\pt} [\tilde{S}_{\projflav}(\pt)]^{(1)}\right)\notag
  \\
&+ \left(\abarmu{\pt}\right)^2\left[\frac{\mathd\sigma_{\scriptscriptstyle\rm FJ}}{\mathd\PhiBJ}\right]^{(2)}_{\flavBJ}\bigg\} + \bigg\{\sum_{\flavB} \exp[-\tilde{S}_{\flavB}(\pt)]\,\mathcal{D}_{\flavB}(\pt)\bigg\}\, F^{\tmop{corr}}_{\flavBJ}(\PhiBJ)\Bigg\}\,,
\end{align}
where $F^{\tmop{corr}}_{\flavBJ}(\PhiBJ)$ guarantees a proper spreading of
the Born-like NNLO corrections in the full $\PhiBJ$ phase space, as discussed 
in detail in Section 3 of \citere{Monni:2019whf} and briefly recalled in the next section.

\subsection{Practical details of the implementation within \POWHEGBOXRES{}}
\label{sec:details}
We have applied the \minnlo{} formalism discussed in the previous section
to our implementation of the $Z\gamma$+jet generator in \POWHEGBOXRES{}.
In the following we discuss practical aspects of our calculation in that framework
and set $\text{\F}=Z\gamma$ from now on.

We briefly recall how the NNLO corrections, which have Born kinematics 
and additionally depend on $\pt$, are included in $Z\gamma$+jet generator. 
The relevant kinematics is obtained by performing a phase-space projection  
$\PhiZJ\to\PhiZgam$ and by determining $\pt{}$ from the given $\PhiZJ$ phase-space point.
Furthermore, the Born-like NNLO corrections need to be distributed in the $\PhiZJ$ kinematics. 
There is a certain degree of freedom in how to associate 
$(\PhiZgam,\pt)$ with the full $\PhiZJ$ phase space. This is encoded in 
the factor $F^{\rm corr}(\PhiZJ)$ of \eqn{eq:Bbar}
\begin{align}
F^{\rm corr}(\PhiZJ)=\frac{ J_{\flavZgJ} (\PhiZJ) }{  \sum_{\flavZgJ'} \int
  \mathd \PhiZJ' J_{\flavZgJ'} (\PhiZJ') \delta (\pt - \pt')
  \delta (\PhiZgam - \PhiZgam')}\,,
\label{eq:spreading}
\end{align}
and in the details of the $Z\gamma$J $\to Z\gamma$ projection.
The functions $J_{\flavZgJ} (\PhiBJ)$ for $\flavZgJ = \{q\bar{q},qg,gq\}$ have been chosen according to 
the collinear limit of the tree-level matrix element squared of the $Z\gamma$+jet process, using eq.~(A.14) 
of \citere{Monni:2019whf}. 
This is a suitable compromise between computational speed and physically sound results, as
the spreading is performed according to the pseudorapidity distribution of the
radiation described by that approximation. 
With this choice the numerical convergence of the integral in the denominator 
of \eqn{eq:spreading} does not scale with the complexity of the process, which 
is a crucial requirement for multi-leg processes, such as the one at hand.
We stress that the spreading factor in \eqn{eq:spreading} is designed in such a way that the integral over the $\PhiZJ$ phase space of its product with any function of the $\PhiZgam$ Born variables yields exactly 
the integral of that function, when integrated over the $\PhiZgam$ phase space (see eq.~(3.2) of \citere{Monni:2019whf}).
Details of our projection to $Z\gamma$ events are given in~\app{app:UUB}.
Note that this projection does not preserve the full Born kinematics. While it keeps 
all invariant masses and  the rapidity of the $Z\gamma$ system unchanged, it does alter for instance the 
transverse momentum of the photon.
As a result, the photon transverse momentum after the $\PhiZJ\to\PhiZgam$ projection
is neither bounded from below by the technical generation cuts nor controlled by the phase-space suppression factor 
introduced for the $\PhiZJ$ kinematics in \sct{sec:photon}. This induces a singular behaviour 
through the Born and virtual amplitudes in both the Sudakov form factor and the luminosity factor 
defined in \neqn{eq:luminosity} and \eqref{eq:minnlops-Sudakov}. We have therefore added 
the requirement $\ptg\ge10$\,GeV in the projected $\PhiZgam$ kinematics. This technical cut is
below the $\ptg{}$ threshold used at analysis level and it can be controlled through the input card.
Its effect is strictly beyond accuracy, affecting only regular contributions at large $\ptzg{}$.
In fact, as discussed in ~\app{app:UUB}, our projection can lead to configurations with $\ptg\to 0$\,GeV only for events where the jet is back to back to the $Z\gamma$ system, and the $Z$ and the photon are aligned with each other. It is then clear that for such events $\ptj > \ptg$ and there are no large logarithms associated to $\ptj$. 
Indeed, we have varied the cutoff down by a factor of ten, finding changes at
the level of the numerical precision of less than a $1\%$. 

Embedding the \minnlo{} corrections within the \POWHEGBOXRES{} framework 
requires some further discussion in respect to the resonance-aware features.
Since the ${\bar B}(\PhiZgam)$ function is modified in an additive way in \eqn{eq:Bbar}, such that 
the NNLO corrections are treated on the same footing as all other contributions with $\PhiZJ{}$ kinematics, 
we decompose them as a weighted sum over $\fullflavBorn$ using the weight functions 
${\cal P}_{\fullflavBorn}$ just like in \eqn{eq:resdecomposition}.
Our final formula for the $Z\gamma$ \minnlo{} generator reads 
\begin{align}
\label{eq:Bbar2}
&{\bar B}(\PhiZJ)\equiv \sum_{\flavZgJ} \Bigg\{\hspace{-0.1cm}\exp[-\tilde{S}_{\projflavZg}(\pt)] \bigg\{\abarmu{\pt}\left[\frac{\mathd\sigma_{\scriptscriptstyle Z\gamma {\rm J}}}{\mathd\PhiZJ}\right]^{(1)}_{\flavZgJ} \hspace{-0.1cm}\left(1+\abarmu{\pt} [\tilde{S}_{\projflavZg}(\pt)]^{(1)}\right)\notag
  \\
&+ \left(\abarmu{\pt}\right)^2\left[\frac{\mathd\sigma_{\scriptscriptstyle\rm Z\gamma {\rm J}}}{\mathd\PhiZJ}\right]^{(2)}_{\flavZgJ}\bigg\} + \hspace{-0.25cm}\sum_{\fullflavZg\in T(\flavZgJ)} \hspace{-0.45cm}{\cal P}_{\fullflavZgJ} \bigg\{\sum_{\flavZg} \exp[-\tilde{S}_{\flavZg}(\pt)]\,\mathcal{D}_{\flavZg}(\pt)\bigg\}\, F^{\tmop{corr}}_{\flavZgJ}(\PhiZJ)\Bigg\}\,.
\end{align}

In the following we provide some details on the treatment of higher-order 
terms and the scale settings. The discussion summarizes briefly the one 
in section 3.2 of \citere{Monni:2020nks}, which we refer to for more details.
In the large transverse-momentum region it is important to switch off
all-order resummation effects to avoid spurious contributions. 
As is standard, we do it by introducing
modified logarithms via the replacement 
\begin{align}
  \log\left(\frac{Q}{\pt}\right) \to L\equiv \frac{1}{p} \log\left(1+\left(\frac{Q}{\pt}\right)^p\right)\,,
  \label{eq:modlog}
\end{align}  
where we set here $p=6$. As pointed out in \citere{Monni:2020nks},
in order to preserve the total derivative of \eqn{eq:start}, this prescription requires three further adjustments. 
The lower integration bound of the Sudakov is to be replaced 
  by $\pt \to Qe^{-L}$; $\mathcal{D}_{\flavB}(\pt)$ (or $\mathcal{D}_{\flavB}(\pt)^{(3)}$) needs to be multiplied by a proper jacobian factor, see eq. (13) and (16) 
of \citere{Monni:2020nks}; the perturbative scales need to be set consistently 
with the scale of the modified logarithms at small transverse momenta.
Additionally, we smoothly approach non-perturbative scales at small $\pt{}$ 
and introduce the non-perturbative parameter $\Q$ to regularize the Landau
singularity, setting the central renormalization and factorization scales to \cite{Monni:2020nks}
\begin{align}
  \label{eq:modifiedscales2}
  \muRc =  Qe^{-L} +\Q\, g(\pt)\,,\qquad
  \muFc =  Qe^{-L} +\Q\, g(\pt)\,,
\end{align}
where we set $\Q=0.5$\,GeV, and $g(\pt)$ has been chosen
so as to suppress the $\Q$ shift at large values of $\pt$:
\begin{align}
g(\pt) = \left(1+\frac{Q}{\Q}e^{-L}\right)^{-1}\,.
\end{align}
The scale setting of \eqn{eq:modifiedscales2} provides a consistent 
treatment in the small $\pt{}$ region and preserves the total derivative
of \eqn{eq:start}. However, at large $\pt{}$ it yields $\muRc\sim\muFc\sim Q$,
while a scale setting of $\muRc\sim\muFc\sim \pt$ might be preferred in 
that region. To this end,
the scales entering the NLO $Z\gamma +$jet cross section in \eqn{eq:Bbar2}
can be set via the flag {\tt largeptscales 1} to
\begin{align}
  \label{eq:modifiedscales3}
  \muRc =  \pt +\Q\, g(\pt)\,,\qquad
  \muFc =  \pt +\Q\, g(\pt)\,,
\end{align}
which we apply in our calculation. It is important that \eqn{eq:modifiedscales3}
matches the scales of the Sudakov form factor and the 
$\mathcal{D}_{\flavB}(\pt)$ terms
at small $\pt$. At the same time it ensures a dynamical scale choice of $\muRc = \muFc \sim \pt$ at large $\pt{}$.

Next, we specify the precise definition of the first and second order hard-virtual function of \eqn{eq:Hdef} 
for $Z\gamma$ production in our subtraction scheme. First, let us recall that all 
tree-level and one-loop amplitudes have been extracted from \noun{MCFM} and for comparison also 
linked through \noun{OpenLoops}, and that the (one-loop and) two-loop 
$q\bar{q}\to \ell\ell\gamma$ amplitudes have been obtained by creating an interface to their implementation 
in \Matrix{}.
At variance with the $2\to1$ processes considered in \citere{Monni:2019whf}, it is not possible to provide 
compact expressions for the hard function of $Z\gamma$ production.
For brevity, we thus start from the expressions of the first and second order
hard-virtual function where IR singularities have been subtracted according to the {\it $\qt{}$-scheme} (more precisely, choosing the {\it hard-scheme} \cite{Catani:2013tia} 
as resummation scheme) that we denote as $H_{\flavB}^{\qt(1)}$ and $H_{\flavB}^{\qt(2)}$
in the following. Those coefficients
are unambiguously defined in eqs. (12) and (62) of \citere{Catani:2013tia} and can be extracted from 
the one-loop and two-loop virtual amplitudes using the expressions of that paper. In fact, the \Matrix{} inteface directly provides the hard-virtual
coefficients in that scheme, so that we only need to perform 
the appropriate scheme conversion to match the \minnlo{} conventions \cite{Catani:2000vq}:\footnote{This scheme conversion follows directly 
from the fact that the hard-scheme is defined in \citere{Catani:2013tia} such that the collinear coefficient functions do not contain any $\delta (1-z)$ terms and have been absorbed into $H_{\flavB}^{\qt}$ instead.}
\begin{align}
\label{eq:Hidef}
  H_{\flavB=cc'}^{(1)} &= H_{\flavB=cc'}^{\qt(1)} - 2C^{(1),\delta}_{cc}\,,\\
  H_{\flavB=cc'}^{(2)} &= H_{\flavB=cc'}^{\qt(2)} -(C^{(1),\delta}_{cc})^2 +2 C^{(2),\delta}_{cc}-2C^{(1),\delta}_{cc}(H_{\flavB=cc'}^{\qt(1)}-2 C^{(1),\delta}_{cc})\,
\end{align}
where $C^{(1),\delta}_{cc}$ and $C^{(2),\delta}_{cc}$ are the terms proportional to $\delta(1-z)$ of the first and second order coefficients of the collinear coefficient function, which in the case of a quark-induced 
process ($cc'=q\bar{q}$) are given by
\begin{align}
\label{eq:Hdef2}
  \tilde C^{(1),\delta}_{qq} &=  -\CF{} \frac{\pi^2}{24}\,,\\
  \tilde C^{(2),\delta}_{qq} &=  \frac{9\CF^2\pi^4+2\CA{}\CF{}(4856-603\pi^2+18\pi^4-2772\zeta_3)+4\nf \CF{}(-328+45\pi^2+252\zeta_3)}{10368}\,,
\end{align}
where $\CF{}=4/3$ and $\CA{}=3$, and $\nf$ is the number of light quark flavours.
Note that $\tilde C^{(1/2)}_{qq}$ in the \minnlo{} convention can be obtained from the ones of \citere{Catani:2013tia}
by simply adding $\tilde C^{(1/2),\delta}_{qq}\times\delta(1-z)$.
For completeness, we also provide the corresponding expressions for the 
gluon-induced case ($cc'=gg$)
\begin{align}
\label{eq:Hdef2}
  \tilde C^{(1),\delta}_{gg} &= - \CA{} \frac{\pi^2}{24}\,,\\
  \tilde C^{(2),\delta}_{gg} &=  -\frac{10718}{864} - \frac{5 \CA{}}{192} - \frac{\CF{}}{24} + \frac{9\CF{}^2}{8} 
   - \frac{1679\pi^2}{192} - \frac{37\pi^4}{64} +   \CA{}\CF{}\biggl(-\frac{145}{48} 
            - \frac{7\pi^2}{16}\biggr) \nonumber \\
            &+ \CA{}^2\biggl(\frac{3187}{576} 
            + \frac{43\pi^2}{36} + \frac{79\pi^4}{1152} - \frac{55\zeta_3}{36}\biggr) +   \CA{}\nf\biggl(-\frac{287}{288} - \frac{5\pi^2}{72} - \frac{2\zeta_3}{9}\biggr) \nonumber \\
            &+ \CF{}\nf\biggl(-\frac{41}{48} 
            + \frac{\zeta_3}{2}\biggr) + \frac{499\zeta_3}{48}\, 
\end{align}

Finally, we conclude this section by reporting two further
non-standard settings related to the showering of the $Z\gamma$
\minnlo{} events.  First, we turn on by default the \POWHEGBOX{}
\texttt{doublefsr} option, which was introduced and discussed in
detail in \citere{Nason:2013uba}.  When this option is turned on, the
emitter-emitted role is exchanged in events for which the emitter
would be softer than the emitted particle. This considerably reduces
the appearance of spikes in distributions due to events with large
weights that pass fiducial cuts after showering.
Furthermore, for the \PYTHIA{8} shower \cite{Sjostrand:2014zea}, we set
the flag \texttt{SpaceShower:dipoleRecoil 1} (see
\citere{Cabouat:2017rzi}). Its effect is to have a global recoil
(i.e. a recoil which affects all final state particles including the
color-singlet system) only for initial-initial colour-dipole
emissions, while for initial-final ones a local recoil scheme is
used. The reason for this choice is that,
as shown in
\citere{Monni:2020nks},
a local recoil for initial-final emissions 
reduces the effect of the
shower on the colour-singlet kinematics, in particular in large
rapidity regions.

\section{Phenomenological results}
\label{sec:phenomenology}
In this section, we present NNLO+PS accurate predictions for $Z\gamma$ production. We consider different 
fiducial selections discussing both integrated cross sections and differential distributions in presence of fiducial cuts. 
\minnlo{} predictions are compared to NNLO and \minlo{} results. This allows us to validate the accuracy 
of our predictions for observables inclusive over QCD radiation and observables requiring the presence of jets.
At the same time, we demonstrate the importance of both NNLO accuracy in the event simulation and the inclusion 
of parton-shower effects. Finally, we compare \minnlo{} predictions to a recent ATLAS measurement~\cite{Aad:2019gpq}.

\subsection{Input parameters, settings and fiducial cuts}
\label{sec:setup}
We present predictions for 13 TeV collisions at the LHC. Our results have been obtained by using 
the $G_\mu$-scheme, where the electroweak coupling is defined as
\begin{align}
  \alpha_{G_\mu}=\frac{\sqrt{2}G_\mu \mw^2 \sin^2\thW}{\pi}\,,
  \label{eq:EW}
\end{align}
where $\cos^2\thW=\mw^2/\mz^2$ and the input parameters are set to 
\begin{align}
 \mw &= 80.385~{\rm GeV}\,, \quad \mz = 91.1876~{\rm GeV}\,, \,\,\,\, \, \, G_\mu = 1.16637 \times 10^{-5}~{\rm GeV}^{-2}\,,  \nonumber \\
\GW &= 2.085~{\rm GeV}\,, \quad \quad \GZ = 2.4952~{\rm GeV}\,.
\end{align}
We use $\nf=5$ massless quark flavours, and we choose the corresponding NNLO PDF set of NNPDF3.0~\cite{Ball:2014uwa} with
a strong coupling constant of $\as(\mz)=0.118$. In the case of the fixed-order predictions we use the PDF set at the respective order
in QCD perturbation theory. To be precise, for \minlo{} and \minnlo{} we read the PDF grids using the \noun{lhapdf} interface~\cite{Buckley:2014ana}, copy them into
\noun{hoppet} grids~\cite{Salam:2008qg} and  then use the \noun{hoppet}
evolution code to perfom the DGLAP evolution of the PDFs,  also for
scales below the internal PDF infrared cutoff, as pointed out in
\sct{sec:details}.  
The calculation of  $\mathcal{D}_{\flavB}(\pt)$ in \eqn{eq:Dnew} requires 
the evaluation of different PDF convolutions
and the computation of polylogarithms.
For the latter we made use of the \noun{hplog} package~\cite{Gehrmann:2001pz}. 

Our setting of the central renormalization and factorization scales ($\muRc$, $\muFc$), which is in line with the \minnlo{} (\minlo{})
method, has been discussed at length in \sct{sec:details}, see \neqn{eq:modifiedscales2} and \eqref{eq:modifiedscales3}.
In all fixed-order results presented throughout this section we adopt the following setting of the central renormalization and factorization scales: 
\begin{align}
   \muRc=\muFc=\mllg{}\,,
  \label{eq:facrenscale}
\end{align}
where $\mllg{}$ is the invariant mass of the $Z\gamma$ system.
The different scale choice between a NLO/NNLO fixed-order and a \minlo{}/\minnlo{} calculation 
induces effects beyond the nominal accuracy, in addition to the different treatment 
of higher-order terms, see comments close to \eqn{eq:Dnew}. As a results, minor differences between 
the fixed-order and matched predictions are expected even for more inclusive cross-sections. Nevertheless, 
the results should largely agree within scale uncertainties, at least in cases where scale uncertainties are 
expected to be a reliable estimate of missing higher-order corrections.

We employ $7$-point scale variations as an estimate of the theoretical uncertainties. 
Accordingly, the minimum and maximum of the cross section have been taken over a set of scales
$(\muR,\muF)$ obtained by varying $\KR$ and $\KF$  in $\muR=\KR\, \muRc$ and
$\muF=\KF\, \muFc$, respectively, within the values:
  \begin{equation}
\label{eq:sevenpoint}
(\KR,\KF)=\left\{(2,2), (2,1), (1,2), (1,1), \left(1,1/2\right),\left(1/2,1\right),\left(1/2,1/2\right)\right\},
  \end{equation}
  We reiterate that, when performing scale variations for \minlo{} and \minnlo{},
  we also include the scale dependence of the Sudakov form factor to
  account for additional sources of uncertainties~\cite{Monni:2019whf}, which
  are absent in a fixed-order calculation.

For all technical details and choices made for our implementation of the $Z\gamma$+jet
generator as well as 
for the treatment of the NNLO corrections through the \minnlo{} method 
we refer the reader to \sct{sec:photon} and in \sct{sec:details}, respectively.
In particular, the settings discussed in \sct{sec:photon} are essential
to get a good convergence of the Monte Carlo integration and an
efficient event generation, by adopting generation cuts and individual suppression 
factors at Born level, for the singular real contributions and the remnant contributions
(cf. also \app{app:cuts}).
In addition, the folding of the radiation variables $(\xi, y, \phi)$ \cite{Alioli:2010qp,Alioli:2010xd} has been 
used (with a choice of $(1,5,1)$ for the folding parameters) to evaluate the double-real correction ($Z\gamma$+2-jet) more often, which further 
improves the numerical convergence.
Despite all those efforts to achieve a better numerical performance, we had to produce
$\sim 100$ million $Z\gamma$ events with our \minnlo{} generator to obtain 
acceptable statistical uncertainties and predict integrated cross sections in the fiducial setups 
considered here at the level of a few permille. Still, our comparison of differential distributions to 
NNLO predictions suffers from some fluctuations. We note, however, that with $Z\gamma$ production 
we have picked probably the most involved diboson processes, featuring various complications, 
in particular its substantial complexity with respect to the QED singularity structure. We therefore 
expect other diboson processes to have a much lower demand for numerical resources.

As pointed out before, we omit the loop-induced gluon-gluon component
in our implementation and throughout this paper. This contribution is relatively
small, being only $\sim 1\%$ of the NNLO cross section, and it can be 
incoherently added to our predictions through a dedicated calculation, which, 
however, is beyond the scope of this paper. Consequently, we drop the 
loop-induced gluon fusion contribution also from the fixed-order calculation, 
when comparing our \minnlo{} against NNLO results, to warrant a meaningful
comparison.

We employ the \PYTHIA{8} parton shower~\cite{Sjostrand:2014zea} with the one of the 
A14 tunes~\cite{TheATLAScollaboration:2014rfk} (specifically {\tt py8tune 21}) to dress the hard event with further soft/collinear QCD radiation
with the default \POWHEG{} setting for the parton-shower starting scale. 
  The showered results do not include any effects from hadronization or underlying event models. 
  Moreover, the photon is required to be generated only at the hard scattering level: contributions 
  from a QED shower or the decay of unstable particles is not included. Finally, we keep photons stable by preventing any photon conversion effect, i.e.\ no $\gamma \to \ell^+ \ell^-$ or $\gamma \to \bar{q} q$ splitting.

\begin{table}[t]
\centering
\begin{tabular}{| c || c | c |}
\hline 
  & \setupone~\cite{Aad:2013izg} & \setuptwo~\cite{Aad:2019gpq} \\
\hline
\hline 
Lepton cuts & $\ptl>25$\,GeV\quad$|\etal|<2.47$& $\ptlone>30$\,GeV$\quad \ptltwo>25$\,GeV\\
&$\mll>40$\,GeV & $|\etal|<2.47$\quad$\mll>40$\,GeV\\
&--&$\mll+\mllg>182$\,GeV\\
\hline
Photon cut & $\ptg>15$\,GeV\quad$|\etag|<2.37$ & $\ptg>30$\,GeV\quad$|\etag|<2.37$ \\
\hline
Separation cuts & $\drlg>0.7$  & $\drlg>0.4$ \\
& $\drlj>0.3\quad\drgj>0.3$  &--\\
\hline
Jet definition & anti-$k_T$ algorithm with $R=0.4$ &  --\\
&$\ptj>30$\,GeV\quad$|\etaj|<4.4$& --\\
\hline
Photon Isolation & Frixione isolation with & Frixione isolation with \\
&$n=1\quad
\epsilon_{\text{\scalefont{0.77}$\gamma$}}
=0.5\quad\delta_{\text{\scalefont{0.77}$0$}}=0.4$&$n=2\quad\epsilon_{\text{\scalefont{0.77}$\gamma$}}=0.1\quad\delta_{\text{\scalefont{0.77}$0$}}=0.1$\\
&&$+\quad\fcl<0.07$\\
\hline 
\end{tabular}
\caption{\label{tab:setup} Fiducial cuts in two different ATLAS setups denoted as \setupone{} and \setuptwo{}. See text for details.}
\end{table}
We present results for two sets of fiducial cuts, which are summarized in \tab{tab:setup}.
The first one is identical to that used in
\citeres{Grazzini:2015nwa,Grazzini:2017mhc} and motivated by in an earlier ATLAS analysis~\cite{Aad:2013izg}.
We refer to it as \setupone{} in the following.
The second one was instead used in the most recent ATLAS 13\,TeV measurement of \citere{Aad:2019gpq}
using the full Run-2 data and named \setuptwo{} in the following.
We make use of \setupone{} for validation purposes and to show the importance of NNLO+PS matching, 
while \setuptwo{} is also used to compare \minnlo{} predictions with the most updated
experimental measurement available for $Z\gamma$ production.
Both setups in \tab{tab:setup} involve standard transverse momentum and pseudorapidity thresholds 
to identify leptons and photons as well as a lower invariant-mass cut on the lepton pair. \setuptwo{}
places an additional requirement on the sum of the invariant masses of the $Z\gamma$ system and of 
the lepton pair. This cut suppresses the contribution from $\ell$-type diagrams, where the photon is emitted from 
the final state leptons (cf.~\fig{subfig:ltypeZgam}), enhancing $t$-channel production through $q$-type diagrams 
(cf.~\fig{subfig:qtypeZgam}).
Moreover, separation cuts between two particles ($i$, $j$) are applied in  
$\Delta R_{ij}=\sqrt{\Delta\eta_{ij}^2+\Delta\phi_{ij}^2}$, where $\Delta\eta_{ij}$ and $\Delta\phi_{ij}$
are their differences in the pseudorapidity and the azimuthal angle, respectively.
In both setups leptons are separated from the isolated photon, while 
only \setupone{} imposes an additional separation of jets from leptons and from 
the isolated photon, which in turn requires a jet definition.
As a consequence, we employ \setupone{} to study jet observables
and show NLO/LO accuracy of \minnlo{} 
predictions for $Z\gamma$+jet/$Z\gamma$+2-jet configurations.
Finally, isolation criteria for the photon are needed, as detailed in \sct{sec:photon},
which is done by means of Frixione isolation in both setups. 
In \setuptwo{}, Frixione isolation in a smaller cone and a second isolation criterium is applied by requiring
the scalar sum of the transverse energy of all stable particles
(except neutrinos and muons) within a cone around the photon of 
size $R = 0.2$ to be less than $7\%$ of the photon transverse momentum (see~\cite{Aad:2019gpq} for more details). Note that we apply the latter 
isolation criterium only when analyzing events after parton showering, 
but not at Les-Houches-Event (LHE) or fixed-order level.

\subsection{Fiducial cross sections}
\label{sec:inclusive}

\renewcommand\arraystretch{1.3}
\begin{table}[t]
\begin{center}
\begin{tabular}{| l  ||   c | c || c|  c |}
\hline 
  & \multicolumn{2}{c||}{ATLAS setup $1$ ~\cite{Aad:2013izg}}
& \multicolumn{2}{c|}{ATLAS setup $2$ ~\cite{Aad:2019gpq}} \\
\hline
& $\sigma_{\rm inclusive}$ [pb]
& $\sigma/\sigma_{\rm NNLO}$
& $\sigma_{\rm inclusive}$ [fb]
& $\sigma/\sigma_{\rm NNLO}$ \\
\hline
LO   & $1.5032(1)_{-11.9\%\phantom{1}}^{+11.2\%\phantom{1}}$ & 0.656 & $271.83(2)_{-7.8\%}^{+6.8\%}$  & 0.508 \\
NLO  & $2.1170(5)_{-4.3\%\phantom{1}}^{+2.8\%\phantom{1}}$ & 0.924 & $456.6(1)_{-3.0\%\phantom{1}}^{+3.6\%\phantom{1}}$ & 0.853 \\
NNLO &  $2.290(3)_{-1.0\%\phantom{1}}^{+0.9\%\phantom{1}}$ &  1.000 & $535.3(6)_{-2.5\%\phantom{1}}^{+2.7\%\phantom{1}}$ & 1.000 \\
\minlo{}  & $2.222(8)_{-11.0\%\phantom{1}}^{+8.8\%\phantom{1}}$ & 0.970 & $516(4)_{-6.5\%}^{+8.8\%}$ & 0.964 \\
\minnlo{}  & $2.276(5)_{-1.1\%}^{+1.2\%}$ & 0.994 & $533(3)_{-3.1\%\phantom{1}}^{+3.9\%\phantom{1}}$ & 0.996 \\
\hline
ATLAS & \multicolumn{2}{c||}{--} & \multicolumn{2}{c|}{$533.7 \pm 2.1 \text{\scalefont{0.55}(stat)}\pm 12.4 \text{\scalefont{0.55}(syst)} \pm 9.1\text{\scalefont{0.55}(lumi)}$} \\
\hline 
\end{tabular}                                                                                                                                                                                      
\end{center}
\renewcommand{\baselinestretch}{1.0}
\caption{ \label{tab:crosssection} Predictions for fiducial cross sections of $Z\gamma$
  production at LO, NLO, and NNLO, as well as using the \minlo{} and \minnlo{} calculations,
  in the two ATLAS setups. For comparison also a column with the ratio to the NNLO cross section is shown.
  In the last row, the ATLAS measurement of \citere{Aad:2019gpq} is reported.
}
\end{table}
\renewcommand\arraystretch{1}

In \tab{tab:crosssection} we report predictions for the $Z\gamma$ cross section in the two fiducial setups
at LO, NLO and NNLO, and for \minlo{} and \minnlo{} matched to \PYTHIA{8}. 
The fixed-order results have been obtained with \Matrix{}.
Although \minnlo{} and NNLO calculations entail a different treatment of terms beyond accuracy, 
in both setups the agreement of their predicted cross sections is remarkably good.
One should bear in mind, however, that in \setuptwo{} there is a slight difference in the treatment
of the isolated photon at fixed order, which does not include the $\fcl<0.07$ cut, as discussed in the previous section.
We further notice from \tab{tab:crosssection}  
that the scale uncertainties of the theoretical predictions are increased in \setuptwo{}.
This is caused by the additional $\mll+\mllg$ cut and
the larger cut on the photon transverse momentum in that 
fiducial setup, rendering the predictions 
more sensitive to additional QCD radiation that is described at a lower perturbative accuracy.

Comparing \minnlo{} and \minlo{} predictions, 
the inclusion of NNLO corrections through \minnlo{} 
has a rather moderate effect for the fiducial cross section of $+2\%$ in \setupone{} and $+0.5\%$ in \setuptwo{}. 
In fact, in both cases (and particularly evident 
for the latter setup) \minlo{} predictions are actually closer to the 
NNLO results than to the NLO ones. After all, 
the Sudakov form factor in \eqn{eq:minnlops-Sudakov} is exactly the same for \minnlo{} and \minlo{}, and \minlo{} predictions 
already contain various contributions beyond NLO accuracy,
including all real corrections at NNLO
through the merging of NLO corrections to $Z\gamma$+jet production. 
Still, by reaching NNLO accuracy through \minnlo{} the predictions get even closer to the NNLO results and the uncertainty bands  substantially decrease, by almost a factor ten in \setupone{} and 
by more than a factor 2 in \setuptwo{}. 
Indeed, the \minnlo{} scale uncertainties are 
comparable with the NNLO ones. The fact that they are 
slightly larger is expected since the \minnlo{} procedure probes lower scales both in 
the PDFs and in $\as$, and it includes scale variations also for the Sudakov form factor.

Finally, we find a very good agreement of our NNLO+PS accurate 
\minnlo{} predictions with the cross section measured by
ATLAS in \citere{Aad:2019gpq}, which are perfectly compatible 
within the quoted experimental errors.

\subsection{Comparison of differential distributions against \minlo{} and \fnnlo{}}
\label{sec:cmpnnlo}

\begin{figure}[ht]

\begin{tabular}{cc}\hspace{-0.5cm}
\includegraphics[width=.33\textheight]{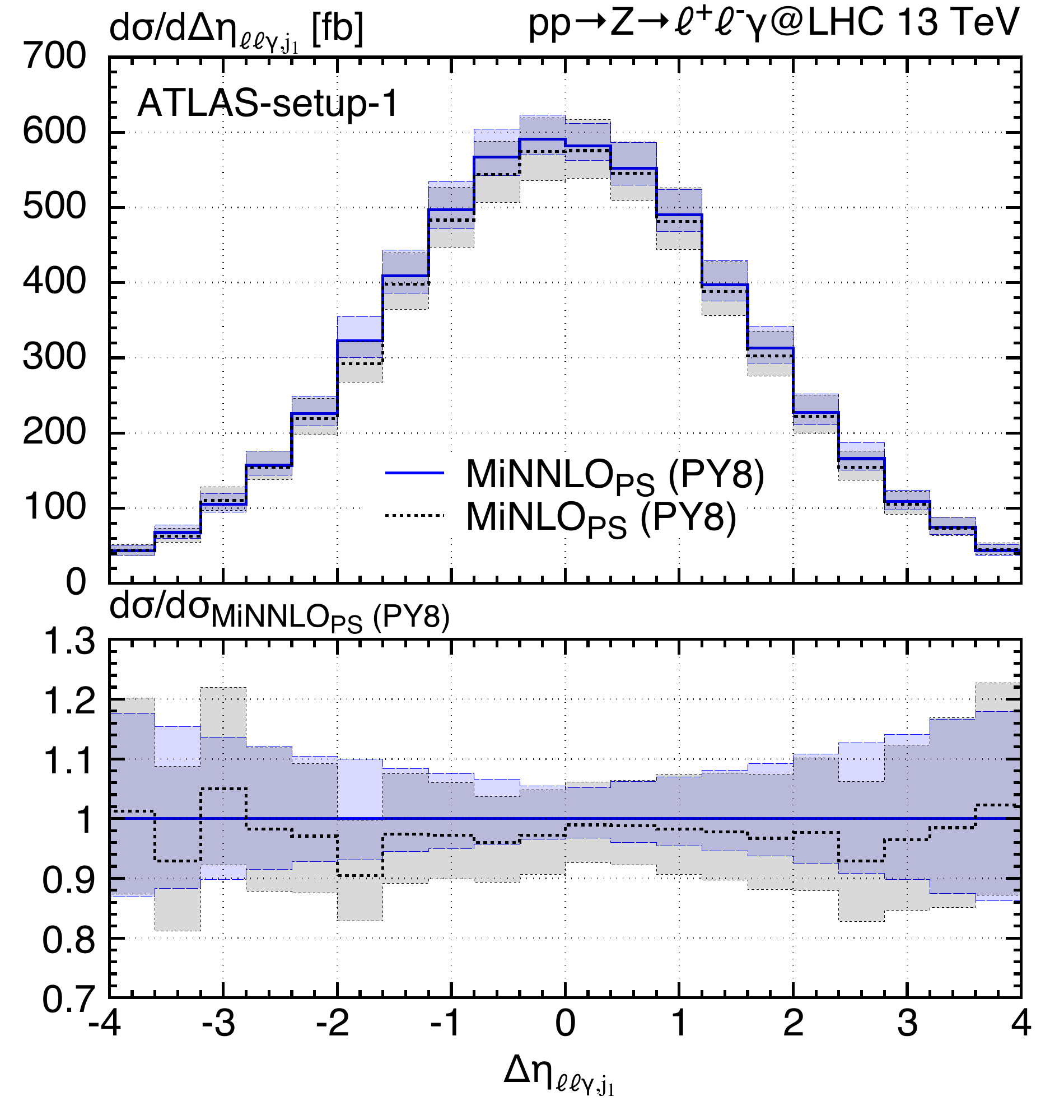} 
&
\includegraphics[width=.33\textheight]{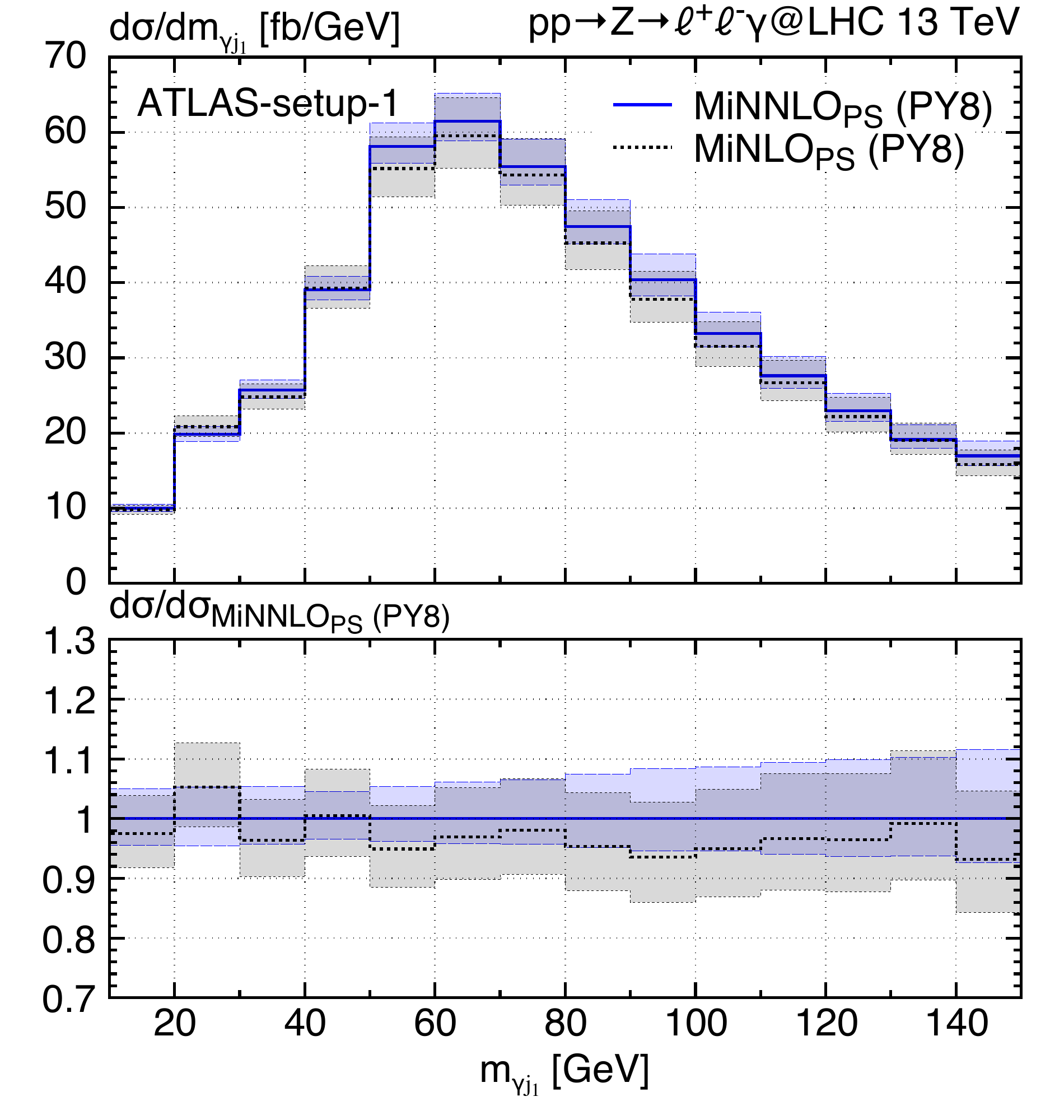}
\end{tabular}\vspace{-0.15cm}
\caption{\label{fig:minlo1} Distribution in the pseudorapidity difference of the color-singlet and the
  hardest jet (left plot) and in the invariant mass of the photon and the hardest jet (right plot) for
\minnlo{} (blue, solid line) and \minlo{} (black, dotted line).}
\end{figure}
\begin{figure}[ht]
\begin{tabular}{cc}\hspace{-0.5cm}
\includegraphics[width=.33\textheight]{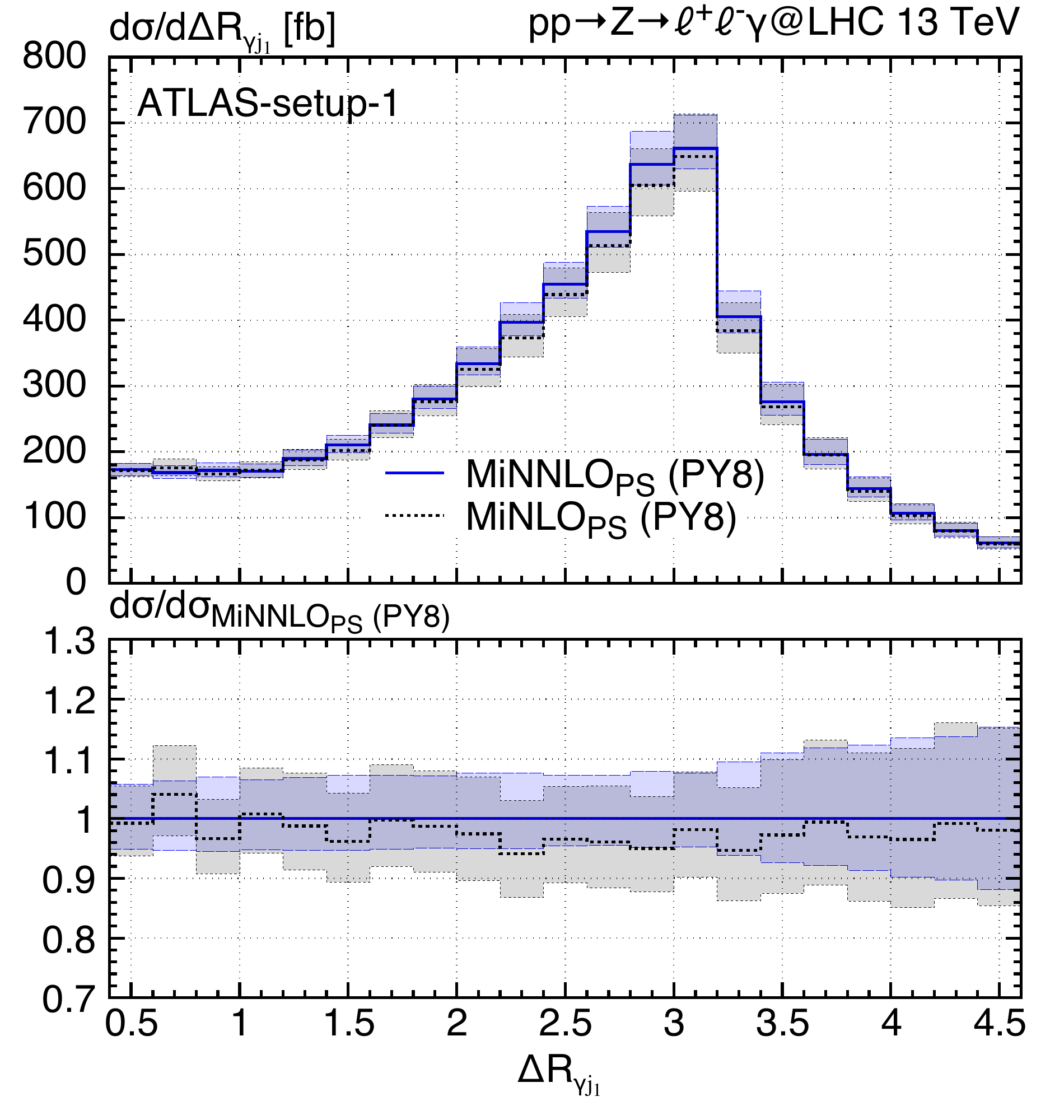} 
&
\includegraphics[width=.33\textheight]{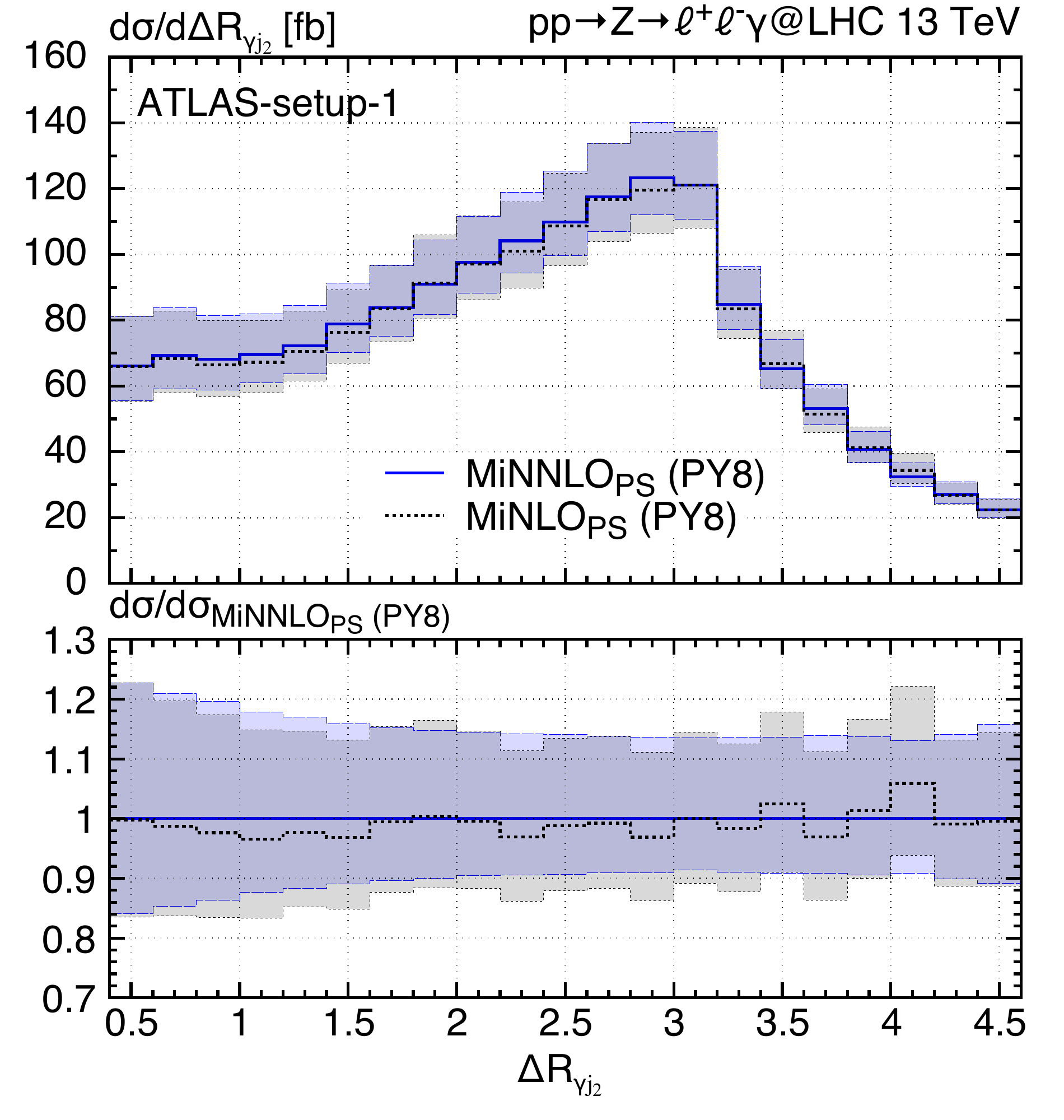}
\end{tabular}
  \caption{\label{fig:minlo2} Distribution in the $\Delta R$ separation between the photon and
  hardest jet (left plot), and between the photon and the second-hardest jet (right plot) for
\minnlo{} (blue, solid line) and \minlo{} (black, dotted line).}
\end{figure}

We now turn to discussing differential distributions in the fiducial phase space. In this section
we compare our \minnlo{} predictions with \minlo{} and NNLO results. This serves two 
purposes. On the one hand, \minnlo{} distributions are validated for one-jet and two-jet observables against the ones obtained with \minlo{} and for Born-level observables
(inclusive over QCD radiation) against NNLO predictions. On the other hand, this allows 
us to show the importance of NNLO+PS matching with respect to less accurate results.
To these ends, we discuss selected distributions which are particularly significant 
to show the performance of \minnlo{} predictions.
The figures of this and the upcoming sections are organized as follows: the
main frame shows the predictions from \minnlo{} (blue, solid line), together with all 
other results relevant for the given comparison. 
In an inset we display the bin-by-bin ratio of
all the histograms that appear in the main frame to the \minnlo{} one. The bands indicate  
the theoretical uncertainties that are computed from scale variations.

We start by discussing quantities that involve jets in the final state in \fig{fig:minlo1} and \fig{fig:minlo2}, where \minnlo{} (blue, solid line) and \minlo{} (black, dotted line) 
are compared at the un-showered LHE level in \setupone{}.
Since for these observables \minlo{} and \minnlo{} have the same 
accuracy the two predictions are expected not to
differ from each other significantly (i.e.\ not beyond uncertainties), 
both in terms of shapes and size of scale uncertainty bands. 
In particular, such agreement serves as a validation that
NNLO corrections are properly spread by the factor in
\eqn{eq:spreading} in the jet-resolved phase space of $Z\gamma$+jet production
without altering the NLO accuracy.
As a matter of fact, the left plot of \fig{fig:minlo1} shows that \minnlo{} and \minlo{} 
predictions agree
well within uncertainties for both the pseudorapidity difference between the 
$Z\gamma$ system and the hardest jet ($\detallgj$). 
Furthermore, the size of the uncertainty bands are comparable
 over the whole pseudorapidity range.
In a similar manner,
the ratio between \minnlo{} and \minlo{} is nearly flat for the invariant mass of the photon and
the hardest jet ($\mgj$) in the right plot of \fig{fig:minlo1}. Here, we further observe the effect of the Frixione isolation,
which dampens the distribution in the photon-jet collinear limit.
Also in \fig{fig:minlo2} \minnlo{} and \minlo{} predictions agree well
for the distance between the photon and
the leading and subleading jet in the $\eta$-$\phi$ plane 
($\drgjone$ and $\drgjtwo$).
As $\drgjtwo$ involves the second-hardest jet,
both \minnlo{} and \minlo{} are only LO accurate, which is 
also evident from the broadening of the uncertainty bands. 
We have examined a large number of other quantities involving jets
(not shown here) observing a similar behaviour in all cases.

\begin{figure}
\begin{center}
\begin{tabular}{cc}\hspace{-0.5cm}
\includegraphics[width=.33\textheight]{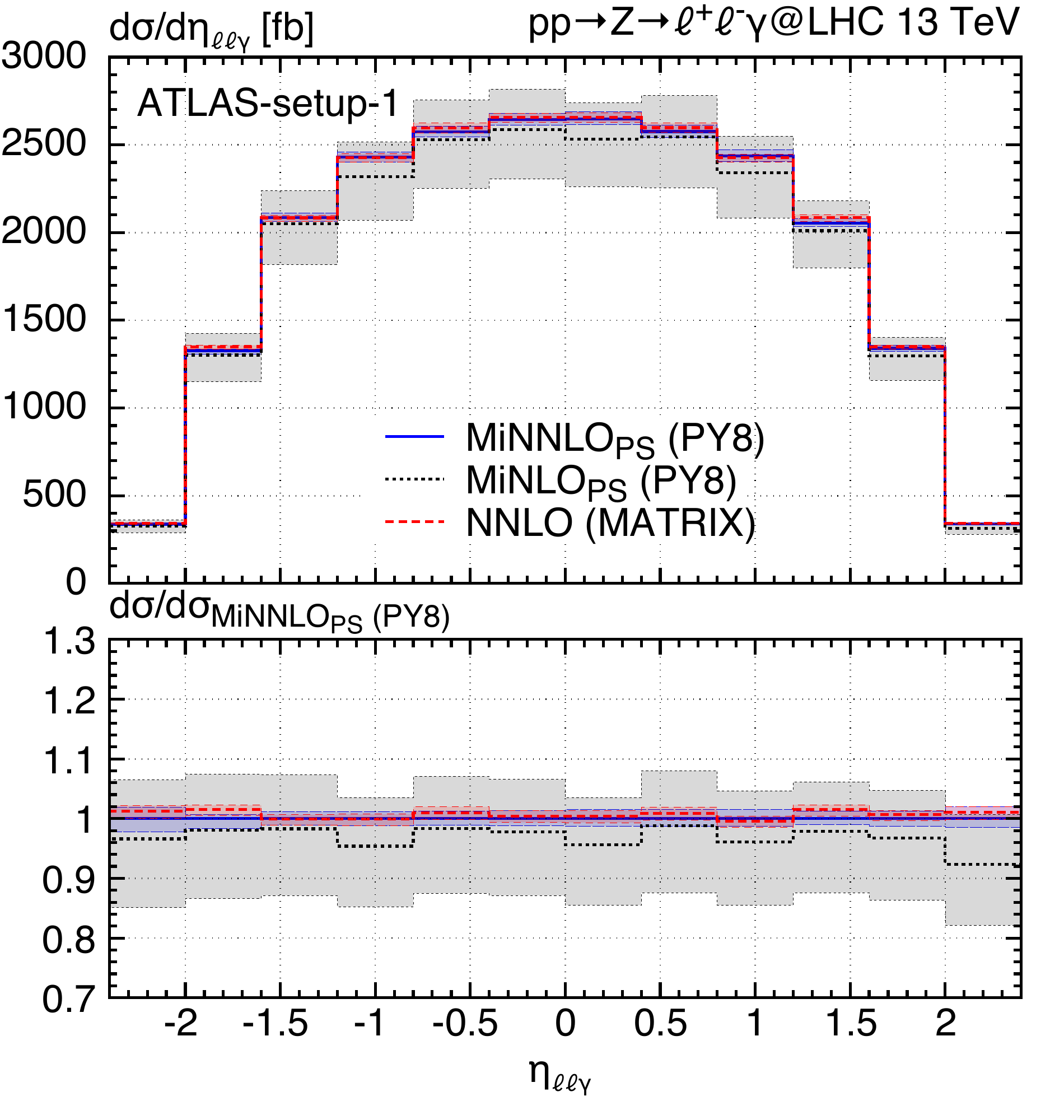}
&
\includegraphics[width=.33\textheight]{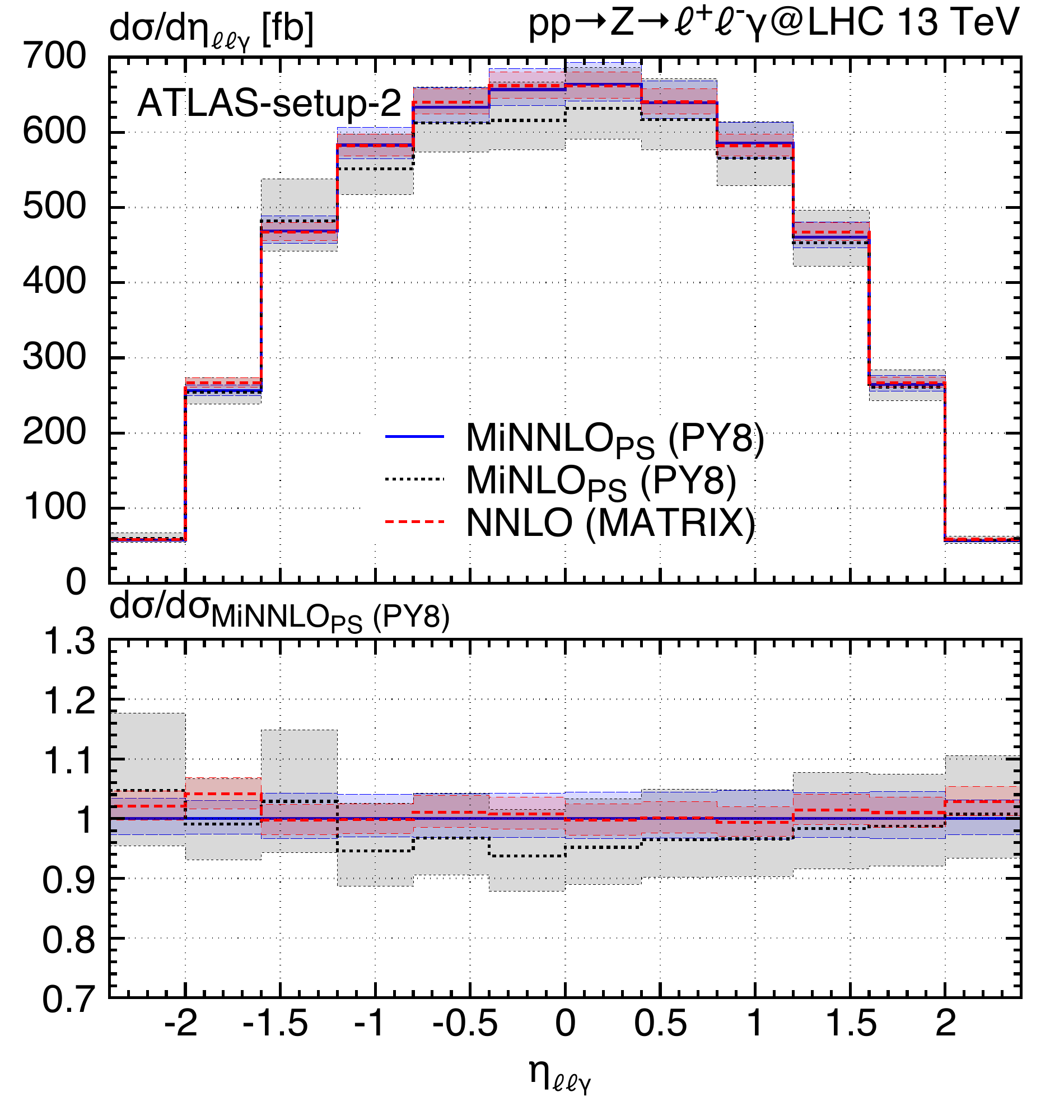}
\end{tabular}\vspace{-0.15cm}
\caption{\label{fig:rap12} Distribution in the pseudorapidity of the $Z\gamma$ system 
in \setupone{} (left plot) and in \setuptwo{} (right plot) for
\minnlo{} (blue, solid), \minlo{} (black, dotted) and NNLO (red, dashed). }
\end{center}
\end{figure}

\begin{figure}[ht]                                                                                                                                                                          
\begin{tabular}{cc}\hspace{-0.5cm}
\includegraphics[width=.33\textheight]{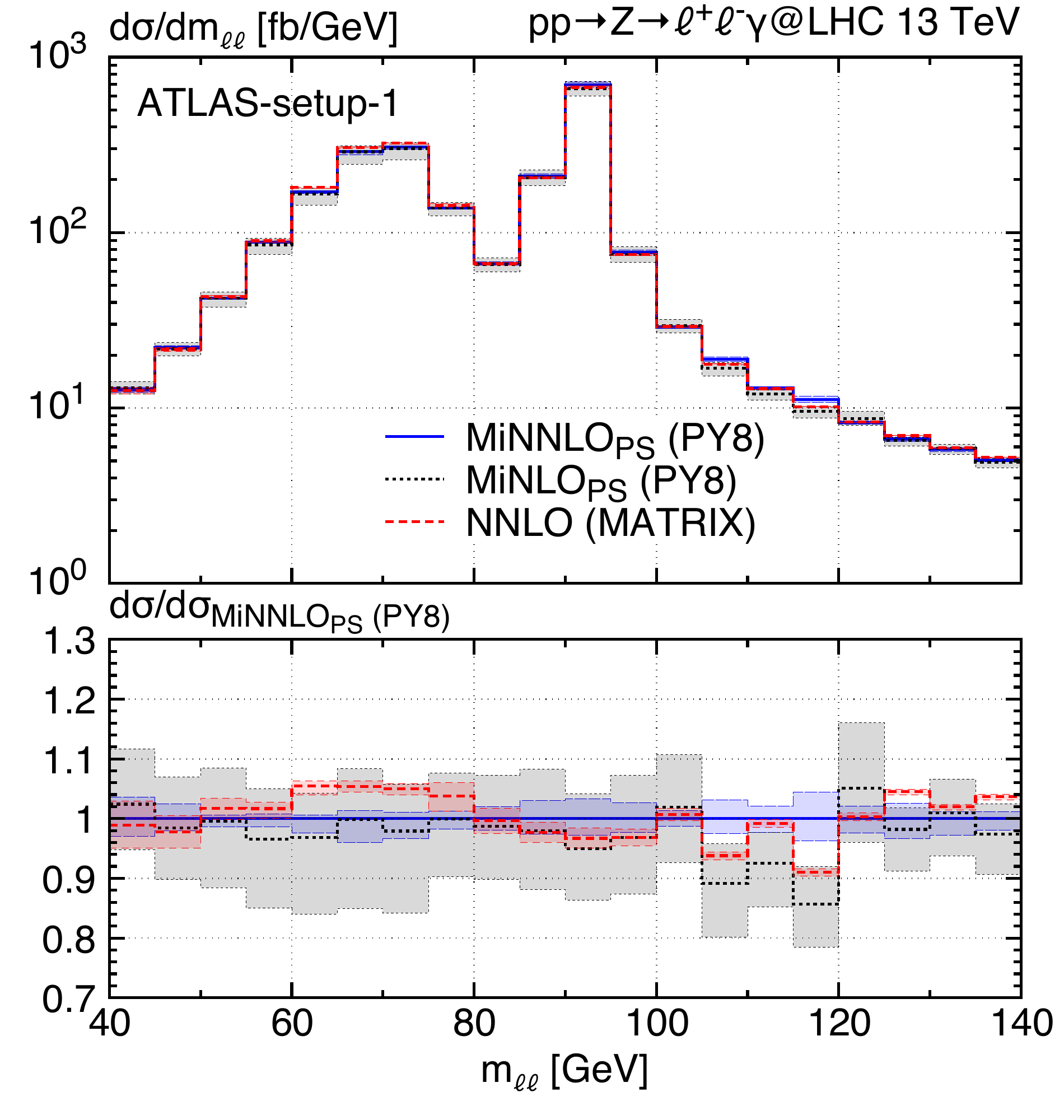}
&
\includegraphics[width=.33\textheight]{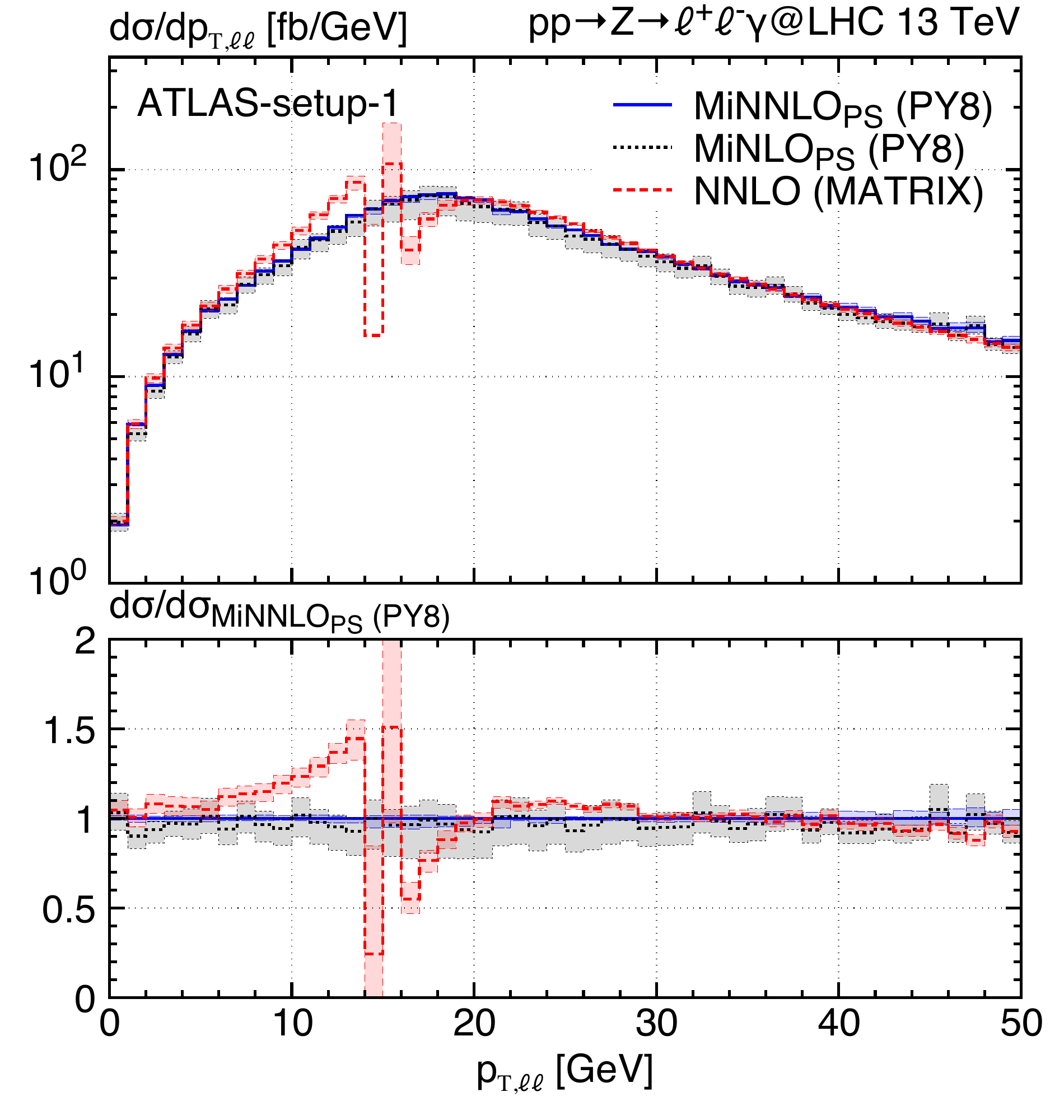}
\end{tabular}\vspace{-0.15cm}
\caption{\label{fig:nnlo1} Distribution in the invariant mass (left plot) and
  in the transverse momentum (right plot) of the lepton pair for
\minnlo{} (blue, solid line), \minlo{} (black, dotted line) and  NNLO (red, dashed ine).}
\end{figure}

\begin{figure}[ht]                                                                                                                                                                                                                                                                                                                                                            
\begin{tabular}{cc}\hspace{-0.5cm}
\includegraphics[width=.33\textheight]{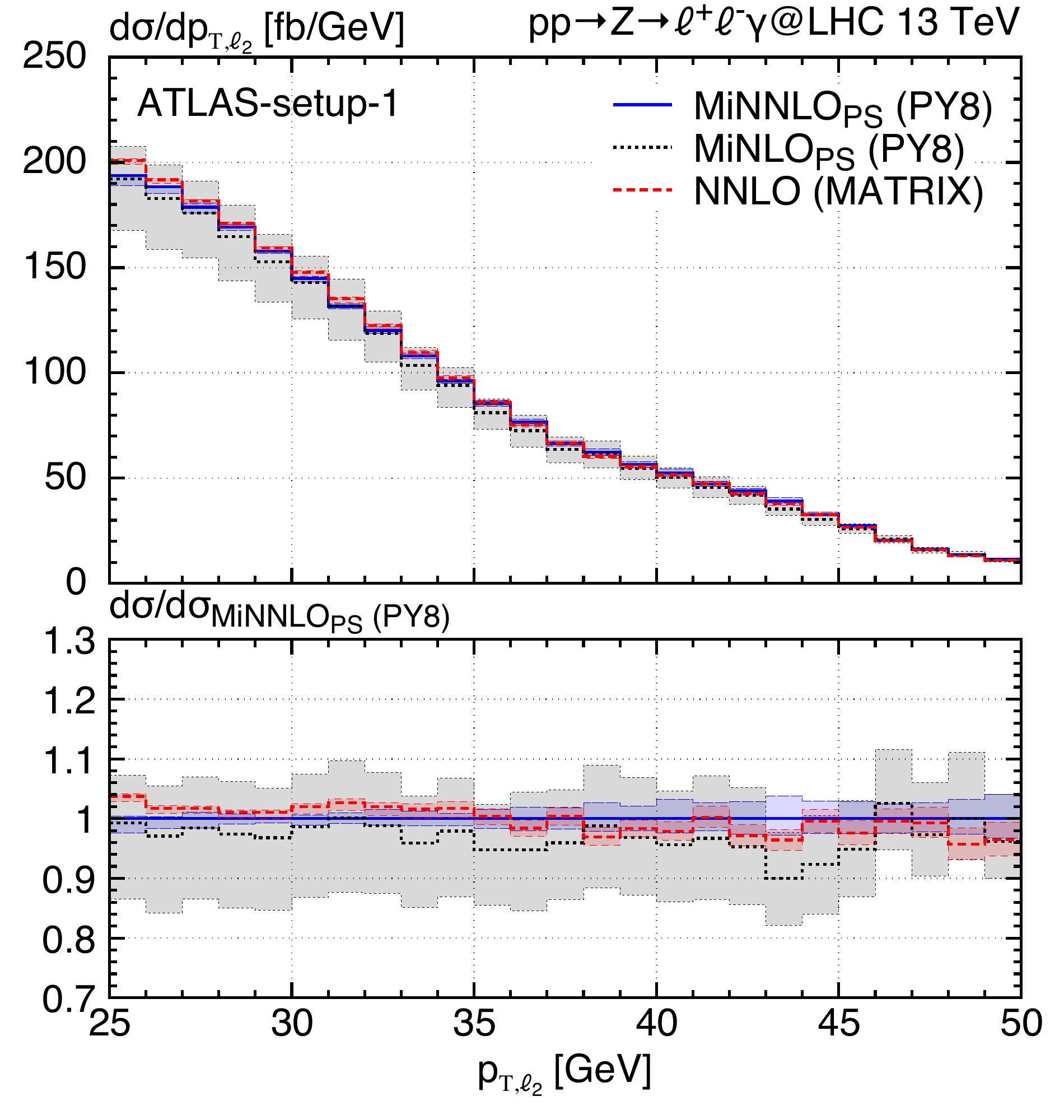}
&
\includegraphics[width=.33\textheight]{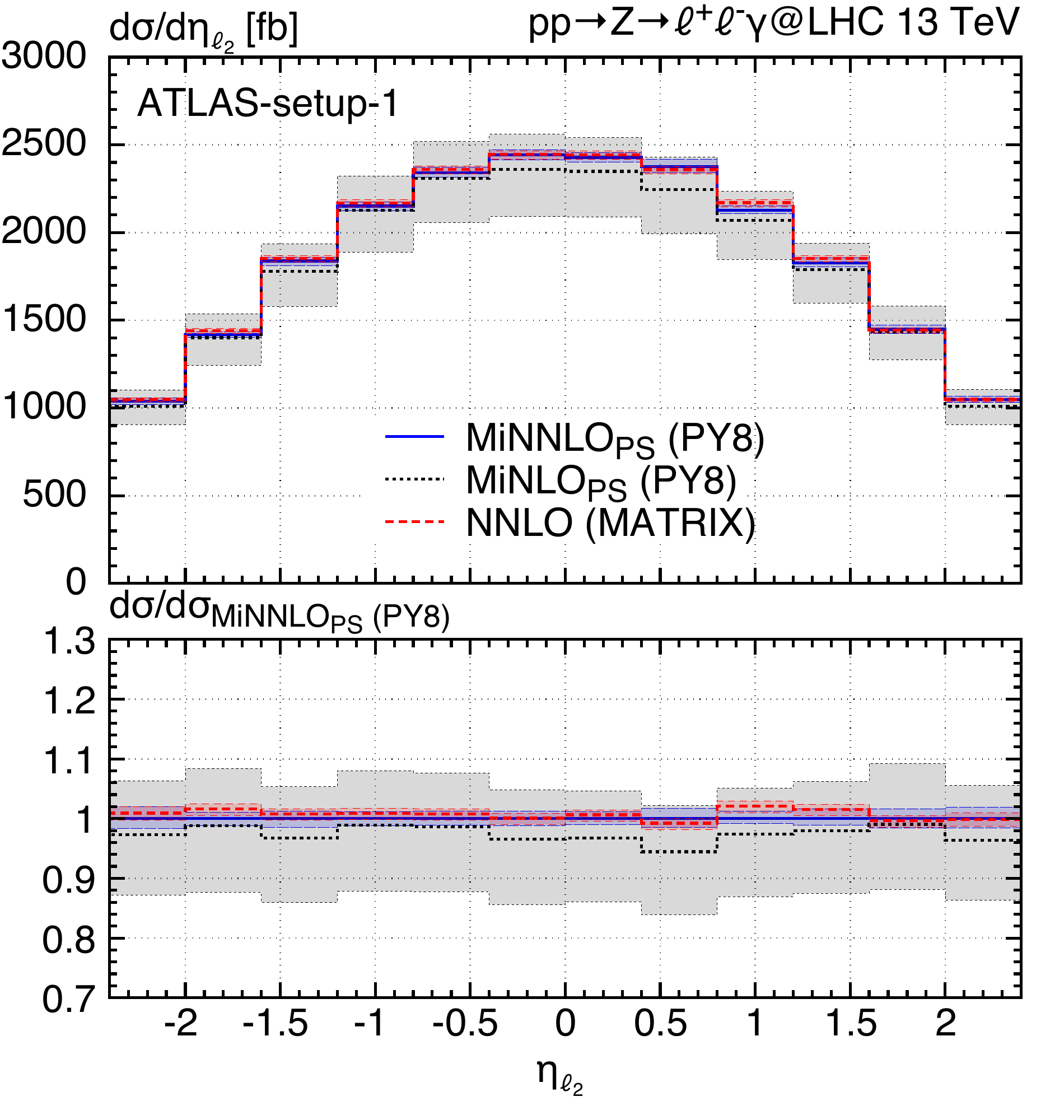}
\end{tabular}
  \caption{\label{fig:nnlo2} Distribution in the transverse momentum (left plot) and
  in the pseudorapidity (right plot) of the second-hardest lepton for
\minnlo{} (blue, solid), \minlo{} (black, dotted) and  NNLO (red, dashed).}
\end{figure}

Next, in \fig{fig:rap12}, \fig{fig:nnlo1} and \fig{fig:nnlo2}, we compare \minnlo{}
against \minlo{} (black, dotted) and NNLO predictions from \Matrix{} (red, dashed) 
for Born-level observables (inclusive over QCD radiation) after showering with \PYTHIA{8}.
By and large, we observe a very good agreement of
\minnlo{} and NNLO predictions, especially considering the fact that they differ from each other
in the choice of the renormalization and factorization scales,
and in the treatment of higher-order contributions. 
What can be appreciated is the clear reduction of the scale
uncertainties of \minnlo{} predictions with respect to the \minlo{} ones
up to a size which is comparable to the NNLO ones.


In particular, \fig{fig:rap12} displays the pseudorapidity distribution of 
the $Z\gamma$ system ($\etallg$) in each of the two fiducial setups. 
The ratio of NNLO over \minnlo{} is close to one in both cases, with 
uncertainty bands of one to two percent in \setupone{}. 
In \setuptwo{}, on the other hand, the bands are roughly twice as large, as already 
observed for the integrated cross section due to the higher sensitivity
to phase-space regions related to real QCD radiation.

In \fig{fig:nnlo1} we show the distributions in the invariant mass ($\mll$) and 
transverse momentum ($\ptll$) of the lepton pair in \setupone{}. 
We can appreciate the $Z$-boson resonance in the $\mll{}$ distribution as well 
as a broader, but smaller, enhancement around $\mll{} \sim 70$\,GeV, caused 
by the $Z$-boson resonance in $\mllg{}$. 
The qualitative behaviour of \minnlo{} with respect to \minlo and 
NNLO predictions in the ratio inset are relatively similar to the one of the $Z\gamma$ rapidity distribution.
However, there seems to be a small effect induced by the resummation 
of the parton shower reducing the enhancement due to $\mllg{}$, where the NNLO prediction is slightly larger.
In the $\ptll$ distribution on the right we observe an interesting behaviour of the NNLO
prediction. The NNLO result develops a perturbative instability (Sudakov shoulder)~\cite{Catani:1997xc}
around $\ptll\sim 15$\,GeV 
caused by an incomplete cancellation of virtual and real contributions from soft gluons, which
is logarithmically divergent, but integrable. 
The reason is the fiducial cut $\ptg>15$\,GeV (see \tab{tab:setup}) that for LO kinematics 
implies $\ptll=\ptg>15$\,GeV, so that the $\ptll{}$ distribution is not filled below $15$\,GeV at LO.
Thus, the fixed-order result is NNLO accurate only for $\ptll>\ptg$, while for
$\ptll<\ptg$ at least one QCD emission is necessary, which is
described only at NLO accuracy. At the same time, at threshold the prediction becomes 
sensitive to soft-gluon effects, resulting in an instability at fixed order.
Indeed, the parton shower cures this behaviour and yields a physical prediction at threshold for 
both \minlo{} and \minnlo{}.
This is one example where an NNLO calculation is insufficient and NNLO+PS matching is required.

In \fig{fig:nnlo2} we consider distributions in the second-hardest lepton, showing its transverse momentum ($\ptltwo{}$)
in the left and its rapidity ($\etaltwo$) in the right plot. Similar conclusions as made before for $\mll$ and $\etallg$
apply also for these observables, so no further comments are needed. We reiterate however that, while the central 
predictions of \minlo{} and \minnlo{} are generally close to each other, since \minlo{} already includes many terms 
beyond NLO accuracy for $Z\gamma$ production, scale uncertainties are substantially reduced in case of 
\minnlo{} down to the level of the NNLO ones.

\subsection{Comparison of $Z\gamma$ transverse-momentum spectrum against NNLO+N$^3$LL}
\label{sec:cmpNNLON3LL}

\begin{figure}[ht]
\begin{center}
\begin{tabular}{cc}\hspace{-0.5cm}
\includegraphics[width=.33\textheight]{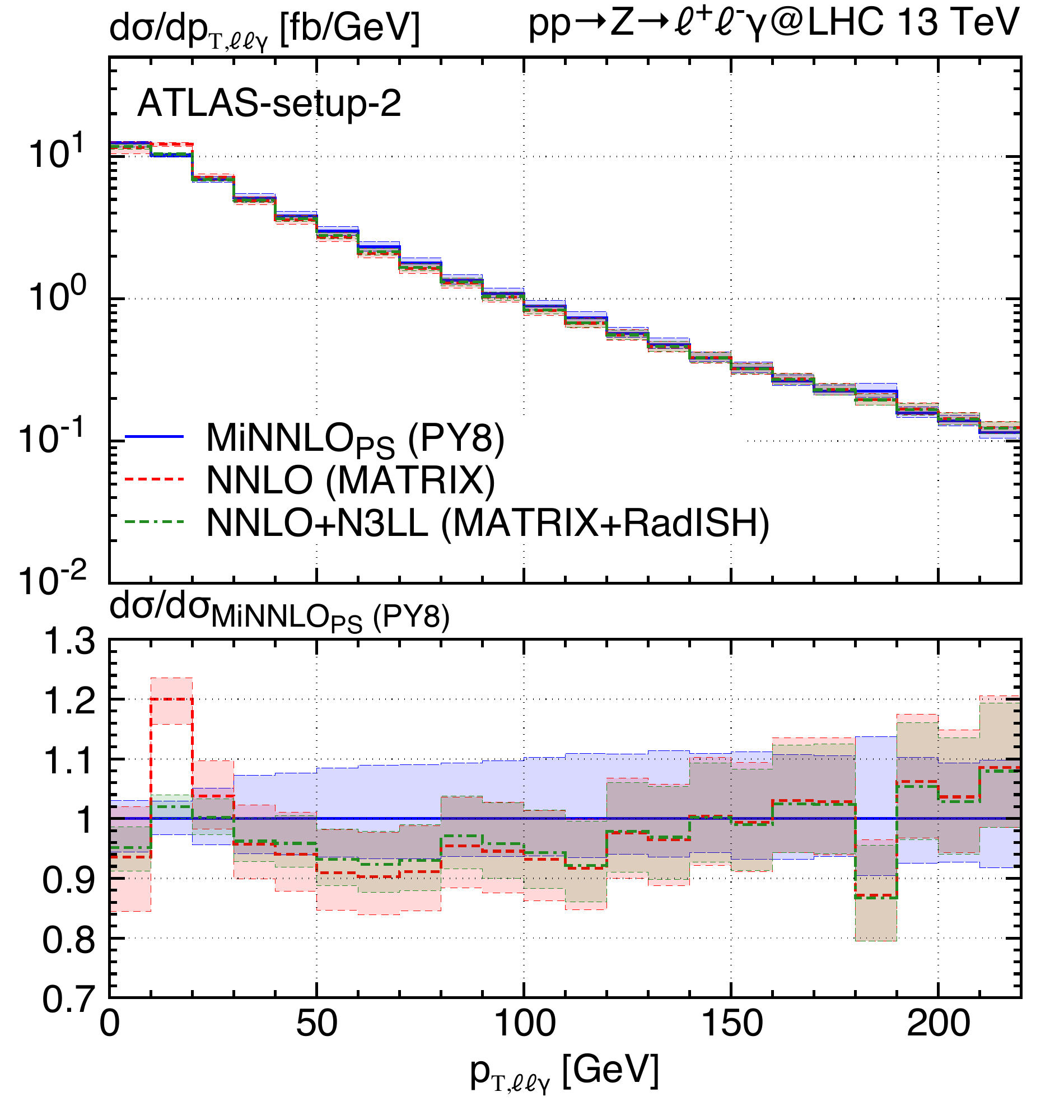} 
&
\includegraphics[width=.33\textheight]{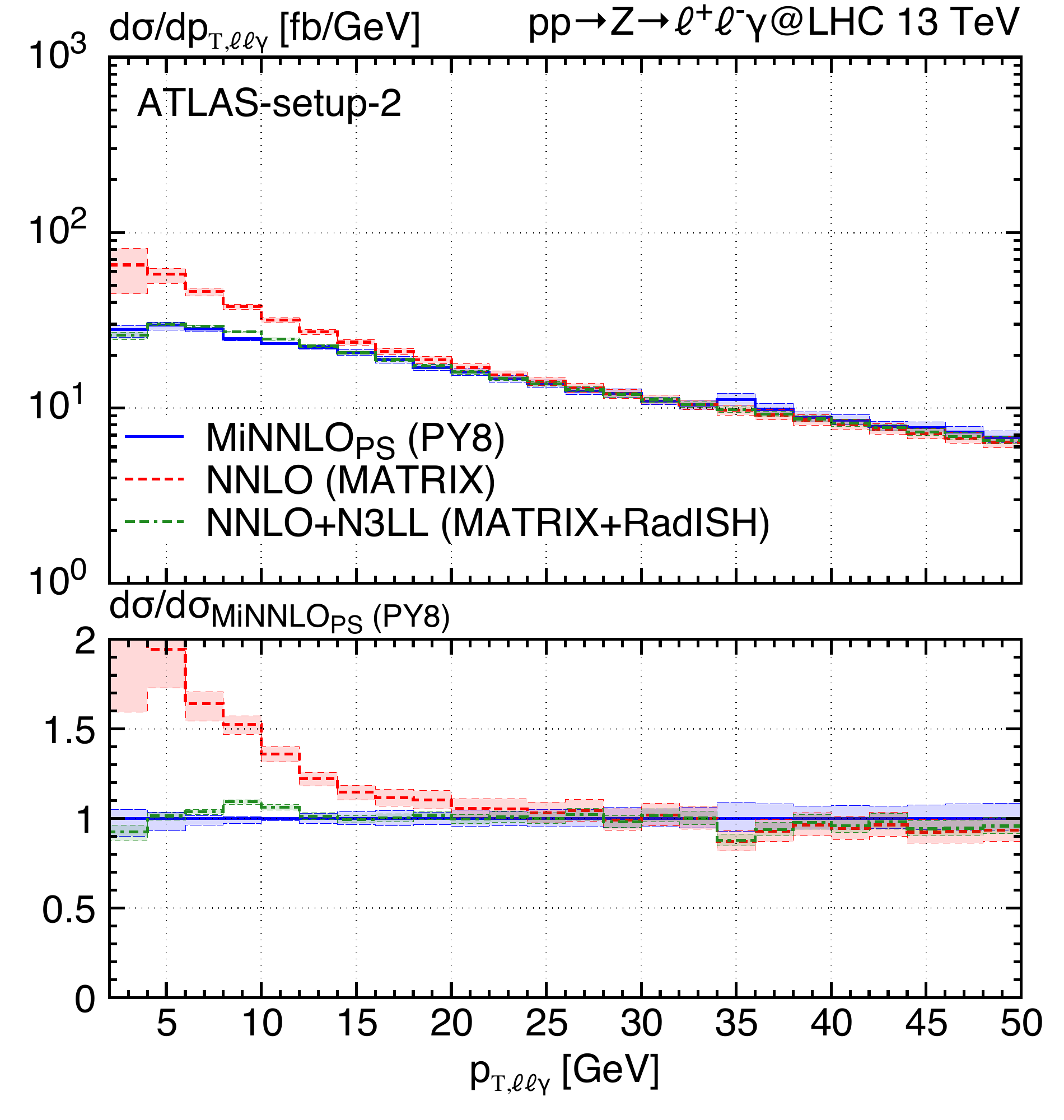}
\end{tabular}
\caption{\label{fig:nnnlo} Distribution in the transverse momentum of the
  $Z\gamma$ system in a wider range (left plot) and at small $\ptllg$ (right plot) for
\minnlo{} (blue, solid), NNLO (red, dashed) and NNLO+N$^3$LL (green, double-dash-dotted).}
\end{center}
\end{figure}

{\sloppy
We continue our discussion of differential distributions with the transverse-momentum
spectrum of the $Z\gamma$ system ($\ptzg$). 
In \fig{fig:nnnlo} we compare the $\ptzg$ distribution in \setuptwo{} obtained with \minnlo{} 
against a more accurate prediction at NNLO+N$^3$LL (green, dashed curve), using the analytic 
resummation of large logarithmic contributions within \noun{Matrix}$+$\noun{RadISH}~\cite{Kallweit:2020gva,Wiesemann:2020gbm}.
For comparison we also show the NNLO result, which is effectively only NLO accurate for this distribution.
The NNLO+N$^3$LL prediction uses \cite{Wiesemann:2020gbm}
\begin{align}
\label{eq:scaleN3LL}
\muRc=\muFc=\sqrt{\mll^2+\ptg^2}\quad \text{and} \quad \Qc=\frac12 \mllg
\end{align}
as central scales, where $\Qc$ is the central resummation scale, which is 
varied by a factor of two up and down, while taking the envelope together with the $7$-point $\muR$ and $\muF$ variation
for the total scale uncertainty.}

The $\ptzg{}$ spectrum is shown in two different ranges in \fig{fig:nnnlo}.
From the wider range in the left plot we notice that despite the different scale
settings in the three calculations their predictions are in reasonable agreement 
at large $\ptllg$ values. In this region all predictions are only effectively NLO accurate,
which is indicated by the enlarged scale-uncertainty bands. At small $\ptllg{}$ the fixed-order
result becomes unphysical, as the distribution is logarithmically divergent in the 
$\pt\to 0$ limit, which is visible already in the left plot of \fig{fig:nnnlo}, but can be
appreciated morein the zoomed version on the right.
In this region, only the calculations that properly account for the resummation of
soft QCD radiation by means of an analytic procedure (NNLO+N$^3$LL) or
through parton shower simulations (\minnlo{}) provide a meaningful description. 
Even though at small $\ptllg{}$ the \noun{Matrix}$+$\noun{RadISH} computation is N$^3$LL accurate, 
while the parton shower has a lower logarithmic accuracy, \minnlo{} 
and NNLO+N$^3$LL predictions are in excellent agreement down to 
the transverse-momentum values (almost) in the non-perturbative regime.

\subsection{Comparison of differential distributions against ATLAS data}
\label{sec:data}

\afterpage{\clearpage}
\begin{figure}[p]
\begin{center}\vspace{-0.2cm}
\begin{tabular}{cc}
\includegraphics[width=.31\textheight]{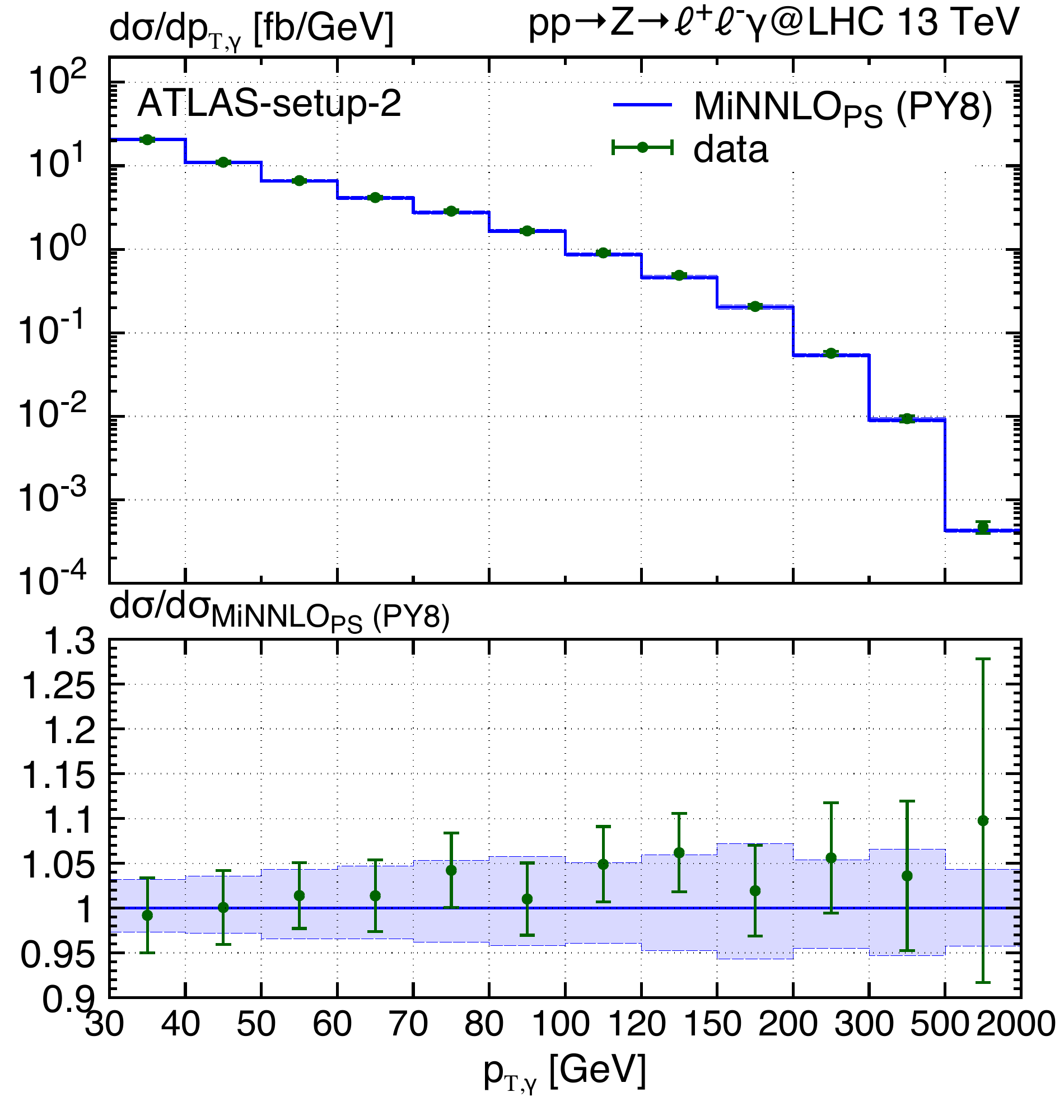} 
&
\includegraphics[width=.31\textheight]{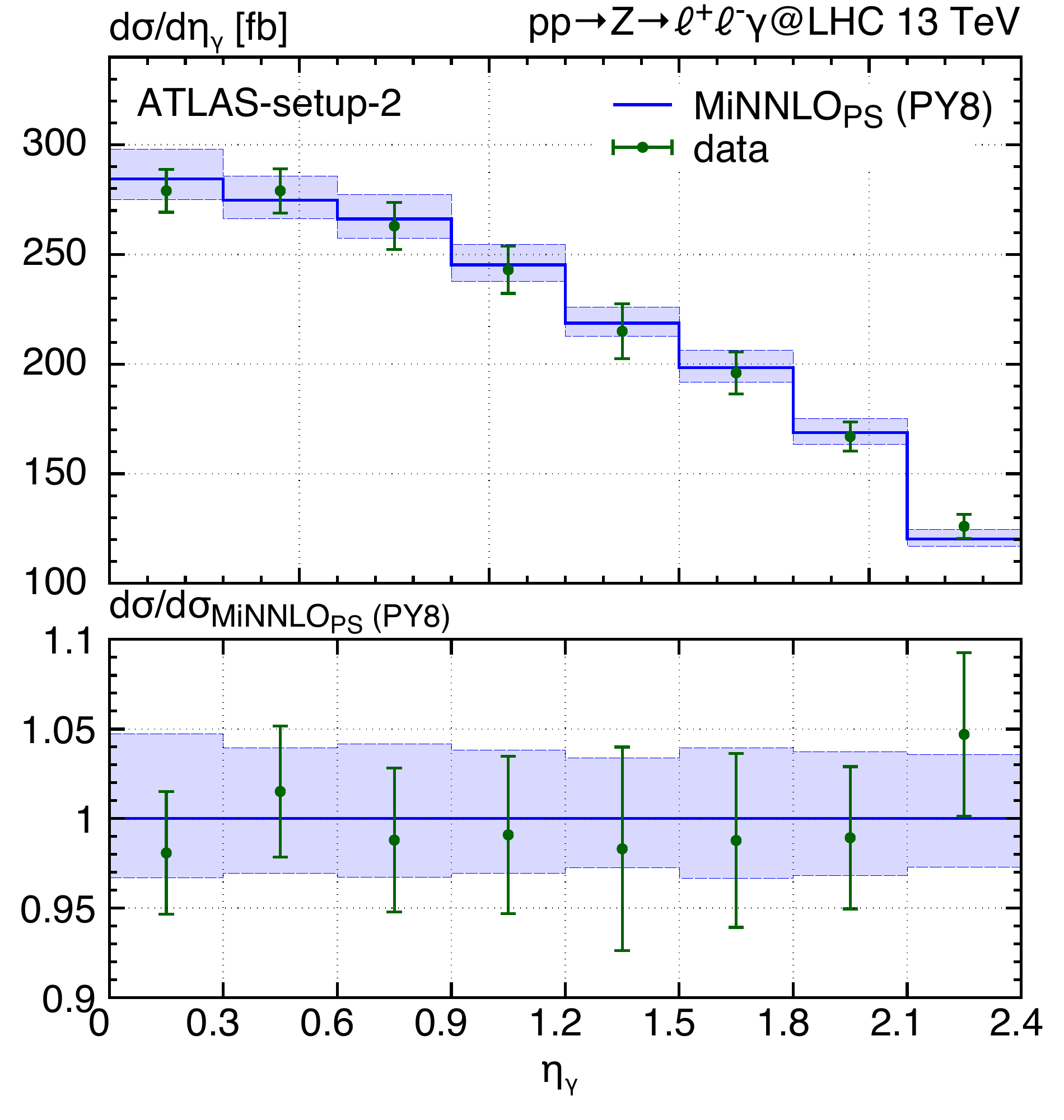}
\end{tabular}\vspace{-0.15cm}
\begin{tabular}{cc}
\includegraphics[width=.31\textheight]{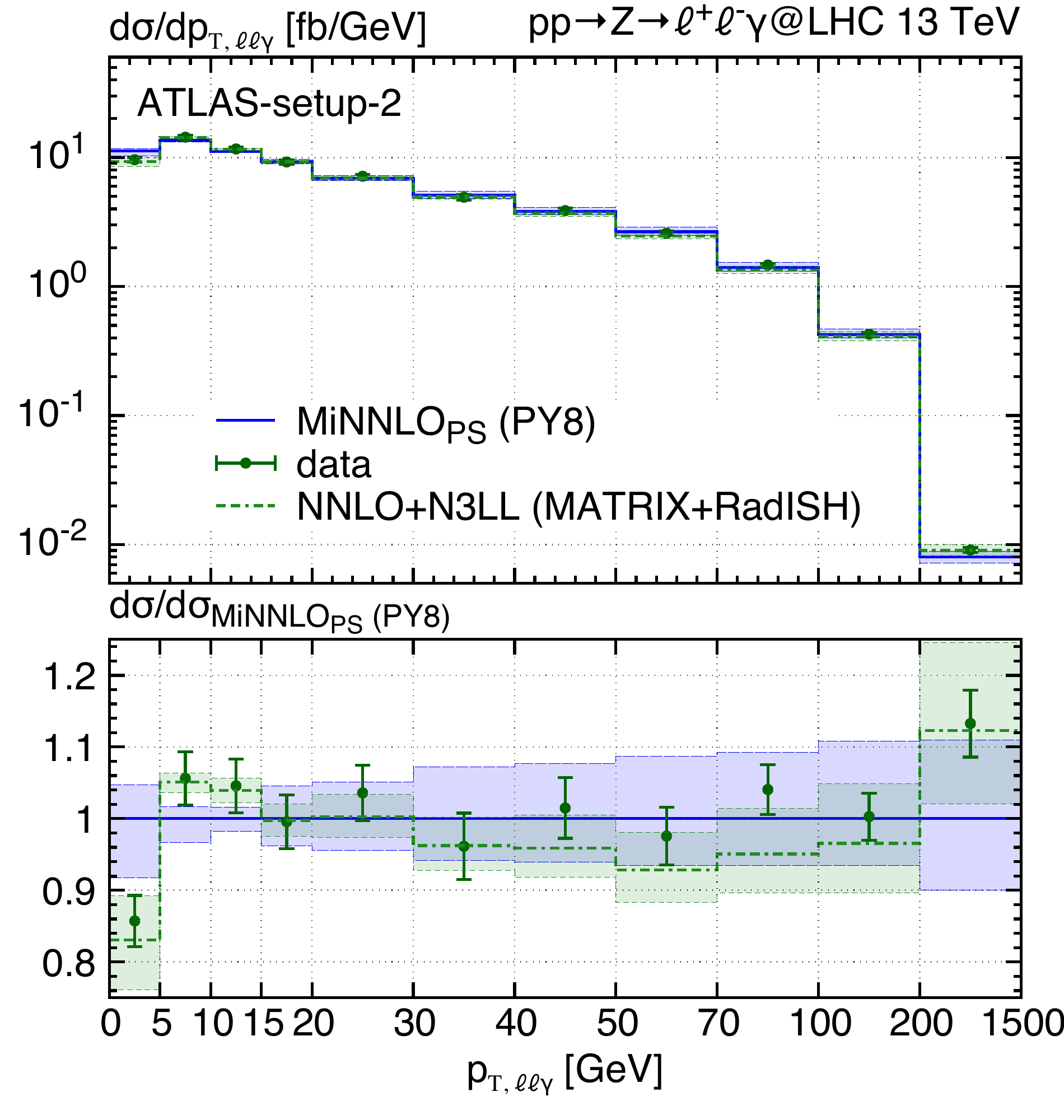}
&
\includegraphics[width=.31\textheight]{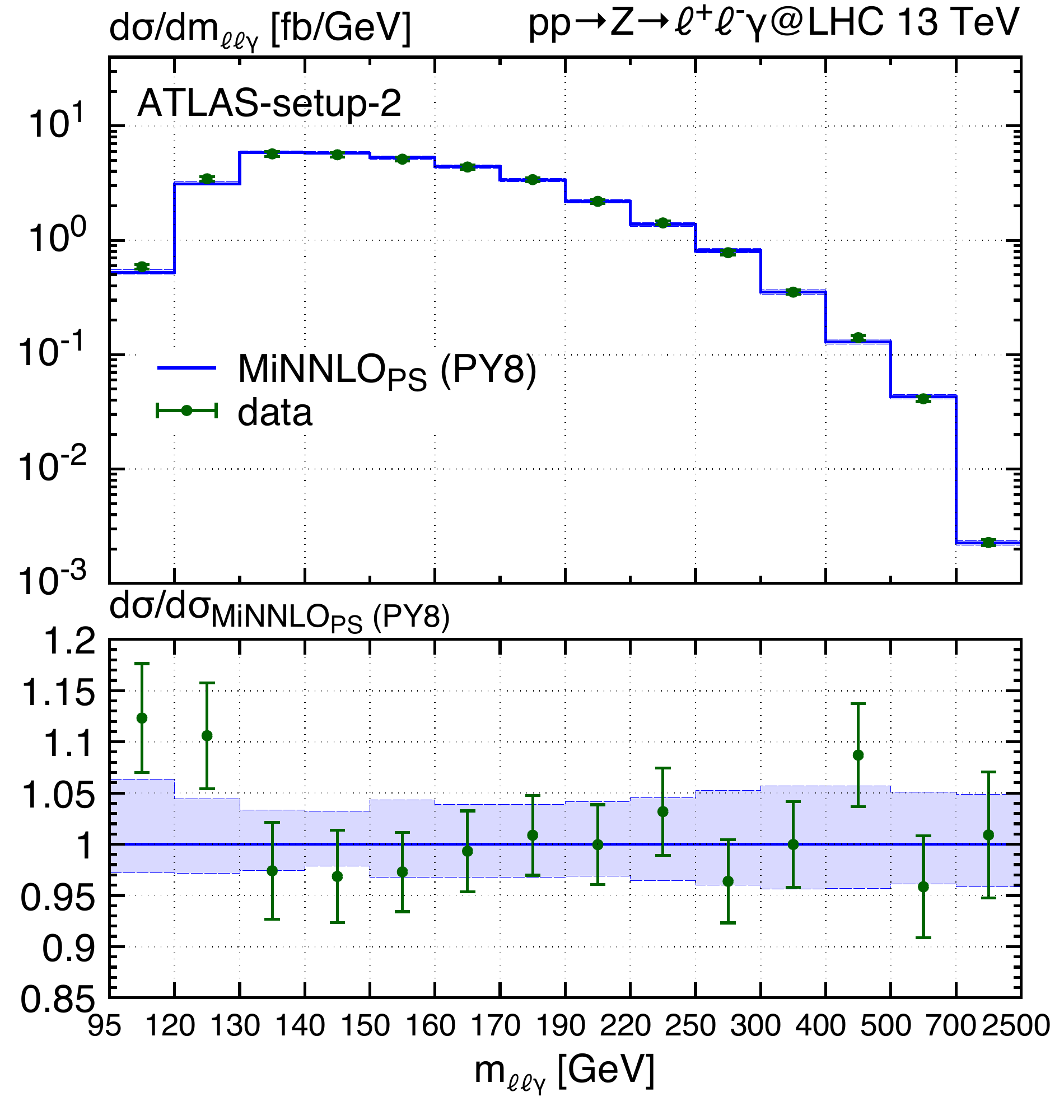}
\end{tabular}\vspace{-0.15cm}
\begin{tabular}{cc}
\includegraphics[width=.31\textheight]{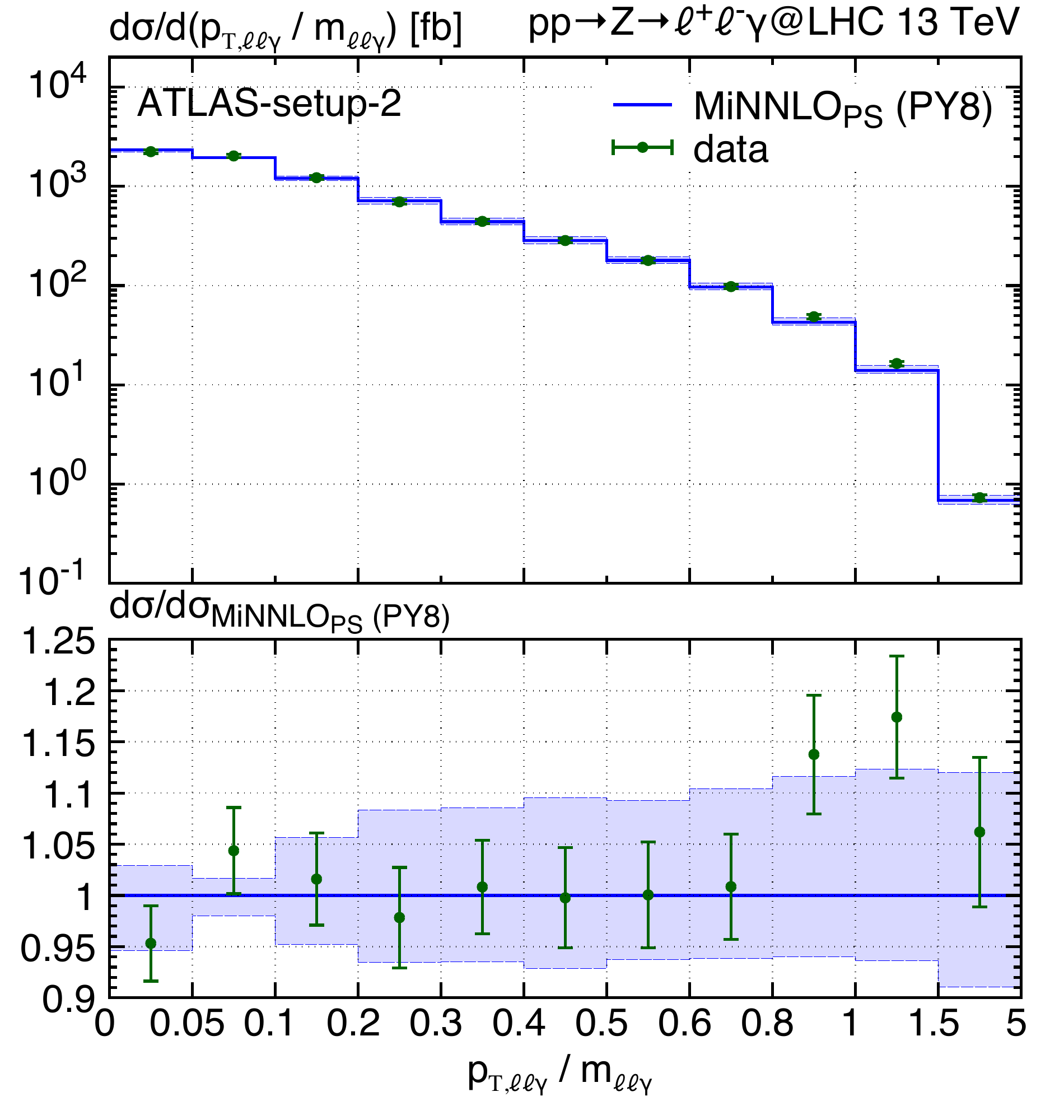}
&
\includegraphics[width=.31\textheight]{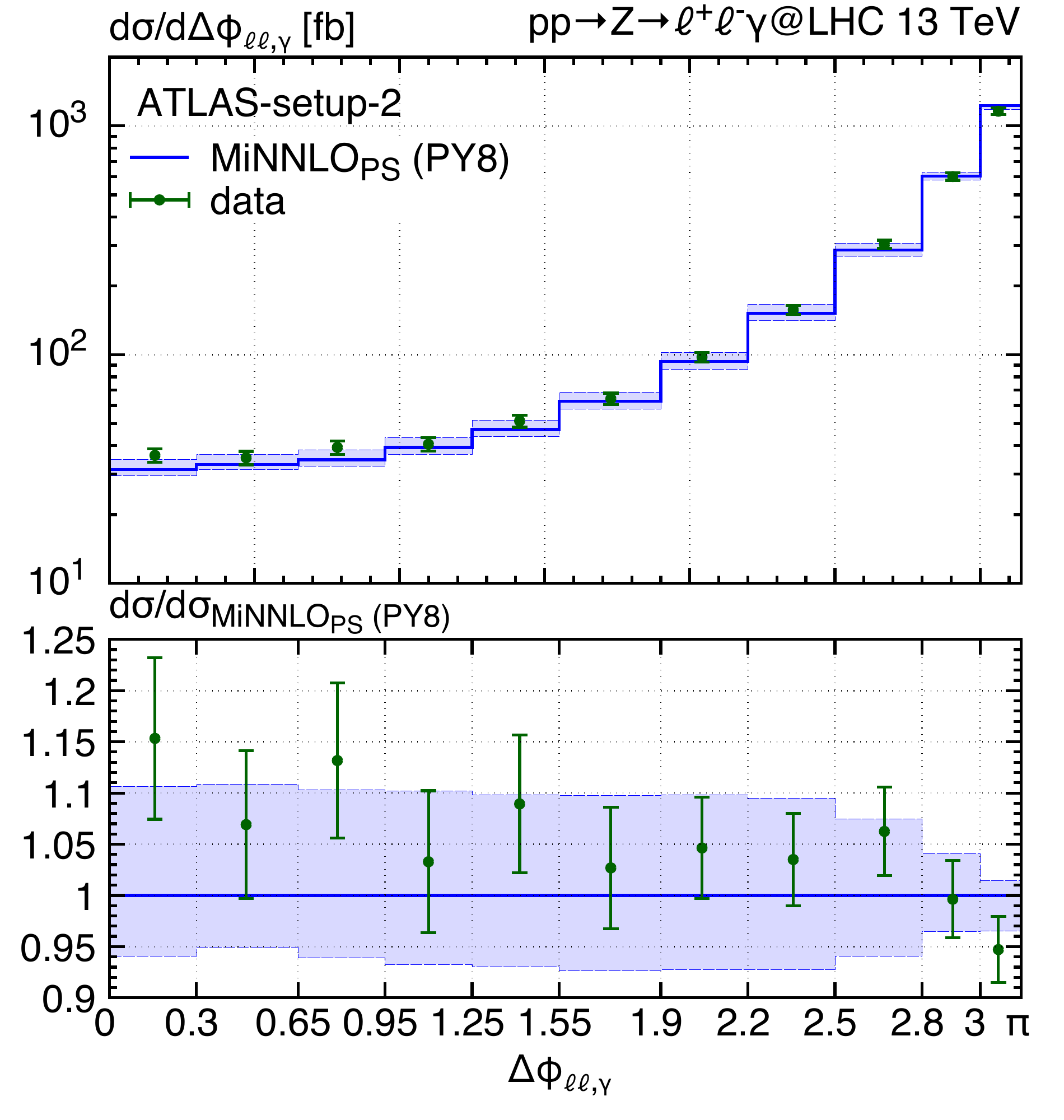}
\end{tabular}
\caption{\label{fig:13TeV} \minnlo{} predictions (blue, solid) compared to
  ATLAS 13\,TeV data (green points with error bars). For $\ptllg$
  also NNLO+N$^3$LL (green, double-dash-dotted) is shown.
}
\end{center}
\end{figure}

Finally, we employ our \minnlo{} generator to compare NNLO+PS accurate 
predictions directly to ATLAS results from the recent 13\,TeV measurement of \citere{Aad:2019gpq}, which relies on the full 
$139\,{\rm fb}^{-1}$ Run-2 data.  The comparison,
carried out in \setuptwo{}, is presented in \fig{fig:13TeV}. The
experimental data are given as green points, with error
bars that refer to the experimental uncertainty.
Six observables are shown: the transverse momentum
($\ptg$) and the pseudorapidity ($\etag$) of the photon, the transverse momentum
($\ptllg$) and the invariant mass ($\mllg$) of the $Z\gamma$ system,
together with their ratio $\ptllg/\mllg$ and the 
difference in the azimuthal angle 
between the lepton pair and the photon ($\dphillg$).

Overall, we observe a remarkably good agreement both in the predicted
shapes of the distributions and in the normalization, especially 
given the fact that the theoretical and the experimental uncertainties are
at the few-percent level only. All data points agree with our 
predictions within the experimental error bars, with the
exception of only very few bins, where the agreement is reached within twice
the experimental error. 
This is a clear improvement over the NLO-accurate event simulations 
employed in the data-theory comparison in figure 6 of \citere{Aad:2019gpq},
both in terms of accuracy (i.e.\ to describe the data) and in terms of precision
(i.e.\ regarding theoretical uncertainties). Moreover, looking at the comparison 
of NNLO-accurate predictions to data in figure 7 of \citere{Aad:2019gpq}, 
it is clear that some (more inclusive) observables are equally well described at fixed 
order, while for observables sensitive to QCD radiation, such as $\ptllg{}$ and
$\dphillg$, NNLO predictions are not sufficient, and the matching to a 
parton shower is essential. In conclusion, our \minnlo{} calculation combines
the two most important aspects (NNLO and parton-shower effects) to
provide the most accurate and most precise $Z\gamma$ predictions to date, 
which will be essential to find potential deviations from the SM for this process in future.

Let us discuss in more detail the $\ptllg{}$ distribution in \fig{fig:13TeV}. In this plot
we have also added the more-accurate NNLO+N$^3$LL prediction, as introduced 
in \sct{sec:cmpNNLON3LL} with the scale setting of \eqn{eq:scaleN3LL} that induces
differences also at large $\ptllg{}$.
Despite the good agreement of \minnlo{} with data, 
the analytically resummed result is performing even better, especially in the first 
few bins, where the higher accuracy in the resummation of large 
logarithmic contributions is important.
Although \minnlo{} and NNLO+N$^3$LL agree quite well (cf. the discussion in \sct{sec:cmpNNLON3LL}),
this shows that for an observable like $\ptllg{}$ it can be very useful to resort to
tools that predict a single distribution more accurately, if available.
Nevertheless, it is reassuring that our accurate multi-purpose \minnlo{} 
simulation, with all its flexibility to predict essentially any IR-safe observable,
provides a very good description of such distributions as well.

We further notice that the deviation at small $\mllg{}$ is due to missing QED effects 
as shown in \citere{Aad:2019gpq}. Our \minnlo{} computation renders  the inclusion of 
such effects with a NNLO prediction feasible  
by using a QED shower within \PYTHIA{8}, which could be very useful  
in an experimental analysis. This is however beyond the scope of this paper and left for future studies.

\section{Summary}
\label{sec:summary}

We have presented a novel calculation of NNLO+PS accurate predictions to 
$Z\gamma$ production at the LHC. This is the first calculation of a genuine $2\to 2$
process at this accuracy that does not require an a-posteriori multi-differential reweighting. To this end, we have extended the \minnlo{} method \cite{Monni:2019whf}
to $2\to 2$ colour-singlet production, with non-trivial one-loop and two-loop corrections, 
and applied it to the $Z\gamma$ process.
More precisely, we have considered all resonant and non-resonant contributions 
to the hard-scattering process $pp\to \ell^+\ell^-\gamma$, including off-shell effects 
and spin correlations.

As a starting point we have implemented NLO+PS generators for both 
$Z\gamma$ and $Z\gamma$+jet production 
within the \POWHEGBOXRES{} framework~\cite{Jezo:2015aia}. The $Z\gamma$+jet generator 
has then been extended to include NNLO corrections to $Z\gamma$ production
by means of the \minnlo{} method. The two-loop virtual corrections \cite{Gehrmann:2011ab} are linked 
through their implementation within the \Matrix{} code~\cite{Grazzini:2017mhc}, which we treat as 
a library. A substantial amount of work has been devoted 
to render those calculations numerically efficient, and we 
have discussed the technical and physical aspects in detail.
All our computations will be made
publicly available within \POWHEGBOXRES{}.\footnote{Instructions to 
download the $Z\gamma$ and $Z\gamma$+jet generators (including the \minnlo{} features) will be provided on {\href{http://powhegbox.mib.infn.it.}{http://powhegbox.mib.infn.it}} and are already available upon request.}

We have presented NNLO+PS accurate predictions using our \minnlo{} calculation 
in hadronic collisions at 13\,TeV. Observables exclusive in the final state jets have 
been used to validate the way we include NNLO corrections through the \minnlo{} 
procedure by comparing \minnlo{} with \minlo{} predictions. Observables 
inclusive over QCD radiation are generally well described by a
fixed-order NNLO calculation. Although \minnlo{} predictions differ
in the treatment of terms beyond accuracy, we found them to be 
in very good agreement with NNLO results,
both for fiducial cross sections and for differential distributions.
Moreover, we have shown the importance of NNLO+PS accurate predictions.
On the one hand, they render predictions physical for observables 
sensitive to soft-gluon effects, where NNLO results fail to provide 
a suitable description. On the other hand, the inclusion of NNLO
corrections through \minnlo{} achieves a substantial improvement
over \minlo{} results in terms of scale uncertainties for inclusive observables.
We also compared \minnlo{} predictions for the $Z\gamma$ transverse-momentum
spectrum against a more accurate analytically resummed calculation at 
NNLO+N$^3$LL and found a remarkable agreement down to transverse-momentum 
values close to the non-perturbative regime.
Finally, we have shown that \minnlo{} predictions are in excellent agreement with 
the latest ATLAS 13\,TeV data, with various improvements over lower-order
simulations.

This calculation paves the way for NNLO+PS predictions for all diboson processes
in the future. Moreover, the contribution 
from the loop-induced gluon fusion component, despite being rather small for $Z\gamma$
production, could be calculated separately at (N)LO+PS and added to our 
predictions. A rather straightforward advancement would also be to consider 
$Z\to \nu\bar\nu$ decays in future, which will be highly relevant also for 
dark-matter searches. In this context, also suitable modifications 
of the SM could be considered, for instance by introducing effective 
 $Z^\star Z\gamma$ and $\gamma^\star \gamma Z$ couplings.

Finally, we reckon that the presented results as well as our \minnlo{} generator 
for $Z\gamma$ production will be a useful advancement over previous Monte Carlo 
predictions and tools. 
Especially since the \minnlo{} generator can be directly applied in 
experimental analyses of $Z\gamma$ production or searches, 
the improved theoretical predictions in terms 
of accuracy and precision might stimulate further studies 
of this process in future.

\section*{Acknowledgements}

We are indebted to Pier Monni, Paolo Nason, Emanuele Re for several 
fruitful discussions. 
We are grateful to Carlo Oleari for his suggestion to move the QED singularities 
into the remnant contribution via a suitable damping factor.
We thank John Cambell for clarifications regarding the implementation of 
top- and bottom-mass effects in the MCFM amplitudes.

\clearpage

\appendix

\section{Generation cuts and suppression factors}
\label{app:cuts}

Since our Born process involves a number of QED and QCD singularities,
we make use of Born and remnant suppression factors to sample the
phase space. Additionally, we introduce a number of small technical
cuts in the phase space generation. In this Appendix we give all
details about the generation cuts and suppression factors that we have
used to obtain the results presented in this paper. 

We start by outlining the generation cuts that we employ.  First, we
introduce a lower cut $\ptgcut{}=5$\,GeV on the photon transverse
momentum, which is required to avoid QED singularities related to
collinear photon emissions from the initial state. We also impose a
similar cut of $\ptjcut=1$\,GeV on the transverse momentum of the
outgoing QCD parton. 
A lower cut $m_{\ell\ell}^{\rm cut} = 40$\,GeV on the invariant mass of
the lepton pair is imposed to avoid singular configurations in
$\gamma^* \to \ell\ell$ splittings.  Note that, since the invariant
mass of the resonances are preserved when radiation is generated, any
value $m_{\ell\ell}^{\rm cut}$ equal or below the cut used in the
analysis is allowed.  Furthermore, we require that the photon is
isolated from leptons and QCD partons in the final state. For this
purpose, we introduce a cut $m^2_{l \gamma} = 0.1$ GeV$^2$, where
$l=\{e^+,e^-\}$ and we introduce a smooth isolation as in
\eqn{eq:frixione} with $E_T^{\rm ref} = \epsilon_{\gamma} \ptg$ with $\delta_{\text{\scalefont{0.77}$0$}} = 0.05$, $\epsilon_\gamma = 0.5$ and $n=1$.
All these generation cuts can be modified via the input card, provided
their values is much smaller than the values used in the calculation of
the fiducial cross sections.

The Born suppression factor that we adopt is constructed in factorized form
\begin{equation}
  B_{\rm supp} = F_{\rm supp}(\ptg)\cdot
  G_{\rm supp}(\Delta R_{\gamma,l^+})\cdot
  G_{\rm supp}(\Delta R_{\gamma,l^-})\cdot
  H_{\rm supp}(\Delta R_{\gamma,j})  \,,
\label{eq:Bsupp}
\end{equation}  
with
\begin{equation}
F_{\rm supp}(\ptg) = \frac{(\ptg)^2}{(\ptg)^2+(\ptgcut)^2}\,,\quad {\rm with}\quad \ptgcut = 10\,{\rm GeV}\,,  
\end{equation}  
\begin{equation}
G_{\rm supp}(\Delta R) = \frac{(\Delta R)^2}{(\Delta R)^2 +(\Delta R_{\rm min})^2}\,,\quad {\rm with}\quad \Delta R_{\rm min}  = 0.5 \,,  
\end{equation}  
and
\begin{equation}
H_{\rm supp}(\Delta R) = \frac{(\Delta R)^2}{(\Delta R)^2 +(\Delta R_{\rm min})^2}\,,\quad {\rm with}\quad \Delta R_{\rm min}  = 0.2 \,. 
\end{equation}
Since we apply a \minlosimple{} procedure, we do not need any
suppression related to the Born outgoing parton.  It is clear that,
whenever a singularity is approched, the Born suppression factor in
\eqn{eq:Bsupp} will vanish in such a way that the cross section
times Born suppression factor itself will remain finite.

As discussed in \sct{sec:photon}, we have a remnant contribution
which is QCD regular but QED singular. Accordingly, we introduce a
remnant suppression factor of the form
\begin{align}
  R_{\rm supp} &= F_{\rm supp}(\ptg)\cdot
  G_{\rm supp}(\Delta R_{\gamma,l^+})\cdot
  G_{\rm supp}(\Delta R_{\gamma,l^-})\cdot
  H_{\rm supp}(\Delta R_{\gamma,j}) \\& \cdot
  H_{\rm supp}(\Delta R_{\gamma,j2})  \cdot
  H_{\rm supp}(\Delta R_{j1,j2})  \cdot
  L_{\rm supp}(p_{\rm t, j2})  \cdot      
  \,,
\end{align}  
with
\begin{equation}
L_{\rm supp}(p_{\rm t, j2}) = \frac{(p_{\rm t, j2})^2}{(p_{\rm t, j2})^2+(\ptjcut)^2}\,,\quad {\rm with}\quad \ptjcut = 20\,{\rm GeV}\,.   
\end{equation}  

As usual in \POWHEG{}, the Born suppression is evaluated using the Born
kinematics, while the remnant suppression is evaluated using the real
radiation partonic kinematics.
We note that in our case it is not necessary to introduce the
additional factor $L_{\rm supp}(p_{\rm t, j2})$, however we found that
results converged more quickly with the introduction of this
additional suppression factor.

\section{Projection to the $Z\gamma$ underlying Born configuration}
\label{app:UUB}
The evaluation of the last term in Eq.~\eqref{eq:Bbar2} requires a
projection from the $Z\gamma$+jet to $Z\gamma$ kinematics. Here we
give details about this projection and comment on configurations,
which, after projection, have $\ptg$ close to zero.

We denote by $p_1$ and $p_2$ the two incoming momenta, and
by $p_{\gamma}$, $p_Z$ and $p_j$ the momenta of the photon, $Z$ boson
and jet in the final state.  We define $p_{\rm tot} = p_1+p_2-p_j=p_\gamma+p_Z$.
Our projection to the underlying Born configuration consists of a longitudinal boost(by $\beta_L$),
such that, after boosting, $p_{\rm tot}$ has no $z$ component. Then, a second
boost (by $\vec \beta_T$) in the transverse plane, such that $p_{\rm tot}$ after both
boosts has no transverse component, is applied, followed by a final boost back in the
longitudinal direction (by $-\beta_L$).
We add a prime to all quantities after the first longitudinal boost and a
double prime to those after the second one.
The boost vector of the transverse boost is then given by 
\begin{align}
\vec \beta_T = \frac{\vec p_{T, j}}{E_{\rm tot}'}\,, 
\label{eq:betat1}
\end{align}
After the second boost, the transverse momentum of the photon becomes
\begin{align}
\vec p''_{T, \gamma} = \vec p_{T, \gamma} \gamma_T \left(\vec p_{T, \gamma} + \vec \beta_T E'_{\gamma}\right)\,, \quad \mbox{with} \quad \gamma_T=\frac{1}{\sqrt{1-\beta_T^2}}\,. 
\end{align}
Therefore, after this boost, the condition  $\vec p''_{T, \gamma}=0$ is met if
\begin{align}
\vec \beta_T = -\frac{\vec p_{T, \gamma}}{E_{\gamma}'}\,. 
\label{eq:betat2}
\end{align}
By comparing Eq.~\eqref{eq:betat1} and~\eqref{eq:betat2} we see that
$\vec p_{T, \gamma}$ and $\vec p_{T, j}$ must be
anti-aligned. Furthermore, since $E_{\rm tot}' = E_{\gamma}'+E_{Z}' >
E_{\gamma}'$ we have that $p_{T, j} > p_{T, \gamma}$. Accordingly, any
boost that leads to a vanishing transverse momentum of the photon in the
underlying Born configuration has a jet that is harder than $p_{T, \gamma}$ in the
${\rm Z\gamma J}$ configuration. Since for the photon we impose a
transverse momemtum cut, this region is free of any large logarithm,
and corrrections such as $\mathcal{D}_{\flavB}(\pt)$ in
\neqn{eq:Dnew} and \eqref{eq:Bbar} are beyond our accuracy and can be dropped. 

\newpage
\addcontentsline{toc}{section}{References}
\bibliography{MiNNLO}
\bibliographystyle{JHEP}

\end{document}